\newif\ifbw\bwfalse
\newif\ifacm\acmfalse
\newif\ifdeletednew\deletednewfalse
\newif\ifuseeps\useepstrue
\makeatletter
\@ifundefined{bwtrue}{\newif\ifbw\bwfalse}{}
\@ifundefined{acmtrue}{\newif\ifacm\acmtrue}{}
\makeatother
\ifacm
\documentclass[prodmode,acmtoplas]{acmsmall} 
\else
\documentclass[prodmode]{noacm}
\fi
\makeatletter
\let\@period=\,
\makeatother

\usepackage{xspace}
\usepackage{color}
\usepackage[leqno]{amsmath}
\usepackage{amssymb}
\usepackage{tikz}
\usepackage{array}
\usepackage{tabularx}
\usepackage{url}
\usepackage{hyperref}
\usepackage{breakurl}
\usepackage{wrapfig}
\usepackage{graphicx}
\usepackage{subfigure}
\usepackage{epsfig}

\usepackage{listings}
\newif\ifcomments
\commentsfalse
\usepackage{url}
\usepackage{mathpartir}
\makeatletter
\def\url@leostyle{%
  \@ifundefined{selectfont}{\def\UrlFont{\small\sf}}{\def\UrlFont{\small\sf}}}
\makeatother
\usepackage{xspace}

\newcommand{\myfig}{Fig.~}
\newcommand{\mytab}{Tab.~}
\newcommand{\mythm}{Thm.~}
\newcommand{\mylem}{Lem.~}

\newcommand{\mysec}{Sec.~}

\newcommand{\wrt}{w.r.t.\xspace}
\newcommand{\eg}{e.g.\xspace}
\newcommand{\ie}{i.e.\xspace}
\newcommand{\etc}{etc\xspace}
\newcommand{\cf}{see\xspace}

\makeatletter
\@ifundefined{deletednewtrue}{\newif\ifdeletednew\deletednewtrue}{}
\makeatother
\ifdeletednew
\let\innew\relax
\def\InNew{IN}
\newcommand{\NEW}[1]
{\ifx\innew\relax\textcolor{blue}{\let\innew\InNew{}#1}\else
\textcolor{red}{#1}\fi}

\newenvironment{new}{\color{blue}}{}
\newcommand{\UPDATENOTE}[1]{\textcolor{blue}{(#1)}}
\usepackage[normalem]{ulem}
\newcommand{\DELETED}[1]{\sout{#1}}
\else
\newcommand{\NEW}[1]{#1}

\newenvironment{new}{}{}
\newcommand{\UPDATENOTE}[1]{}
\newcommand{\DELETED}[1]{}
\fi

\let\mo\mathord

\newcommand{\rstar}[1]{\mathop{{#1}^{*}}}
\newcommand{\transc}[1]{\mathop{{#1}^{+}}}

  \newcommand{\dfn}[2]{{#1}~\triangleq~{#2}}
  
  \newcommand{\e}{e}

  \let\setset\setbf

  \newcommand{\evts}{\setset{E}}

  \newcommand{\mfence}{\textsf{mfence}}
  \newcommand{\cfence}{\textsf{cfence}}
  \newcommand{\lwf}{\textsf{lwf}}
  \newcommand{\lwfence}{\textsf{lwfence}}
  \newcommand{\ff}{\textsf{ff}}
  \newcommand{\ffence}{\textsf{ffence}}
  \newcommand{\fences}{\textsf{fences}}
  \newcommand{\cumul}{\textsf{cumul}}
  \newcommand{\Acumul}{\textsf{A-cumul}}

  \newcommand{\pr}{\operatorname{proc}}

  \newcommand{\lo}{\ell}
  \newcommand{\loc}{\operatorname{addr}}

  \newcommand{\val}{\operatorname{val}}

  \newcommand{\init}{\operatorname{\textsf{i}}}
  \newcommand{\commit}{\operatorname{\textsf{c}}}

\newcommand{\stacklabel}[1]
{\stackrel{\smash{\scriptstyle\textnormal{#1}}}}
  \newcommand{\rln}[1]{\ensuremath{\xrightarrow{#1}}}

  \newcommand{\iico}{\textsf{iico}}
  
  \newcommand{\po}{\textsf{po}}
  \newcommand{\modord}{\textsf{mo}}
  \newcommand{\syncbef}{\textsf{sb}}

  \newcommand{\iiz}{\textsf{ii}_0}
  \newcommand{\icz}{\textsf{ic}_0}
  \newcommand{\ciz}{\textsf{ci}_0}
  \newcommand{\ccz}{\textsf{cc}_0}
  \newcommand{\ii}{\textsf{ii}}
  \newcommand{\ic}{\textsf{ic}}
  \newcommand{\ci}{\textsf{ci}}
  \newcommand{\cc}{\textsf{cc}}
  \newcommand{\detour}{\textsf{detour}}
  \newcommand{\rdw}{\textsf{rdw}}
  \newcommand{\addr}{\textsf{addr}}
  \newcommand{\data}{\textsf{data}}
  \newcommand{\ctrl}{\textsf{ctrl}}
  \newcommand{\ctrlcfence}{\textsf{\DELETED{(}ctrl+cfence\DELETED{)}}}
  \newcommand{\dep}{\textsf{dp}}

  \newcommand{\rf}{\textsf{rf}}
  \newcommand{\rfreg}{\textsf{rf-reg}}

  \newcommand{\rfe}{\textsf{rfe}}

  \newcommand{\rfi}{\textsf{rfi}}

  \newcommand{\co}{\textsf{co}}
  \newcommand{\coi}{\textsf{coi}}
  \newcommand{\ws}{\textsf{co}}

  \newcommand{\ews}{\textsf{coe}}

  \newcommand{\fr}{\textsf{fr}}
  \newcommand{\ifr}{\textsf{fri}}
  \newcommand{\efr}{\textsf{fre}}

  \newcommand{\com}{\textsf{com}}
  \newcommand{\hb}{\textsf{hb}}

  \newcommand{\ppo}{\textsf{ppo}}

  \newcommand{\polocllh}{\operatorname{\textsf{po-loc-llh}}}
  \newcommand{\poloc}{\operatorname{\textsf{po-loc}}}

  \newcommand{\ddreg}{\textsf{dd-reg}}
  \newcommand{\dmb}{\textsf{dmb}}
  \newcommand{\dsb}{\textsf{dsb}}
  \newcommand{\dmbst}{\textsf{dmb.st}}
  \newcommand{\dsbst}{\textsf{dsb.st}}
  \newcommand{\isb}{\textsf{isb}}
  \newcommand{\isync}{\textsf{isync}}
  
  \newcommand{\acyclic}{\operatorname{acyclic}}
  \newcommand{\reflexive}{\operatorname{reflexive}}
  \newcommand{\irrefl}{\operatorname{irreflexive}}

  \newcommand{\udr}{\operatorname{\textsf{udr}}}

    \newcommand{\cmp}{\operatorname{\textsf{cmp}}}

  \newcommand{\WW}{\operatorname{\textsf{{WW}}}}
  
  \newcommand{\RM}{\operatorname{\textsf{RM}}}
  \newcommand{\RB}{\operatorname{\textsf{RB}}}
  
  \newcommand{\RW}{\operatorname{\textsf{RW}}}
  
  \newcommand{\RR}{\operatorname{\textsf{RR}}}

  \newcommand{\WR}{\operatorname{\textsf{WR}}}

\newcommand{\import}{\operatorname{import}}

\let\prog\textsf
\let\as\texttt

\makeatletter
\@ifundefined{useepstrue}{\newif\ifuseeps\useepsfalse}{}
\makeatother
\newlength{\fmtlength}
\ifuseeps

\else

\fi
\ifuseeps

\else

\fi
\newcommand{\sync}{\textsf{sync}}
\newcommand{\lwsync}{\textsf{lwsync}}
\newcommand{\eieio}{\textsf{eieio}}

\newcommand{\lb}[4]{\((#1)\)\,#2#3#4}

\newcommand{\Delw}{\operatorname{c}}
\newcommand{\Delr}{\operatorname{s}}
\newcommand{\Flushw}{\operatorname{cp}}
\newcommand{\Flushr}{\operatorname{c}}

\newcommand{\buff}{\operatorname{\textsf{cw}}} 
\newcommand{\rcp}{\operatorname{\textsf{cpw}}} 
\newcommand{\queue}{\operatorname{\textsf{sr}}} 
\newcommand{\cread}{\operatorname{\textsf{cr}}} 

\newcommand{\visible}{\operatorname{visible}}
\newcommand{\propbase}{\operatorname{\textsf{prop-base}}}

\newcommand{\prop}{\operatorname{\textsf{prop}}}
\newcommand{\pltol}{\operatorname{\textsf{pl2l}}}

\newcommand{\pstos}{\operatorname{\textsf{ps2s}}}
\iftrue
\makeatletter
\gdef\SetFigFontNFSS#1#2#3#4#5{%
  \reset@font\fontsize{#1}{#2pt}%
  \fontfamily{\sfdefault}\fontseries{#4}\fontshape{#5}%
  \selectfont}%
\gdef\SetFigFont#1#2#3#4#5{%
  \reset@font\fontsize{#1}{#2pt}%
  \fontfamily{\sfdefault}\fontseries{#4}\fontshape{#5}%
  \selectfont}
\makeatother

\newcommand{\myth}[1]{\textsf{T}$_#1$}

\ifbw\newcommand{\bw}{-bw}\else\newcommand{\bw}{}\fi
\ifuseeps

\else

\fi
\fi
\newcommand{\cwcoww}{(\textsc{cw: \DELETED{uniproc}\NEW{sc per location}/}\textsf{coWW})}
\newcommand{\cwprop}{(\textsc{cw: propagation})}
\newcommand{\cwciwr}{(\textsc{cw:} $\fences \cap \WR$)}

\newcommand{\cpwcom}{(\textsc{cpw: write is committed})}
\newcommand{\cpwaccord}{(\textsc{cpw: }\DELETED{$\buff$}\NEW{$\poloc$} \textsc{and} $\rcp$ \textsc{are in accord})}
\newcommand{\cpwprop}{(\textsc{cpw: propagation})}
\newcommand{\srokw}{(\textsc{sr: write is either local or committed})}
\newcommand{\sriirr}{(\textsc{sr: ppo/}$\iiz \cap \RR$)}
\newcommand{\srcaus}{(\textsc{sr: \DELETED{causality}\NEW{observation}})}
\newcommand{\crsat}{(\textsc{cr: read is satisfied})}
\newcommand{\cruni}{(\textsc{cr: \DELETED{uniproc}\NEW{sc per location}/} \textsf{coWR, coRW\{1,2\}, coRR})}

\newcommand{\crci}{(\textsc{cr: ppo/}${(\ciz \cup \ccz)} \cap {\RR}$)}
\newcommand{\crccrw}{(\textsc{cr: ppo/}$\ccz \cap \RW$)}

\newcommand{\noop}{\textsf{noop} }

\newcommand{\fixmeskip}[1]{}

\newcolumntype{Y}{@{}r@{\,}X}
\newcommand{\instab}[2]{\ \(#2\)  & \as{#1}}

\newcommand{\pset}[2]{\(\as{#1} \leftarrow \as{#2}\)}
\newcommand{\pstore}[2]{\pset{#2}{#1}}
\newcommand{\pload}[2]{\pset{#1}{#2}}

\newcommand{\plwarx}[3]{\as{lwarx #1,#2,#3}}
\newcommand{\pstwcx}[3]{\as{stwcx. #1,#2,#3}}
\newcommand{\pstw}[3]{\as{stw #1,#2,#3}}
\newcommand{\pbne}[1]{\as{bne #1}}
\newcommand{\pcmp}[2]{\as{cmpw #1,#2}}
\newcommand{\haut}{\rule{0ex}{2ex}}
\newcommand{\bas}{\rule[-1ex]{0.5ex}{0ex}}

\newcommand{\Rmw}[1][40ex]
{\begin{tabular}{rl}
\haut \instab{\as{loop:}}{} \\
\instab{\plwarx{r1}{0}{r5}}{(a_1)} \\
\instab{[\dots]}{}\\
\instab{\pstwcx{r2}{0}{r5}}{(a_2)} \\
\bas\instab{\pbne{loop}}{(b)} \\ 
\end{tabular}}
\newcommand{\Atom}[1][40ex]
{\begin{tabular}{rl}
\haut \instab{\plwarx{r1}{0}{r5}}{(a_1)} \\
\instab{[\dots]}{}\\
\bas\instab{\pstwcx{r2}{0}{r5}}{(a_2)} \\
\end{tabular}}
\newcommand{\Lo}[1][40ex]
{\begin{tabular}{rl}
\haut\instab{\as{loop:}}{} \\
\instab{\plwarx{r6}{0}{r3}}{(a_1)} \\
\instab{\pcmp{r4}{r6}}{(b)} \\
\instab{\pbne{loop}}{(c)} \\
\instab{\pstwcx{r5}{0}{r3}}{(a_2)} \\
\instab{\pbne{loop}}{(d)} \\
\instab{\as{isync}}{(e)} \\
\bas\instab{[\dots]}{}\\ 
\end{tabular}}
\newcommand{\ULo}[1][40ex]
{\begin{tabular}{rl}
\instab{[\dots]}{} \\
\haut\instab{\as{lwsync}}{(f)} \\ 
\bas\instab{\pstw{r4}{0}{r3}}{(g)} \\
\end{tabular}}

\newcommand{\timepldipower}{14922996}
\newcommand{\npldipower}{4704}

\newcommand{\timemodelpower}{321}
\newcommand{\nmodelpower}{8117}

\newcommand{\timecavpower}{2846}
\newcommand{\ncavpower}{8117}
\newcommand{\npower}{8117}
\newcommand{\nmodelpowerunseen}{1182}
\newcommand{\nmodelpowerinvalid}{0}

\newcommand{\narm}{9761}
\newcommand{\nmodelarmunseen}{1820}
\newcommand{\nmodelarminvalid}{1502}

\newcommand{\narmreluctant}{33}
\newcommand{\nrelaxedinvalid}{33}

\bgroup

\egroup
\bgroup

\global\let\acerelaxedSinvalid\aceSinvalid

\global\let\acerelaxedTinvalid\aceTinvalid

\global\let\acerelaxedPinvalid\acePinvalid

\global\let\acerelaxedSTinvalid\aceSTinvalid

\global\let\acerelaxedSCinvalid\aceSCinvalid

\global\let\acerelaxedSPinvalid\aceSPinvalid

\global\let\acerelaxedCPinvalid\aceCPinvalid

\global\let\acerelaxedSTCinvalid\aceSTCinvalid

\global\let\acerelaxedSCPinvalid\aceSCPinvalid

\global\let\nrelaxedinvalid\nALLinvalid
\global\let\acerelaxedinvalid\aceALLinvalid
\egroup
\bgroup

\global\let\acemodelarmSinvalid\aceSinvalid

\global\let\acemodelarmTinvalid\aceTinvalid

\global\let\acemodelarmPinvalid\acePinvalid

\global\let\acemodelarmSTinvalid\aceSTinvalid

\global\let\acemodelarmSCinvalid\aceSCinvalid

\global\let\acemodelarmSPinvalid\aceSPinvalid

\global\let\acemodelarmCPinvalid\aceCPinvalid

\global\let\acemodelarmSTCinvalid\aceSTCinvalid

\global\let\acemodelarmSCPinvalid\aceSCPinvalid

\global\let\nmodelarminvalid\nALLinvalid
\global\let\acemodelarminvalid\aceALLinvalid
\egroup
\newcommand{\nnodetourpower}{24}

\newcommand{\nnodetourarm}{8}

  \renewenvironment{thebibliography}[1]{%
    \begin{oldthebibliography}{#1}%
      \setlength{\parskip}{0ex}%
      \setlength{\itemsep}{0ex}%
      \small
  }%
  {%
    \end{oldthebibliography}%
  }

\mathchardef\ordinarycolon\mathcode`\:
\mathcode`\:=\string"8000
\begingroup \catcode`\:=\active
\gdef:{\mathrel{\mathop\ordinarycolon}}
\endgroup

\begin{document}
\ifacm This is \NEW{the revision of} a manuscript entitled \emph{Herding cats,
\NEW{Modelling, simulation, testing, and data-mining for weak memory}} destined
for the TOPLAS journal. The authors are Jade Alglave, Luc Maranget and Michael
Tautschnig. \NEW{We give the details of our modifications in the text file
``cats-response.txt''.}

We are interested in the special call associated with PLDI (deadline September
1, 2013 as indicated at \url{http://toplas.acm.org/index.html}). \NEW{We
submitted the revision on December 31, 2013.} 

The paper is best viewed in color; we will provide a black and white version
online at \url{http://diy.inria.fr/herd/}. 
\fi

\newpage

\markboth{J. Alglave et al.}{Herding cats}

\title{Herding cats \\
\NEW{Modelling, simulation, testing, and data-mining for weak memory}}
\author{Jade Alglave
\affil{University College London}
Luc Maranget
\affil{INRIA}
Michael Tautschnig
\affil{Queen Mary University of London}}

\begin{abstract}
We propose an axiomatic generic framework for modelling weak memory. We show
how to instantiate this framework for SC, TSO, C++ restricted to
release-acquire atomics, and Power.  For Power, we compare our model to a
preceding operational model in which we found a flaw. To do so, we define an
operational model that we show equivalent to our axiomatic model.

We also propose a model for ARM. Our testing on this architecture revealed a
behaviour later acknowledged as a bug by ARM, and more recently
$\nrelaxedinvalid$ additional anomalies.

We offer a new simulation tool, called \prog{herd}, which allows the user to
specify the model of his choice in a concise way. Given a specification of a
model, the tool becomes a simulator for that model. The tool relies on an
axiomatic description; this choice allows us to outperform all previous
simulation tools. Additionally, we confirm that verification time is vastly
improved, in the case of bounded model-checking.

Finally, we put our models in perspective, in the light of empirical data
obtained by analysing the C and C++ code of a Debian Linux distribution. We
present our new analysis tool, called \prog{mole}, which explores a piece of
code to find the weak memory idioms that it uses.

\end{abstract}




\acmformat{Jade Alglave, Luc Maranget, and Michael Tautschnig, 2013. Herding cats.}

%

\maketitle

\section{Introduction}

There is a joke where a physicist and a mathematician are asked to herd cats.
The physicist starts with an infinitely large pen which he reduces until it is
of reasonable diameter yet contains all the cats. The mathematician builds a
fence around himself and declares the outside to be the inside. Defining memory
models is akin to herding cats: both the physicist's or mathematician's
attitudes are tempting. 

Recent years have seen many formalisations of memory models
emerge; see for
example~\cite{aas03,am06,ba08,ci08,bp09,ssz09,afi09,ams10,bos11,ssa11,ams12,smo12,alg12,mms12,bps12}.
Yet we feel the need for more work in the area of \emph{defining} models.
There are several reasons for this.

\paragraph{On the hardware side, \!\!\!} all existing models of Power (some of
which we list in~\myfig\ref{fig:power-models}) \DELETED{are broken in one way
or another}\NEW{have some flaws} (\cf~\mysec\ref{sec:rw}). This calls for
reinvestigating the model, for the sake of repairing it of course, but for
several other reasons too\NEW{, that we explain below}. 

One important reason is that Power underpins C++'s \DELETED{so-called}
\emph{atomic} concurrency features~\cite{ba08,bos11,smo12}: implementability on
Power has had a major influence on the design of the C++ model. Thus modelling
flaws in Power \DELETED{transmit to}\NEW{could affect} C++.  

\DELETED{In addition,}\NEW{Another important reason is that} a good fraction of
the code in the wild (see our experiments in~\mysec\ref{sec:search} on
\DELETED{the} release 7.1 of the Debian Linux distribution) still does not use
the C++ atomics. Thus, we believe that programmers have to rely on what the
hardware does, which requires descriptive models of the hardware.

\paragraph{On the software side, \!\!\!\!} recent work shows that the C++ model
allows behaviours that break modular reasoning (\cf the \emph{satisfaction
cycles} issue in~\cite{bdg13}), whereas Power does not, since it prevents
\emph{out of thin air} values (\cf\mysec\ref{sec:model}). Moreover, C++
requires the irreflexivity of a certain relation (\cf the \textsc{HBvsMO} axiom
in~\cite{bdg13}), whereas Power offers a stronger acyclicity guarantee, as we
show in this paper.

\NEW{Ideally,} we believe that these models would benefit from stating
\emph{principles} that underpin weak memory as a whole, not just one particular
architecture or language. Not only would it be aesthetically pleasing, but it
would allow more informed decisions on the design of high-level memory models,
ease the conception and proofs of compilation schemes, and allow the
reusability of simulation and verification techniques from one model to
another.

Models roughly fall into two classes: \emph{operational} and \emph{axiomatic}.
Operational models, \eg the Power model of~\cite{ssa11}, are abstractions of
actual machines, composed of idealised hardware components such as buffers and
queues. Axiomatic models, \eg the C++ model of~\cite{bos11}, distinguish
allowed from forbidden behaviours, usually by constraining various relations on
memory accesses.   

We now list a few criteria that we believe our models should meet; we do not
claim to be exhaustive\NEW{, nor do we claim that the present work fully meets
all of them, although we discuss in the conclusion of this paper to what extent
it does. Rather, we see this list as enunciating some wishes for works on weak
memory (including ours of course), and more generally weak consistency, as can
be found for example in distributed systems.}

\medskip
\centerline{***}

\paragraph{\textsf{Stylistic proximity of models, \!\!\!\!}} whether hardware (\eg x86,
Power or ARM) or software (\eg C++), would permit the statement of general
principles spanning several models of weak memory. It should be easier to find
principles common to Power and C++, amongst others, if their respective models
were described in the same terms.  

\paragraph{\textsf{Rigour \!\!\!}} is not our only criterion: for example, all the
recent Power models enjoy a considerable amount of rigour, yet are still
somewhat flawed.

\paragraph{\textsf{Concision \!\!\!}} of the model seems crucial to us: we
want to specify a model concisely, to grasp it and modify it rapidly, without
needing to dive into a staggering number of definitions.

One could tend towards \emph{photorealistic} models and account for each and
every detail of the machine. \NEW{We find that operational models in general
have such traits, although some do more than others.} \DELETED{To some extent,
this characterises}\NEW{For example, we find that} the work of Sarkar et
al.~\cite{ssa11,smo12} \NEW{is too close to the hardware, and, perhaps
paradoxically, too precise, to be easily amenable to pedagogy, automated
reasoning and verification}.  \DELETED{Admirable as such a \emph{tour de force}
can be,}\NEW{Although we do recognise the effort and the value of this work,
without which we would not have been able to build the present work,} we
believe that we need models \NEW{that are more} \emph{rationally descriptive}
(as coined by Richard Bornat). \NEW{We go back to the meaning of ``rational''
at the end of the introduction.}

\paragraph{\textsf{Efficient simulation and verification \!\!\!\!}} have been
relatively neglected by previous modelling work, except
for~\cite{mam10,akn13,akt13}. These works show that simulation~\cite{mam10} and
verification~\cite{akt13} (for bounded model-checking) can be orders of
magnitude faster when it relies on axiomatic models rather than operational
ones. 

Yet operational models are often considered more intuitive than axiomatic
models. Perhaps the only tenable solution to this dilemma is to propose both
styles in tandem, and show their equivalence. As exposed in~\cite{hl74},
\emph{``a single formal definition is unlikely to be equally acceptable to both
implementor and user, and [\dots] at least two definitions are required, a
constructive one [\dots] for the implementor, and an
implicit one for the user [\dots].''}

\paragraph{\textsf{Soundness \wrt hardware \!\!\!}} is mandatory regardless of
the modelling style. Ideally, one would prove the soundness of a model \wrt the
formal description of the hardware, \eg at the RTL level~\cite{gor02}. However,
we do not have access to these data because of commercial confidentiality.
\DELETED{thus we resort to testing the hardware itself, an approach which has
already been adopted in the past (see \eg~\cite{ams12,ssa11,mms12}).} \NEW{To
overcome this issue, some previous work has involved experimental testing of
hardware (see \eg\cite{col92,ssz09,ams12,ssa11,mms12}), with increasing
thoroughness over time}.  \footnote{Throughout this paper, we refer to online
material to justify our experimental claims; the reader should take these as
bibliography items, and refer to them when details are needed.}

A credible model cannot forbid behaviours exhibited on hardware, unless the
hardware itself is flawed. Thus models should be extensively tested against
hardware, and retested regularly: this is how we found flaws in the model
of~\cite{ssa11} (\cf~\mysec\ref{sec:rw}). Yet, we find the experimental flaws
themselves to be less of an issue than the fact that the model does not seem to
be easily fixable.

\paragraph{\textsf{Adaptability \!\!\!}} of the model, \ie setting the model in
a flexible formalism, seems crucial, if we want stable models.  By stable we
mean that even though we might need to change parameters to account for an
experimental flaw, the general shape of the model, its principles, should not
need to change. 

Testing (as extensive as it may be) cannot be sufficient, since it cannot
guarantee that an unobserved behaviour might not be triggered in the future.
Thus one needs some guarantee of \emph{completeness} of the model.

\paragraph{\textsf{Being in accord with the architectural intent \!\!\!\!}} might
give some guarantee of completeness. We should try to create models that
respect or take inspiration from the architects' intents.  This is one of the
great strengths of the model of~\cite{ssa11}. However, this cannot be the only
criterion, as the experimental flaw in~\cite{ssa11} shows.  Indeed the
architects' intents might not be as formal as one might need in a model, two
intents might be contradictory, or an architect might not realise all the
consequences of a given design. 

\paragraph{\textsf{Accounting for what programmers do \!\!\!\!}} seems a sensible
criterion. One cannot derive a model from programming patterns, since some of
these patterns might rely on erroneous understanding of the hardware. Yet to
some extent, these patterns should reflect part of the architectural intent,
since systems programmers or compiler writers communicate relatively closely
with hardware designers. 

Crucially, we have access to open-source code, as opposed to the chips'
designs. Thus we can analyse the code, and derive some common programming
patterns from it.

\medskip
\centerline{***}

\paragraph{Rational models \!\!\!\!} is what we advocate here. We believe that
a model should allow a \emph{rational explanation} of what programmers can rely
on. We believe that by balancing all the criteria above, one can provide such a
model. This is what we set out to do in this paper.
 
\NEW{By rational we mean the following: we think that we, as architects,
semanticists, programmers, compiler writers, are to understand concurrent
programs. Moreover we find that to do so, we are to understand some particular
patterns (\eg the message passing pattern given in \myfig\ref{fig:mp}, the very
classical store buffering pattern given in \myfig\ref{fig:sb}, or the
controversial load buffering pattern given in \myfig\ref{fig:lb}). We believe
that by being able to explain a handful of patterns, one should be able to
generalise the explanation and thus to understand a great deal of weak memory.}

\NEW{To make this claim formal and precise, we propose a generic model of weak
memory, in axiomatic style. Each of our four axioms has a few canonical
examples, that should be enough to understand the full generality of the axiom.
For example, we believe that our \textsc{no thin air} axiom is fully explained
by the load buffering pattern of \myfig\ref{fig:lb}. Similarly our
\textsc{observation} axiom is fully explained by the message passing, write to
read causality and Power ISA2 patterns of \myfig\ref{fig:mp},\ref{fig:wrc} and
\ref{fig:isa2} respectively.}

\NEW{On the modelling front, our main stylistic choices and contributions are
as follows: to model the propagation of a given store instruction to several
different threads, we use only one memory event per instruction (see
\mysec\ref{sec:model}), instead of several subevents (one per thread for
example, as one would do in Itanium~\cite{intel:itanium} or in the Power model
of~\cite{mms12}). We observe that this choice makes simulation much faster (see
\mysec\ref{sec:testing}).}

\NEW{To account for the complexity of write propagation, we introduce the novel
notion of \emph{propagation order}. This notion is instrumental in describing
the semantics of fences for instance, and the subtle interplay between fences
and coherence (see \mysec\ref{sec:model}).} 

\NEW{We deliberately try to keep our models concise, as we aim at describing
them as simple text files that one can use as input to an automated tool (\eg{}
a simulation tool, or a verification tool). We note that we are able to
describe IBM Power in less than a page (see \myfig\ref{fig:herd-ppc-model}).}
 
\paragraph{Outline:} we present related works in~\mysec\ref{sec:rw}.  We
describe our new generic model of weak memory in~\mysec\ref{sec:model}, and
show how to instantiate it to describe Sequential Consistency
(SC)~\cite{lam79}, Total Store Order (TSO) (used in Sparc~\cite{sparc:94} and
x86's~\cite{oss09} architectures) and C++ restricted to release-acquire
atomics.

\mysec\ref{sec:instr-sem} presents examples of the semantics of instructions,
which are necessary to understand~\mysec\ref{sec:power}, where we explain how
to instantiate our model to describe Power and ARM\NEW{v7}. We compare formally
our Power model to the one of~\cite{ssa11} in~\mysec\ref{sec:pldi}. To do so,
we define an operational model that we show equivalent to our axiomatic model.

We then present our experiments on Power and ARM hardware
in~\mysec\ref{sec:testing}, detailing the anomalies that we observed on ARM
hardware. We also describe our new \prog{herd} simulator, which allows the user
to specify the model of his choice in a concise way. Given a specification of a
model, the tool becomes a simulator for that model. 

Additionally, we demonstrate in the same section that our model is suited for
verification by implementing it in the bounded model-checker
\prog{CBMC}~\cite{ckl04} and comparing it with the previously implemented
models of~\cite{ams12} and \cite{mms12}.

In~\mysec\ref{sec:search}, we present our \prog{mole} analysis tool, which
explores a piece of code to find the weak memory behaviours that it contains.
We detail the data gathered by \prog{mole} by analysing the C and C++ code in a
Debian Linux distribution. This gives us a pragmatic perspective on the models
that we present in \mysec\ref{sec:model}.  Additionally, \prog{mole} may be
used by programmers to identify areas of their code that may be (unwittingly)
affected by weak memory, or by static analysis tools to identify areas where
more fine-grained analysis may be required.

\paragraph{Online companion material:} we provide the source and documentation
of \prog{herd} at \url{http://diy.inria.fr/herd}. We provide all our
experimental reports \wrt hardware at \url{http://diy.inria.fr/cats}. We
provide our Coq scripts at \url{http://www0.cs.ucl.ac.uk/staff/j.alglave/cats}.
We provide the source and documentation of \prog{mole} at
\url{http://diy.inria.fr/mole}, as well as our experimental reports \wrt
\DELETED{the} release 7.1 of the Debian Linux distribution. 




\section{Related work}
\label{sec:rw}
\DELETED{As an echo to our introduction, we refer the reader to}\NEW{Our
introduction echoes} position papers by Burckhardt and
Musuvathi~\cite{bm08:ec2}, Zappa Nardelli et al.~\cite{zss09} and Adve and
Boehm~\cite{ab10,ba12}, which all formulate criteria, prescriptions or wishes
as to how memory models should be defined.

Looking for general principles of weak memory, one might look at the hardware
documentation: we cite Alpha~\cite{alpha:02}, ARM~\cite{arm:arm08},
Intel~\cite{intel:rev30}, Itanium~\cite{intel:itanium},
IBMPower~\cite{ppc:2.06} and Sun~\cite{sparcboth}. Ancestors of our
\DELETED{\textsc{uniproc}}\NEW{\textsc{sc per location}} and \textsc{no thin
air} axioms (\cf~\mysec\ref{sec:model}) appear notably in Sun and Alpha's
documentations.

We also refer the reader to work on modelling particular instances of weak
memory, \eg ARM~\cite{ci08}, TSO~\cite{bp09} or x86~\cite{ssz09,oss09},
C++~\cite{ba08,bos11}, or Java~\cite{mpa05,cks07}. We go back to the case of
Power at the end of this section.

\NEW{In the rest of this paper, we write TSO for Total Store Order, implemented
in Sparc TSO~\cite{sparc:94} and Intel x86~\cite{oss09}.  We write PSO for
Partial Store Order and RMO for Relaxed Memory Order, two other Sparc
architectures. We write Power for IBM Power~\cite{ppc:2.06}.  }

Collier~\cite{col92}, \NEW{Neiger~\cite{nei00}}, as well as Adve and
Gharachorloo~\cite{ag96} have provided general overviews of weak memory, but in
a less formal style than one might prefer. 

\NEW{Steinke and Nutt provide a unified framework to describe consistency
models. They choose to express their models in terms of the view order of each
processor, and describe instances of their framework, amongst them several
classical models such as PRAM~\cite{ls88} or Cache Consistency~\cite{goo89}.
Meyer et al.\fixmeskip{jade: add bib when their paper is accepted} later
showed how to instantiate Steinke and Nutt's framework to express TSO and PSO.
However, it is unclear to us whether the view order method \`a la Steinke and
Nutt is expressive enough to describe models without store atomicity, such as
Power and ARM, or indeed describe them easily.}

Rational models appear in Arvind and Maessen's work, aiming at weak memory in
general but applied only to TSO~\cite{am06}, and in Batty et al.'s~\cite{bdg13}
for C++. Interestingly, Burckhardt et al.'s work on distributed
systems~\cite{bgy13,bgy14} belongs to the same trend.

Some works on weak memory provide simulation tools: the \prog{ppcmem} tool
of~\cite{ssa11} and  Boudol et al.'s~\cite{bps12} implement their respective
operational model of Power, whilst the \prog{cppmem} tool of~\cite{bos11}
enumerates the axiomatic executions of the associated C++ model.  Mador-Haim et
al.'s tool~\cite{mms12} does the same for their axiomatic model of Power.
MemSAT has an emphasis towards the Java memory model~\cite{tvd10}, whilst Nemos
focusses on classical models such as SC or causal consistency~\cite{ygls04},
and TSOTool handles TSO~\cite{hvm04}. 

To some extent, decidability and verification
papers~\cite{gys04,bam07,abb10,bmm11,abp11,abb12,bdm13,kvy10,kvy11,lnp12,aac12,aac13}
do provide some general principles about weak memory, although we find them
less directly applicable to programming than semantics work. Unlike our work,
most of them are restricted to TSO, or its siblings PSO and RMO, or theoretical
models.  

Notable exceptions are~\cite{akn13,akt13} which use the generic model
of~\cite{ams12}. The present paper inherits some of the concepts
developed in~\cite{ams12}: it adopts the same style in describing executions of
programs, and pursues the same goal of defining a generic model of weak memory.
Moreover, we adapt the tool of~\cite{akt13} to our new models, and reuse the \prog{diy} testing tool
of~\cite{ams12} to conduct our experiments against hardware.

Yet we emphasise that the model that we present here is quite different from
the model of~\cite{ams12}, despite the stylistic similarities: in
particular~\cite{ams12} did not have a distinction between the
\textsc{\DELETED{causality}\NEW{observation}} and \textsc{propagation} axioms
(\cf\mysec\ref{sec:model}), which were somewhat merged into the \emph{global
happens-before} notion of~\cite{ams12}.


\begin{table}[!ht]
\begin{tabular}{p{.33\linewidth}|p{.15\linewidth}|p{.4\linewidth}}
model                           & style & comments \\\hline
\cite{aas03}         & axiomatic
& based on discussion with IBM architects; pre-cumulative barriers \\[.3em]
\cite{afi09}      & axiomatic
& based on documentation; practically untested \\[.3em]
\cite{ams12,am11} & single-event axiomatic
& based on extensive testing; semantics of $\lwsync$ stronger than~\cite{ssa11} on \textsf{r+lwsync+sync}, weaker on \textsf{mp+lwsync+addr} \\[.3em]
\cite{ssa11,smo12} & operational & based on discussion with IBM architects and extensive testing; flawed \wrt Power h/w on \eg \textsf{mp+lwsync+addr-po-detour} (\cf \myfig\ref{fig:mp-detour} and \url{http://diy.inria.fr/cats/pldi-power/\#lessvs}) and ARM h/w on \eg \textsf{mp+dmb+fri-rfi-ctrlisb} (\cf \url{http://diy.inria.fr/cats/pldi-arm/\#lessvs}) \\[.3em]
\cite{mms12}   & multi-event axiomatic & thought to be equivalent to~\cite{ssa11} but not experimentally on \eg \textsf{mp+lwsync+addr-po-detour} (\cf \url{http://diy.inria.fr/cats/cav-power}) \\[.3em]
\cite{bps12} & operational & semantics of $\lwsync$ stronger than~\cite{ssa11} on \eg \textsf{r+lwsync+sync}
\\[.3em]
\cite{akn13} & operational & equivalent to~\cite{ams12}
\end{tabular}
\caption{A decade of Power models in order of publication\label{fig:power-models}}
\end{table}

\paragraph{A decade of Power models \!\!\!\!} is presented in
\mytab\ref{fig:power-models}. Earlier work (omitted for brevity) accounted for
outdated versions of the architecture.  For example in 2003, Adir et
al.~described an axiomatic model~\cite{aas03}, \emph{``developed through
[\dots] discussions with the PowerPC architects''}, with outdated
\emph{non-cumulative barriers}, following the pre-PPC 1.09 PowerPC
architecture. 

Below, we refer to particular weak memory behaviours which serve as test cases
for distinguishing different memory architectures. These behaviours are
embodied by \DELETED{so-called} \emph{litmus tests}, with standardised names in
the style of~\cite{ssa11}, \eg \textsf{mp+lwsync+addr}. All these behaviours
will appear in the rest of the paper, so that the novice reader can refer to
them after a first read through.  We explain the naming convention
in~\mysec\ref{sec:model}.

In 2009, Alglave et al.~proposed an axiomatic model~\cite{afi09}, but this was
not compared to actual hardware. In 2010, they provided another axiomatic
model~\cite{ams10,ams12}, as part of a generic framework. This model is based
on extensive and systematic testing. It appears to be sound \wrt Power
hardware, but its semantics for $\lwsync$ cannot guarantee the
\textsf{mp+lwsync+addr} behaviour (\cf\myfig\ref{fig:mp}), and allows the
\textsf{r+lwsync+sync} behaviour (\cf\myfig\ref{fig:r}), both of which clearly
go against the architectural intent (\cf\cite{ssa11}).  This model (and the
generic framework to which it belongs) has a provably equivalent operational
counterpart~\cite{akn13}.

In 2011, Sarkar et al.~\cite{ssa11,smo12} proposed an operational model in
collaboration with an IBM designer, which might be taken to account for the
architectural intent. Yet, we found this model to forbid several behaviours
observed on Power hardware (\eg~\textsf{mp+lwsync+addr-po-detour}, \cf
\NEW{\myfig\ref{fig:mp-detour} and}
\url{http://diy.inria.fr/cats/pldi-power/\#lessvs}).  Moreover, although this
model was not presented as a model for ARM, it was thought to be a suitable
candidate. Yet, it forbids behaviours (\eg~\textsf{mp+dmb+fri-rfi-ctrlisb}, \cf
\NEW{\myfig\ref{fig:arm-feature}}
\url{http://diy.inria.fr/cats/pldi-arm/\#lessvs}) that are observable on ARM
machines\NEW{, and claimed to be desirable features by ARM designers}.

In 2012, Mador-Haim et al.~\cite{mms12} proposed an axiomatic model, thought to
be equivalent to the one of~\cite{ssa11}. Yet, this model does not forbid the
behaviour of \textsf{mp+lwsync+addr-po-detour} (\cf
\url{http://diy.inria.fr/cats/cav-power}), which is a counter-example to the
proof of equivalence appearing in~\cite{mms12}. The model of~\cite{mms12} also
suffers from the same experimental flaw \wrt ARM hardware as the model
of~\cite{ssa11}. 

More fundamentally, the model of~\cite{mms12} uses several write events to
represent \NEW{the propagation of} one memory store \NEW{to several different
threads}, which in effect mimics the operational transitions of the model
of~\cite{ssa11}. We refer to this style as \emph{multi-event axiomatic}, as
opposed to \emph{single-event axiomatic} (as in \eg~\cite{ams10,ams12})\NEW{,
where there is only one event to represent the propagation of a given store
instruction}.  Our experiments (\cf~\mysec\ref{sec:testing}) show that this
choice impairs the simulation time by up to a factor of ten.  

Later in 2012, Boudol et al.~\cite{bps12} proposed an operational model where
the semantics of $\lwsync$ is stronger than the architectural intent on \eg
\textsf{r+lwsync+sync} (like Alglave et al.'s~\cite{ams10}). 


\section{Preamble on axiomatic models}

\NEW{We give here a brief presentation of axiomatic models in general. The
expert reader might want to skip this section.}

\NEW{Axiomatic models are usually defined in three stages. First, an
\emph{instruction semantics} maps each instruction to some mathematical
objects. This allows us to define the \emph{control flow semantics} of a
multi-threaded program.  Second, we build a set of \emph{candidate executions}
from this control-flow semantics: each candidate execution represents one
particular data-flow of the program, \ie which \emph{communications} might
happen between the different threads of our program. Third, a \emph{constraint
specification} decides which candidate executions are valid or not.}

\NEW{ We now explain these concepts in a way that we hope to be intuitive.
Later in this paper, we give the constraint specification part of our model in
\mysec\ref{sec:model}, and an outline of the instruction semantics in
\mysec\ref{sec:instr-sem}.}

\paragraph{\NEW{Multi-threaded programs, \!\!\!}}\NEW{such as the one given in
\myfig\ref{fig:mp-prog}, give one sequence of \emph{instructions} per thread.
Instructions can come from a given assembly language instruction set, \eg Power
ISA, or be pseudo-code instructions, as is the case in
\myfig\ref{fig:mp-prog}.}

\begin{figure}[!h]
\begin{center}
\begin{tabularx}{.35\linewidth}{Y|Y}
\multicolumn{4}{c}{\textsf{mp}} \\ \hline
\multicolumn{4}{l}{initially \as{x=0; \as{y=0}}}\\ \hline 
\multicolumn{2}{c|}{\haut\myth{0}} &
\multicolumn{2}{c}{\haut\myth{1}} \\ \hline
\haut\instab{\pstore{1}{x}}{(a)} & \instab{\pload{r1}{y}}{(c)} \\
\bas\instab{\pstore{1}{y}}{(b)} & \instab{\pload{r2}{x}}{(d)} 
\end{tabularx}
\end{center}
\caption{\NEW{A multi-threaded program implementing a message passing pattern}\label{fig:mp-prog}}
\end{figure}

\NEW{In \myfig\ref{fig:mp-prog}, we have two threads \myth{0} and \myth{1} in
parallel. These two threads communicate via the two memory locations $x$ and
$y$, which hold the value $0$ initially. On \myth{0} we have a store of value
$1$ into memory location $x$, followed in program order by a store of value $1$
into memory location $y$. On \myth{1} we have a load of the contents of memory
location $y$ into register $r_1$, followed in program order by a load of the
contents of memory location $x$ into register $r_2$. Memory locations, \eg $x$
and $y$, are shared by the two threads, whereas the registers are private to
the thread holding them, here \myth{1}.}

\NEW{The snippet in \myfig\ref{fig:mp-prog} is at the heart of a message
passing pattern, where \myth{0} would write some data into memory location $x$,
then set a flag in $y$. \myth{1} would then check if he has the flag, then read
the data in $x$.}

\paragraph{\NEW{Control-flow semantics}}
\NEW{The instruction semantics, in our case, translates instructions into
\emph{events}, that represent \eg \emph{memory or register accesses} (\ie reads
and writes from and to memory or registers), \emph{branching decisions} or
\emph{fences}.}  

\begin{figure}[!h]
\begin{center}
\input{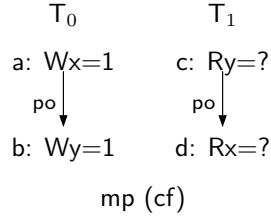}
\end{center}
\caption{\NEW{Control-flow semantics for the message passing pattern of \myfig\ref{fig:mp-prog}\label{fig:mp-cf}}}
\end{figure}

\NEW{Consider \myfig\ref{fig:mp-cf}: we give a possible control-flow semantics
to the program in \myfig\ref{fig:mp-prog}. To do so, we proceed as follows:
each store instruction, \eg \pstore{1}{x} on \myth{0}, corresponds to a write
event specifying a memory location and a value, \eg Wx=1. Each load
instruction, \eg \pload{r1}{y} on \myth{1} corresponds to a read event
specifying a memory location and a undetermined value, \eg Ry=?.  Note that the
memory locations of the events are determined by the program text, as well as
the values of the writes. For reads, the values will be determined in the next
stage.}

\NEW{Additionally, we also have fictitious write events Wx=0 and Wy=0
representing the initial state of $x$ and $y$, that we do not depict here.}

\NEW{The instruction semantics also defines \emph{relations} over these events,
representing for example the \emph{program order} within a thread, or
\emph{address, data or control dependencies} from one memory access to the
other, via computations over register values.} 

\NEW{Thus in \myfig\ref{fig:mp-cf}, we also give the program order relation,
written \po{}, in \myfig\ref{fig:mp-prog}, which lifts the order in which
instructions have been written to the level of events. For example, the two
stores on \myth{0} in \myfig\ref{fig:mp-prog} have been written in program
order, thus their corresponding events Wx=1 and Wy=1 are related by \po{} in
\myfig\ref{fig:mp-cf}.}

\NEW{We are now at a stage where we have, given a program such as the one in
\myfig\ref{fig:mp-prog}, several \emph{event graphs}, such as the one in
\myfig\ref{fig:mp-cf}. Each graph gives a set of events representing accesses
to memory and registers, the program order between these events, including
branching decisions, and the dependencies.}

\paragraph{\NEW{Data-flow semantics}}
\NEW{The purpose of this step is to define which \emph{communications}, or
\emph{interferences}, might happen between the different threads of our
program. To do so, we need to define two relations over memory events: the
\emph{read-from} relation \rf{}, and the \emph{coherence order} \co{}.}

\begin{figure}[!h]
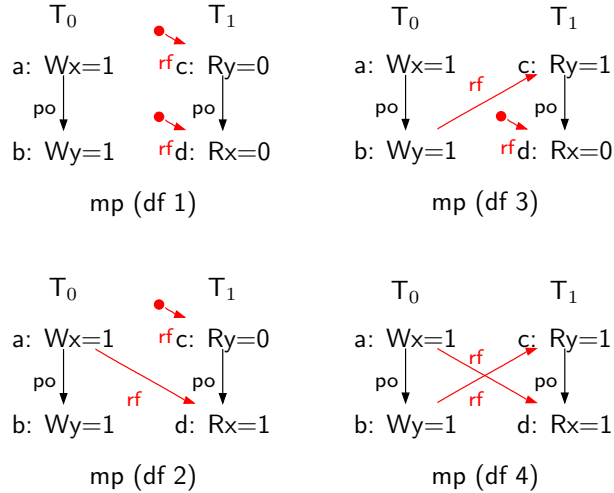

\begin{center}
\begin{tabular}{cc}
\input{img/mp-df1\bw.pstex_t} & \input{img/mp-df3\bw.pstex_t} \\
\input{img/mp-df2\bw.pstex_t} & \input{img/mp-df4\bw.pstex_t} 
\end{tabular}
\end{center}
\caption{\NEW{One possible data-flow semantics per control-flow semantics as given in \myfig\ref{fig:mp-cf}\label{fig:mp-df}}}
\end{figure}

\NEW{The read-from relation \rf{} describes, for any given read, from which
write this read could have taken its value. A read-from arrow with no source,
as in the top left of \myfig\ref{fig:mp-cf}, corresponds to reading from the
initial state.} 

\NEW{For example in \myfig\ref{fig:mp-df}, consider the drawing at the bottom
left-most corner. The read $c$ from $y$ takes its value from the initial state,
hence reads the value $0$. The read $d$ from $x$ takes its value from the
update $a$ of $x$ by \myth{0}, hence reads the value $1$.}

\NEW{The coherence order gives the order in which all the memory writes to a
given location have hit that location in memory. For example in
\myfig\ref{fig:mp-df}, the initial write to x (not depicted) hits the memory
before the write $a$ on \myth{0}, by convention, hence the two writes are
ordered in coherence.}

\NEW{We are now at a stage where we have, given a program such as the one in
\myfig\ref{fig:mp-prog}, several event graphs as given by the control-flow
semantics (see \myfig\ref{fig:mp-cf}), and for each of these, a read-from
relation and a coherence order describing the communications across threads
(see \myfig\ref{fig:mp-df}).}

\NEW{Note that for a given control-flow semantics there
could be several suitable data-flow semantics, if for example there were
several writes to $x$ with value $1$ in our example: in that case there would
be two possible read-from to give a value to the read of $x$ on~\myth{1}.}

\NEW{Each such object (see \myfig\ref{fig:mp-df}, which gathers events, program order,
dependencies, read-from and coherence, is called a \emph{candidate execution}.
As one can see in \myfig\ref{fig:mp-df}, there can be more than one candidate execution for a given program.}

\paragraph{\NEW{Constraint specification}} \NEW{Now, for each candidate execution,
the constraint specification part of our model decides if this candidate
represents a valid execution or not.}

\NEW{Traditionally, such specifications are in terms of acyclicity or
irreflexivity of various combinations of the relations over events given by the
candidate execution. This means for example that the model would reject a
candidate execution if this candidate contains a cycle amongst a certain
relation defined in the constraint specification.}

\NEW{For example in \myfig\ref{fig:mp-df}, the constraints for describing
Lamport's Sequential Consistency~\cite{lam79} (see
also~\mysec\ref{sec:instances}) would rule out the right-most top candidate
execution because the read from $x$ on \myth{1} reads from the initial state,
whereas the read of $y$ on \myth{1} has observed the update of $y$ by
\myth{0}.}

\section{A model of weak memory}
\label{sec:model}

We present our axiomatic model, and show its SC and TSO instances. We also
explain how to instantiate our model to produce C++ R-A, \ie the fragment
of C++ restricted to the use of release-acquire atomics.
\mysec\ref{sec:power} presents Power. 

The inputs to our model are candidate executions of a given multi-threaded
program. Candidate executions can be ruled out by the four axioms of our model,
given in \myfig\ref{fig:model}: \DELETED{\textsc{uniproc}}\textsc{\NEW{sc per
location}, no thin air, \DELETED{causality}\NEW{observation}} and
\textsc{propagation}. 

\subsection{\label{sec:prelim}Preliminaries}

Before explaining each of these axioms, we define a few preliminary notions.

\paragraph{Conventions:} in this paper, we use several notations on
\emph{relations} and \emph{orders}.  We denote the transitive (resp.~\DELETED{reflexive
and transitive}\NEW{reflexive-transitive}) closure of a relation $\textsf{r}$ as $\transc{\textsf{r}}$
(resp.~$\rstar{\textsf{r}}$).  We write $\textsf{r}_1;\textsf{r}_2$ for the
sequence of two relations $\textsf{r}_1$ and $\textsf{r}_2$, \ie $\dfn{(x,y)
\in (\textsf{r}_1;\textsf{r}_2)}{\exists z.  (x,z) \in \textsf{r}_1 \:\:\wedge}$
$(z,y) \in \textsf{r}_2$. We write $\irrefl(\textsf{r})$ to express the
irreflexivity of $\textsf{r}$, \ie $\neg(\exists x. (x,x) \in \textsf{r})$. We
write $\acyclic(\textsf{r})$ to express its acyclicity, \ie $\neg(\exists x.
(x,x) \in \transc{\textsf{r}})$.

A \emph{partial order} is a relation $\textsf{r}$ that is \emph{transitive}
(\ie $\textsf{r} = \transc{\textsf{r}}$), and irreflexive. Note that this
entails that $\textsf{r}$ is also acyclic. A \emph{total order} is a partial
order $\textsf{r}$ defined over a set $\mathbb{S}$ that enjoys the
\emph{totality property}: $\forall x \neq y \in \mathbb{S}. (x,y) \in
\textsf{r} \vee (y,x) \in \textsf{r}$.

\paragraph{Executions \!\!\!} are tuples $(\evts,\po,\rf,\co)$, which consist
of a set of \emph{events} $\evts$, giving a semantics to the instructions, and
three relations over events: $\po,\rf$ and $\co$ (see below). 

\paragraph{Events \!\!\!\!} consist of a unique identifier (in this paper we
use lower-case letters, \eg $a$), the thread holding the corresponding
instruction (\eg \myth{0}), the line number or program counter of the
instruction, and an \emph{action}. 

Actions are of several kinds, which we detail in the course of this paper.  For
now, we only consider read and write events relative to memory locations. For
example for the location $x$ we can have a read of the value $0$ noted R$x=0$,
or a write of the value $1$, noted W$x=1$. 
We write $\pr(e)$ for the thread holding the event $e$, and $\loc(e)$ for its
location.

Given a candidate execution, the events are determined by the program's
instruction semantics -- we give examples in~\mysec\ref{sec:instr-sem}. 

Given a set of events, we write $\WR, \WW, \RR, \RW$ for the set of write-read,
write-write, read-read and read-write pairs respectively. For example $(w,r)
\in \WR$ means that $w$ is a write and $r$ a read. We write $\po \cap \WR$ for
the write-read pairs in program order, and $\po \setminus \WR$ for all the pairs
in program order except the write-read pairs.

\paragraph{Relations over events:} the \emph{program order} $\po$ lifts the
order in which instructions have been written in the program to the level of
events. \NEW{The program order is a total order over the memory events of a
given thread, but does not order events from different threads.} Note that the
program order unrolls the loops and determines the branches taken. 

The \emph{read-from} $\rf$ links a read from a register or a memory location to a
unique write to the same register or location. The value of the read must be
equal to the one of the write. We write $\rfe$ (external read-from) when the
events related by $\rf$ belong to distinct threads, \ie $\dfn{(w,r) \in
\rfe}{(w,r) \in \rf \wedge \pr(w) \neq \pr(r)}$. We write $\rfi$ for internal
read-from, when the events belong to the same thread.

The \emph{coherence} order $\co$ totally orders writes to the same memory
location. We write $\coi$ (resp.~$\ews$) for internal (resp.~external)
coherence. 

We derive the \emph{from-read} $\fr$ from the read-from $\rf$
and the coherence $\co$, as follows: 
\begin{center}
\input{img/fr\bw.pstex_t}\vspace*{-2ex}
\end{center}
That is, a read $r$ is in $\fr$ with a write $w_1$ (resp.~$w_2$) if $r$ reads
from a write $w_0$ such that $w_0$ is in the coherence order before $w_1$
(resp.~$w_2$).  We write $\ifr$ (resp.~$\efr$) for the internal
(resp.~external) from-read.

We gather all \emph{communications} in $\dfn{\com}{\co \cup \rf \cup \fr}$. We
give a glossary of all the relations which we describe in this section
in~\mytab\ref{fig:gloss-rlns}. \NEW{For each relation we give its notation, its
name in English, the directions (\ie write W or read R) of the source and
target of the relation (column ``dirns''), where to find it in the text (column
``reference''), and an informal prose description.} \NEW{Additionally in the
column ``nature'', we give a taxonomy of our relations: are they fundamental
execution relations (\eg $\po, \rf$), architectural relations (\eg $\ppo$), or
derived (\eg $\fr, \hb$)?} 

\begin{table}[!h]
\scalebox{.85}{
\begin{tabular}{c|p{.2\linewidth}|c|p{.1\linewidth}|p{.2\linewidth}|p{.3\linewidth}}
notation & name          & \NEW{nature} & \NEW{dirns} & \NEW{reference} & description \\\hline
$\po$    & program order & \NEW{execution} & \NEW{any, any} & \NEW{\S Relations over events} & instruction order lifted to events \\
$\rf$    & read-from     & \NEW{execution} & \NEW{WR} & \NEW{\S Relations over events} & links a write $w$ to a read $r$ taking its value from $w$ \\
$\co$    & coherence     & \NEW{execution} & \NEW{WW} & \NEW{\S Relations over events} & total order over writes to the same memory location \\\hline
$\ppo$   & preserved program order & \NEW{architecture} & \NEW{any, any} & \NEW{\S Architectures} & program order maintained by the architecture \\
$\ffence, \ff$ & full fence & \NEW{architecture} & \NEW{any, any} & \NEW{\S Architectures}        & \eg \sync~on Power\NEW{, \dmb{} and \dsb{} on ARM} \\
$\lwfence, \lwf$ & lightweight fence & \NEW{architecture} & \NEW{any, any} & \NEW{\S Architectures} & \eg \lwsync~on Power \\
$\cfence$ & control fence & \NEW{architecture} & \NEW{any, any} & \NEW{\S Architectures}            & \eg \isync{} on Power\NEW{, \isb{} on ARM} \\
\NEW{$\fences$} & \NEW{fences} & \NEW{architecture} & \NEW{any, any} & \NEW{\S Architectures}      & \NEW{union of some (depending on the architecture) of the fence relations, \eg $\ffence, \lwfence, \cfence$} \\
$\prop$ & propagation & \NEW{architecture} & \NEW{WW} & \NEW{\S Architectures}               & order in which writes propagate\NEW{, typically enforced by fences} \\\hline
$\poloc$ & program order restricted to the same memory location & \NEW{derived} & \NEW{any, any} & \DELETED{\S \textsc{uniproc}} \NEW{\S \textsc{sc per location}} & $\{(x,y) \mid (x,y) \in \po \wedge \loc(x)=\loc(y)\}$\\
$\com$ & communications & \NEW{derived} & \NEW{any, any} &  \NEW{\S Relations over events}            & $\co \cup \rf \cup \fr$ \\
$\fr$    & from-read   & \NEW{derived}  & \NEW{RW} & \NEW{\S Relations over events} & links a read $r$ to a write $w'$ $\co$-after the write $w$ from which $r$ takes its value \\
$\hb$ & happens before & \NEW{derived} & \NEW{any, any} & \NEW{\S \textsc{no thin air}}              & \DELETED{${\transc{(\ppo \cup \fences)}} \cup {\rfe}$}\NEW{$\ppo \cup \fences \cup \rfe$} \\
\NEW{\textsf{rdw}} & \NEW{read different writes} & \NEW{derived} & \NEW{RR} & \NEW{\myfig\ref{fig:rdw}} & \NEW{two threads; first thread holds a write, second thread holds two reads} \\
\NEW{\textsf{detour}} & \NEW{detour} & \NEW{derived} & \NEW{WR} & \NEW{\myfig\ref{fig:detour}} & \NEW{two threads; first thread holds a write, second threads holds a write and a read} 
\end{tabular}}
\caption{Glossary of relations\label{fig:gloss-rlns}}
\end{table}


\paragraph{Reading notes:} we refer to orderings of events \wrt
several relations.  To avoid ambiguity, given a relation \textsf{r}, we say
that an event $e_1$ is \emph{\textsf{r}-before} another event $e_2$ (or $e_1$
is \NEW{an} \emph{\textsf{r}-predecessor} of $e_1$, or $e_2$ is
\emph{\textsf{r}-after} $e_1$, or $e_2$ is \emph{\textsf{r}-subsequent}, \etc)
when $(e_1,e_2) \in \textsf{r}$. 

In the following we present several examples of executions, in the style
of~\cite{ssa11}. We depict the events of a given thread vertically to represent
the program order, and the communications by arrows labelled with the
corresponding relation. \myfig\ref{fig:mp-naked} shows a classic \emph{message
passing} (\textsf{mp}) example. 

\begin{figure}
\begin{center}
\input{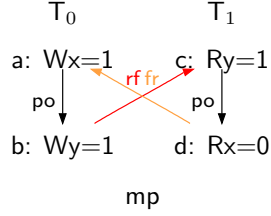}
\end{center}
\caption{\NEW{Message passing pattern}\label{fig:mp-naked}}
\end{figure}

This \DELETED{idiom}\NEW{example} is a communication pattern involving two memory locations $x$ and
$y$: $x$ is a message, and $y$ a flag to signal to the other thread that it can
access the message.

\myth{0} writes the value~$1$ to memory at location~$x$ (see the event~$a$).  In
program order after~$a$ (hence the~$\po$ arrow between~$a$ and~$b$), we have a
write of value~$1$ to memory at location~$y$. \myth{1} reads from~$y$ (see the
event~$c$). In the particular execution shown here, this read takes its value
from the write~$b$ by~\myth{0}, hence the~$\rf$ arrow between~$b$ and~$a$. In
program order after~$d$, we have a read from location~$x$. In this execution, we
suppose that this event~$d$ reads from the initial state (not depicted), which
by convention sets the values in all memory locations and registers to~$0$.
This is the reason why the read~$d$ has the value~$0$. This initial write
to~$x$ is, by convention, $\co$-before the write~$a$ of~$x$ by~\myth{0}, hence
we have an~$\fr$ arrow between~$d$ and~$a$. 
 
 \NEW{Note that, in the following, even if we do not always depict all of the
program order, a program order edge is always implied between each pair of
events ordered vertically below a thread id, \eg $\textsf{T}_0$.}

\paragraph{Convention for naming tests:} we refer to tests following the same
convention as in~\cite{ssa11}. \NEW{We roughly have two flavours of names:
classical names, that are abbreviations of classical litmus test names
appearing in the litterature; and systematic names, that describe the accesses
occurring on each thread of a test.}

\NEW{Classical} patterns, such as the message passing
\DELETED{idiom}\NEW{pattern} above, have an abbreviated name: \textsf{mp}
stands for ``message passing'', \textsf{sb} for ``store buffering'',
\textsf{lb} for ``load buffering'', \textsf{wrc} for ``write-to-read
causality'', \textsf{rwc} for ``read-to-write causality''. 

\NEW{When a pattern does not have a classical name from the literature,}
\DELETED{Sometimes the} \NEW{we give it a} name \NEW{that} simply describes
which accesses occur: for example \textsf{2+2w} means that the test is made of
two threads holding two writes each; \textsf{w+rw+2w} means that we have three
threads: a write on a first thread, a read followed by a write on a second
thread, and then two writes on a last thread. 

Finally when we extend a test (\eg \textsf{rwc}, ``read-to-write causality'')
with an access (\eg a write) on an extra thread, we extend the name
appropriately: \textsf{w+rwc} (\cf\myfig\ref{fig:w+rwc}). We give a glossary of
the test names presented in this paper in \mytab\ref{fig:gloss-litmus}, in the
order in which they appear\NEW{; for each test we give its systematic name, and
its classic name (\ie borrowed from previous works) when there is one.}

Note that in every test we choose the locations so that we can form a cycle in
the relations of our model: for example, \textsf{2+2w} has two threads with two
writes each, such that the first one accesses \eg the locations~$x$ and~$y$ and
the second one accesses~$y$ and~$x$. This precludes having the first thread
accessing~$x$ and~$y$ and the second one~$z$ and~$y$, because we could not link
the locations on each thread to form a cycle.

\begin{table}[!h]
\begin{tabular}{c|c|c|p{.55\linewidth}}
classic & systematic & diagram & description \\\hline
\textsf{coXY} & & \myfig\ref{fig:co} & coherence test involving an access of kind X and an access of
kind Y; X and Y can be either R (read) or W (write)\\
\textsf{lb} & \textsf{rw+rw} & \myfig\ref{fig:lb} & load buffering \NEW{\ie two threads each holding a read then a write} \\
\textsf{mp} & \textsf{ww+rr}& \myfig\ref{fig:mp} & message passing \NEW{\ie two threads; first thread holds two writes, second thread holds two reads} \\
\textsf{wrc} & \textsf{w+rw+rr} & \myfig\ref{fig:wrc} & write to read causality \NEW{\ie three threads; first thread holds a write, second thread holds a read then a write, third thread holds two reads} \\
\textsf{isa2} & \textsf{ww+rw+rr} & \myfig\ref{fig:isa2} & one of the tests appearing in the Power \NEW{ISA} documentation~\cite{ppc:2.06} \NEW{\ie write to read causality prefixed by a write, meaning that the first thread holds two writes instead of just one as in the \textsf{wrc} case} \\
\textsf{2+2w} & \textsf{ww+ww} & \myfig\ref{fig:2+2w} & two threads holding two writes each \\
& \textsf{w+rw+2w} & \myfig\ref{fig:w+rw+2w} & \DELETED{three threads, holding
a write, a read and write, then two writes} \NEW{three threads; first thread
holds a write, second thread holds a read then a write, third thread holds two
writes} \\
\textsf{sb} & \textsf{wr+wr} & \myfig\ref{fig:sb} & store buffering \NEW{\ie two threads each holding a write then a read} \\
\textsf{rwc} & \textsf{w+rr+wr} & \myfig\ref{fig:rwc} & read to write causality \NEW{three threads; first thread holds a write, second thread holds two reads, third thread holds a write then a read} \\
\textsf{r} & \textsf{ww+wr} & \myfig\ref{fig:r} & \DELETED{two threads, holding
two writes, then a write and a read} \NEW{two threads; first thread holds two
writes, second thread holds a write and a read} \\
\textsf{w+rwc} & \textsf{ww+rr+wr} & \myfig\ref{fig:w+rwc} & read to write
causality pattern \textsf{rwc}, prefixed by a write \NEW{\ie the first thread
holds two writes intead of just one as in the \textsf{rwc} case} \\
\textsf{iriw} & \textsf{w+rr+w+rr} & \myfig\ref{fig:iriw} & independent reads of independent writes \NEW{\ie four threads; first thread holds a write, second holds two reads, third holds a write, fourth holds two reads} \\
\end{tabular}
\caption{Glossary of litmus tests names\label{fig:gloss-litmus}}
\end{table}

Given a certain pattern such as \textsf{mp} above, we write
\textsf{mp+lwfence+ppo} for the same pattern, but where the first thread has a
lightweight fence \lwfence{} between the two writes and the second thread
maintains its two accesses in order thanks to some \emph{preserved program
order} mechanism (\ppo, see below). We write \textsf{mp+lwfences} for the
\textsf{mp} pattern with two lightweight fences, one on each thread. We
sometimes specialise the naming to certain architectures and mechanisms, as in
\textsf{mp+lwsync+addr}, where \lwsync{} refers to Power's lightweight fence,
and \addr{} denotes an \emph{address dependency} --- a particular way of
preserving program order on Power.

\paragraph{Architectures \!\!\!\!} are instances of our model.  An
architecture is a triple of functions $(\ppo,\fences,\prop)$, which specifies
the \emph{preserved program order} $\ppo$, the \emph{fences} $\fences$ and the
\emph{propagation order} $\prop$.

The preserved program order gathers the set of pairs of events which are
guaranteed not to be reordered \wrt the order in which the corresponding
instructions have been written. For example on TSO, only write-read pairs can
be reordered, so that the preserved program order for TSO is $\po \setminus
\WR$. On weaker models such as Power or ARM, the preserved program order merely
includes \emph{dependencies}, for example address dependencies, when the
address of a memory access is determined by the value read by a preceding load.
We detail these notions, and the preserved program order for Power and ARM, in
\mysec\ref{sec:power}. 

The function $\ppo$, given an execution $(\evts,\po,\co,\rf)$, returns the
preserved program order. \NEW{For example, consider the execution of the
message passing example given in \myfig\ref{fig:mp}.  Assume that there is an
address dependency between the two reads on \myth{1}. As such a dependency
constitutes a preserved program order relation on Power, the $\ppo$ function
would return the pair $(c,d)$ for this particular execution.}

Fences (or \emph{barriers}) are special instructions which prevent certain
behaviours.  On Power and ARM (\cf \mysec\ref{sec:power}), we distinguish
between control fence (which we write $\cfence$), lightweight fence (\lwfence)
and full fence (\ffence).  On TSO there is only one fence, called \mfence.

\NEW{In this paper, we use the same names for the fence instructions and the
relations that they induce over events. For example, consider the execution of
the message passing example given in \myfig\ref{fig:mp}.  Assume that there is
a lightweight Power fence \lwsync{} between the two writes $a$ and $b$ on
\myth{0}. In this case, we would have $(a,b) \in
\lwsync{}$.}\footnote{\NEW{Note that if there is a fence ``\textsf{fence}''
between two events $e_1$ and $e_2$ in program order, the pair of events
$(e_1,e_2)$ belongs to the eponymous relation ``\textsf{fence}'' (\ie
$(e_1,e_2) \in \textsf{fence}$), regardless of whether the particular fence
``\textsf{fence}'' actually orders these two accesses. For example on Power,
the lightweight fence \lwsync{} does not order write-read pairs in program
order.  Now consider the execution of the store buffering idiom in
\myfig\ref{fig:sb}, and assume that there is an \lwsync{} between the write $a$
and the read $b$ on \myth{0}. In this case, we have $(a.b) \in \lwsync$.
However, the pair $(a,b)$ would not be maintained in that order by the barrier,
which we model by excluding write-read pairs separated by an \lwsync{} from the
propagation order on Power (see~\myfig\ref{fig:fences-power+arm} and
\ref{fig:prop-power+arm}: the propagation order $\prop$ contains $\lwfence$,
which on Power is defined as $\lwsync{} \setminus \WR$ only, not $\lwsync$).}}
 
The function $\fences$ returns the pairs of events in program order which are
separated by a fence, when given an execution. \NEW{For example, consider the
execution of the message passing example given in \myfig\ref{fig:mp}.  Assume
that there is a lightweight Power fence \lwsync{}  between the two writes on
\myth{0}. On Power, the $\fences$ function would thus return the pair $(a,b)$
for this particular execution.}

The propagation order \DELETED{determines}\NEW{constrains} the order in which
writes are propagated to the memory system. \NEW{This order is a partial order
between writes (not necessarily to the same location), which can be enforced by
using fences. For example on Power, two writes in program order separated by an
\lwsync{} barrier (see \myfig\ref{fig:mp}) will be
ordered the same way in the propagation order.} 

\NEW{We note that the propagation order is distinct from the coherence order
$\co$: indeed $\co$ only orders writes to the same location, whereas the
propagation order can related writes with different locations through the use
of fences. However both orders have to be compatible, as expressed by our
\textsc{propagation} axiom, which we explain next (see \myfig\ref{fig:model} and
\myfig\ref{fig:2+2w}). }

The function $\prop$ returns the pairs of writes ordered by the propagation
order, given an execution. \NEW{For example, consider the execution of the
message passing example given in \myfig\ref{fig:mp}.  Assume that there is a
lightweight Power fence \lwsync{}  between the two writes on \myth{0}. On
Power, the presence of this fence forces the two writes to propagate in the
order in which they are written on \myth{0}. The function $\prop$ would thus
return the pair $(a,b)$ for this particular execution.}

\subsection{Axioms of our framework}
We can now explain the axioms of our model (\cf \myfig\ref{fig:model}).
\DELETED{Note that all the executions that we depict below are forbidden by our
model.} \NEW{For each example execution that we present in this section, we
write in the caption of the corresponding figure whether it is allowed or
forbidden by our model.}

\begin{figure}[!t]
  \begin{alignat*}{2}
    \text{\textsf{Input data:}} & &\ \ &(\ppo,\fences,\prop) \text{ and }
  (\evts,\po,\co,\rf)\\[0.7em]
  \text{\textsc{(sc per location)}} &&& \acyclic({\poloc} \cup {\com})  \text{ with}\\
                            &&& \dfn{\poloc}{\{(x,y) \in \po \wedge \loc(x) =
                              \loc(y)\}} \\
                            &&& \NEW{\dfn{\fr}{\{(r,w_1) \mid \exists w_0. (w_0,r) \in \rf
                              \wedge (w_0,w_1) \in \co \}}} \\
                            &&& \dfn{\com}{\co \cup \rf \cup \fr} \\[0.3em]
\text{\textsc{(no thin air)}} &&& \acyclic(\hb) \text{ with} \\
                           &&& \dfn{\hb}{{\ppo \cup \fences} \cup {\rfe}} \\[0.3em]
\text{\textsc{\NEW{(observation)}\DELETED{(causality)}}} &&& \irrefl(\efr;\prop;\rstar{\hb}) \\[0.3em]
\text{\textsc{(propagation)}} &&& \acyclic(\co \cup \prop)
\end{alignat*}
\caption{A model of weak memory\label{fig:model}}
\end{figure}

\subsection{\NEW{\textsc{SC per location}}}

\NEW{\textsc{sc per location}} ensures that the communications $\com$ cannot
contradict  $\poloc$ (program order between events relative to the same memory
location), \ie $\acyclic({\poloc} \cup {\com})$. This requirement forbids
exactly the five \DELETED{scenarios}\NEW{patterns} (as shown in~\cite[A.3
p.~184]{alg10}) given in \myfig\ref{fig:co}.

\begin{figure}[!h]
\begin{center}
\begin{tabular}{m{.3\linewidth} m{.3\linewidth} m{.3\linewidth}}
\input{img/coww\bw.pstex_t}
&
\input{img/corw1\bw.pstex_t}
&
\input{img/corw2\bw.pstex_t}
\end{tabular}
\end{center}
\begin{center}
\begin{tabular}{m{.5\linewidth} m{.5\linewidth}}
\input{img/cowr\bw.pstex_t}
&
\input{img/corr\bw.pstex_t}
\end{tabular}
\end{center}
\vspace*{-5mm}
\caption{The five \DELETED{scenarios}\NEW{patterns} forbidden by \DELETED{\textsc{uniproc}}\NEW{\textsc{sc per location}}\label{fig:co}}
\end{figure}

The \DELETED{idiom}\NEW{pattern} \textsf{coWW} forces two writes to the same memory location $x$ in
program order to be in the same order in the coherence order $\co$. The \DELETED{idiom}\NEW{pattern}
\textsf{coRW1} forbids a read from $x$ to read from a $\po$-subsequent write.
The \DELETED{idiom}\NEW{pattern} \textsf{coRW2} forbids the read $a$ to read from a write $c$ which is
$\co$-after a write $b$, if $b$ is $\po$-after $a$. The \DELETED{idiom}\NEW{pattern} \textsf{coWR}
forbids a read $b$ to read from a write $c$ which is $\co$-before a previous
write $a$ in program order. The \DELETED{idiom}\NEW{pattern} \textsf{coRR} imposes that if a read $b$
reads from a write $a$, all subsequent reads in program order from the same
location (\eg the read $c$) read from $a$ or a $\co$-successor write. 

\subsection{\textsc{No thin air}}

\textsc{no thin air} defines a \emph{happens-before} relation, written $\hb$,
defined as ${\ppo \cup \fences} \cup {\rfe}$, \ie an event $e_1$ happens before
another event $e_2$ if they are in preserved program order, or there is a fence
in program order between them, or $e_2$ reads from $e_1$.

\textsc{no thin air} requires the happens-before relation to be acyclic, which
prevents values from appearing out of thin air. Consider the
\DELETED{following} load buffering \DELETED{idiom}\NEW{pattern}
(\textsf{lb+ppos}) \NEW{in \myfig\ref{fig:lb}}.  
\begin{figure}[!h]
\begin{center}
\input{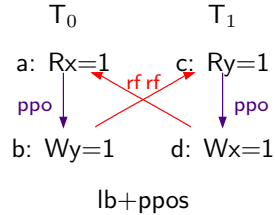}
\end{center}
\vspace*{-5mm}
\caption{The load buffering \DELETED{idiom}\NEW{pattern} \textsf{lb} \NEW{with ppo on both sides} \NEW{(forbidden)}\label{fig:lb}}
\end{figure}

\myth{0} reads from $x$ and writes to $y$, imposing (for example) an address
dependency between the two accesses, so that they cannot be reordered.
Similarly \myth{1} reads from $y$ and (dependently) writes to $x$.  \textsc{no
thin air} ensures that the two threads cannot communicate in a manner that
creates a \DELETED{causal}\NEW{happens-before} cycle, with the read from $x$ on
\myth{0} reading from the write to $x$ on \myth{1}, whilst \myth{1} reads
\myth{0}'s write to $y$. 

\NEW{Using the terminology of~\cite{ssa11}, we say that a read is \emph{satisfied} when it binds its value; note that the value is not yet irrevocable. It becomes irrevocable when the read is \emph{committed}. We say that a write is \emph{committed} when it makes its value available to other threads}.

\NEW{Our happens-before relation orders the point in time where a read is
satisfied, and the point in time where a write is committed.}  

\NEW{The pattern above shows that the ordering due to happens-before applies to
any architecture that does not speculate values read from memory (\ie values
written by external threads as opposed to the same thread as the reading
thread), nor allows speculative writes (\eg{} in a branch) to send their value
to other threads.}

\NEW{Indeed, in such a setting, a read such as the read $a$ on \myth{0} can be
satisfied from an external write (\eg{} the write $d$ on \myth{1}) only after
this external write has made its value available to other threads, and its
value is irrevocable.}

Our axiom forbids this \DELETED{scenario}\NEW{pattern} regardless of the method chosen to maintain the
order between the read and the write on each thread: address dependencies on
both (\textsf{lb+addrs}), a lightweight fence on the first and an address
dependency on the second (\textsf{lb+lwfence+addr}), two full fences
(\textsf{lb+ffences}). If however one or both pairs are not maintained, the
\DELETED{scenario}\NEW{pattern} can happen.

\subsection{\textsc{Observation} \!\!\!\!} 
\textsc{\DELETED{causality}\NEW{observation}} constrains the value of reads.
If a write $a$ to $x$ and a write $b$ to $y$ are ordered by the propagation
order $\prop$, and if a read $c$ \DELETED{witnesses}\NEW{reads from} the write
of $y$, then any read $d$ from $x$ which happens after $c$ (\ie $(c,d) \in
\hb$) cannot read from a write that is $\co$-before the write $a$ (\ie $(d,a)
\not\in \fr$). 

\subsubsection{Message passing (\textsf{mp})} A typical instance of this
\DELETED{scenario}\NEW{pattern} is the message passing
\DELETED{idiom}\NEW{pattern} (\textsf{mp+lwfence+ppo}) given in
\myfig\ref{fig:mp}.

\begin{figure}[!h]
\begin{center}
\input{img/mp+lwfence+ppo\bw.pstex_t}
\end{center}
\vspace*{-5mm}
\caption{The message passing \DELETED{idiom}\NEW{pattern} \textsf{mp} \NEW{with lightweight fence and ppo} \NEW{(forbidden)} \label{fig:mp}}
\end{figure}

\myth{0} writes a message in $x$, then sets up a flag in $y$, so that when
\myth{1} sees the flag (via its read from $y$), it can read the message in $x$.
For this \DELETED{idiom}\NEW{pattern} to behave as intended, following the
message passing protocol described above, the write to $x$ needs to be
propagated to the reading thread before the write to $y$. 

TSO guarantees it, but Power or ARM need at least a lightweight fence (\eg{}
\lwsync{} on Power) between the writes.  We also need to ensure that the two
reads on \myth{1} stay in the order in which they have been written, \eg with
an address dependency.  

The protocol would also be ensured with a full fence on the writing thread
(\textsf{mp+ffence+addr}), or with two full fences (\textsf{mp+ffences}).

\NEW{More precisely, our model assumes that a full fence is at least as
powerful as a lightweight fence. Thus the behaviours forbidden by the means of
a lightweight fence are also forbidden by the means of a full fence.  We insist
on this point since, as we shall see, our ARM model does not feature any
lightweight fence. Thus when reading an example such as the message passing one
in \myfig\ref{fig:mp}, the ``lwf'' arrow should be interpreted as any device
that has at least the same power as a lightweight fence. In the case of ARM,
this means a full fence, \ie \textsf{dmb} or \textsf{dsb}.}

\NEW{From a micro-architectural standpoint, fences order the propagation of
writes to other threads.  A very naive implementation can be by locking a
shared bus; modern systems feature more sophisticated protocols for ordering
write propagation.}

\NEW{By virtue of the fence, the writes to~$x$ and~$y$ on \myth{0} should
propagate to \myth{1} in the order in which they are written on \myth{0}. Since
the reads $c$ and $d$ on \myth{1} are ordered by $\ppo$ (\eg{} an address
dependency), they should be satisfied in the order in which they are written on
\myth{1}. In the scenario depicted in \myfig\ref{fig:mp}, \myth{1} reads the
value $1$ in $y$ (see the read event $c$) and the value~$0$ in~$x$ (see the
read event $d$). Thus \myth{1} observes the write $a$ to~$x$ after the write
$b$ to $y$. This contradicts the propagation order of the writes $a$ and $b$
enforced by the fence on~\myth{0}.}

\NEW{\label{alpha:addr}We note that the Alpha architecture~\cite{alpha:02}
allows the pattern~\textsf{mp+fence+addr} (a specialisation
of~\myfig\ref{fig:mp}). Indeed some implementations feature more than one
memory port per processor, \eg{} by the means of a split cache~\cite{kernel}.}

\NEW{Thus in our \textsf{mp} pattern above, the values written on $x$ and~$y$
by~\myth{0} could reach~\myth{1} on different ports.  As a result, although the
address dependency forces the reads $c$ and~$d$ to be satisfied in order, the
second read may read a stale value of $x$, while the current value of $x$ is
waiting in some queue. One could counter this effect by synchronising memory
ports.}

\subsubsection{\NEW{Cumulativity}} \NEW{To explain a certain number of the
following patterns, we need to introduce the concept of \emph{cumulativity}.}

\NEW{We consider that a fence has a cumulative effect when it ensures a
propagation order not only between writes on the fencing thread (\ie the thread
executing the fence), but also between certain writes coming from threads other
than the fencing thread.}

\begin{figure}[!h]
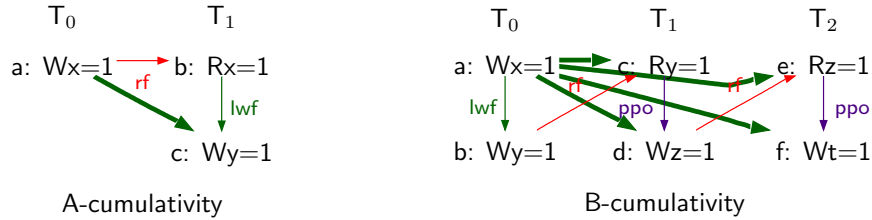

\begin{center}\input{img/A-cumul\bw.pstex_t}\quad\quad\quad\quad\quad\input{img/B-cumulativity\bw.pstex_t}\end{center}
\caption{\NEW{Cumulativity of fences}\label{fig:cumul}}
\end{figure}

\paragraph{\NEW{A-cumulativity}}
\NEW{More precisely, consider a situation as shown on the left of
\myfig\ref{fig:cumul}. We have a write $a$ to $x$ on \myth{0}, which is read by
the read $b$ of $x$ on \myth{1}. On \myth{1} still, we have an access $e$ in
program order after $b$. This event $e$ could be either a read or a write
event; it is a write $c$ in \myfig\ref{fig:cumul}. Note that $b$ and $e$ are
separated by a fence.}

\NEW{We say that the fence between $b$ and $e$ on \myth{1} is
\emph{A-cumulative} when it imposes that the read $b$ is satisfied before $e$
is either committed (if it is a write) or satisfied (if it is a read). Note
that for $b$ to be satisfied, the write $a$ on \myth{0} from which $b$ reads
must have been propagated to \myth{1}, which enforces an ordering between $a$
and $e$. We display this ordering in \myfig\ref{fig:cumul} by the mean of a
thicker arrow from $a$ to $c$.}

\NEW{The vendors documentations describe A-cumulativity as follows.  On Power,
quoting~\cite[Book II, \mysec 1.7.1]{ppc:2.06}: ``[the group] A [of \myth{1}]
includes all [\dots] accesses by any [\dots] processor [\dots] [\eg \myth{0}]
that have been performed with respect to [\myth{1}] before the memory barrier
is created.'' We interpret ``performed with respect to [\myth{1}]'' as a write
(\eg the write $a$ on the left of \myfig\ref{fig:cumul}) having been propagated
to \myth{1} and read by \myth{1} such that the read is \po-before the barrier.}


\NEW{On ARM, quoting~\cite[\mysec 4]{arm:cookbook}: ``[the] group A [of
\myth{1}] contains: All explicit memory accesses [\dots] from observers [\dots]
[\eg \myth{0}] that are observed by [\myth{1}] before the \dmb{} instruction.''
We interpret ``observed by [\myth{1}]'' on ARM as we interpret ``performed with
respect to [\myth{1}]'' on Power (see paragraph above).}

\paragraph{\NEW{Strong A-cumulativity}}

\begin{figure}[!h]
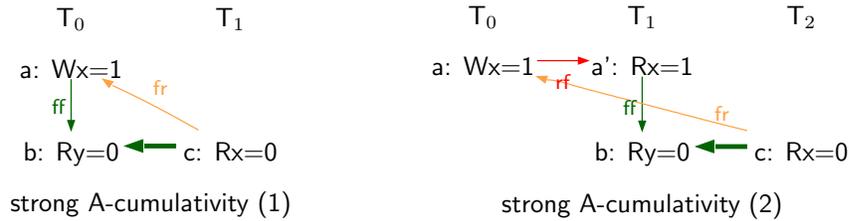

\begin{center}
\begin{center}\input{img/s-A-cumul1\bw.pstex_t}\quad\input{img/s-A-cumul2\bw.pstex_t}\end{center}
\end{center}
\caption{\NEW{Strong A-cumulativity of fences}\label{fig:strong-A-cumul}}
\end{figure}

\NEW{Interestingly, the ARM documentation does not stop there, and includes in
the group A ``[a]ll loads [\dots] from observers [\dots] [(\eg \myth{1})] that
have been observed by any given observer [\eg \myth{0}], [\dots] before
[\myth{0}] has performed a memory access that is a member of group A.''}

\NEW{Consider the situation on the left of \myfig\ref{fig:strong-A-cumul}. We
have the read $c$ on \myth{1} which reads from the intial state for $x$, thus
is $\fr$-before the write $a$ on \myth{0}. The write $a$ is $\po$-before a read
$b$, such that $a$ and $b$ are separated by a full fence. In that case, we
count the read $c$ as part of the group A of \myth{0}. This enforces a (strong)
A-cumulativity ordering from $c$ to $b$, which we depict with a thicker arrow.} 

\NEW{Similarly, consider the situation on the right of
\myfig\ref{fig:strong-A-cumul}. The only difference with the left of the figure
is that we have one more indirection between the read $c$ of $x$ and the read
$b$ of $y$, via the $\rf$ between $a$ and $a'$. In that case too we count the
read $c$ as part of the group A of \myth{1}. This enforces a (strong)
A-cumulativity ordering from $c$ to $b$, which we depict with a thicker arrow.}

\NEW{We take strong A-cumulativity into account only for full fences (\ie
\sync{} on Power and \dmb{} and \dsb{} on ARM). We reflect this in
\myfig\ref{fig:prop-power+arm}, in the second disjunct of the definition of the
$\prop$ relation ($\rstar{\com};\rstar{\propbase};\ffence;\rstar{\hb}$).}

\paragraph{\NEW{B-cumulativity}}
\NEW{Consider now the situation shown on the right of
\myfig\ref{fig:cumul}. On \myth{0}, we have an access $e$ (which could be
either a read or a write event) in program order before a write $b$ to $y$.
Note that $e$ and $b$ are separated by a fence. On \myth{1} we have a read $c$
from $y$, which reads the value written by $b$. }

\NEW{We say that the fence between $e$ and $b$ on \myth{0} is
\emph{B-cumulative} when it imposes that $e$ is either committed (if it is a
write) or satisfied (if it is a read) before $b$ is committed. Note that the
read $c$ on \myth{1} can be satisfied only after the write $b$ is committed and
propagated to \myth{1}, which enforces an ordering from $e$ to $c$. We display
this ordering in \myfig\ref{fig:cumul} by the mean of a thicker arrow from $a$
to $c$.}

\NEW{This B-cumulativity ordering extends to all the events that happen after
$c$ (\ie are $\hb$-after $c$), such as the write $d$ on \myth{1}, the read $e$
on \myth{2} and the write $f$ on \myth{2}. In \myfig\ref{fig:cumul}, we display
all these orderings via thicker arrows.}

\NEW{The vendors documentations describe B-cumulativity as follows.  On Power,
quoting~\cite[Book II, \mysec 1.7.1]{ppc:2.06}: ``[the group] B [of \myth{0}]
includes all [\dots] accesses by any [\dots] processor [\dots] that
are performed after a load instruction [such as the read $c$ on the right of
\myfig\ref{fig:cumul}] executed by [\myth{0}] [\dots] has returned the value
stored by a store that is in B [such as the write $b$ on the right of
\myfig\ref{fig:cumul}].'' On the right of \myfig\ref{fig:cumul}, this includes
for example the write $d$. Then the write $d$ is itself in the group B, and
observed by the read $e$, which makes the write $f$ part of the group B as
well.} 

\NEW{On ARM, quoting~\cite[\mysec 4]{arm:cookbook}: ``[the] 
group B [of \myth{0}] contains [a]ll [\dots] accesses [\dots] by [\myth{0}] that
occur in program order after the \dmb{} instruction. [\dots]'' On the right of \myfig\ref{fig:cumul}, this mean the write $b$.}

\NEW{Furthermore, ARM's group B contains ``[a]ll [\dots] accesses [\dots] by
any given observer [\eg \myth{1}] [\dots] that can only occur after [\myth{1}]
has observed a store that is a member of group B.'' We interpret this bit as we
did for Power's group B.}

\subsubsection{Write to read causality (\textsf{wrc})} 
This \DELETED{idiom}\NEW{pattern} (\textsf{wrc+lwfence+ppo}, given in
\myfig\ref{fig:wrc}) \DELETED{distributes the causality over three threads
instead of two}\NEW{illustrates the A-cumulativity of the lightweight fence on
\myth{1}, namely the $\rfe;\fences$ fragment of the definition illustrated in
\myfig\ref{fig:cumul} above}.

\begin{figure}[!h]
\begin{center}
\input{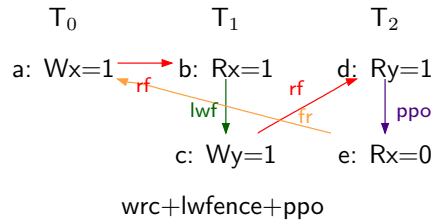}
\end{center}
\vspace*{-5mm}
\caption{The write-to-read causality \DELETED{idiom}\NEW{pattern} \textsf{wrc} \NEW{with lightweight fence and ppo} \NEW{(forbidden)}\label{fig:wrc}}
\end{figure}

There are now two writing threads, \myth{0} writing to $x$ and \myth{1} writing
to $y$ after reading the write of \myth{0}.  TSO still \DELETED{guarantees
causality}\NEW{enforces this pattern} without any help. But on Power and ARM we
need to place (at least) a lightweight fence on \myth{1} between the read of
\myth{0} (the read $b$ from $x$) and the write $c$ to $y$. The barrier will
force the write of $x$ to propagate before the write of $y$ to the \myth{2}
even if the writes are not on the same thread. 

\NEW{A possible implementation of A-cumulative fences, discussed in~\cite[Sec.
6]{gha90}, is to have the fences not only wait for the previous read ($a$ in
\myfig\ref{fig:wrc}) to be satisfied, but also for all stale values for~$x$ to
be eradicated from the system, \eg the value $0$ read by $e$ on~\myth{2}.}
 
\bigskip

\subsubsection{Power ISA2 (\textsf{isa2})} This \DELETED{idiom}\NEW{pattern}
(\textsf{isa2+lwfence+ppos}, given in \myfig\ref{fig:isa2}), distributes the
\DELETED{causality}\NEW{message passing pattern} over three threads like
\textsf{wrc+lwfence+ppo}, but keeping the writes to $x$ and $y$ on the same
thread. 

\begin{figure}[!h]
\begin{center}
\input{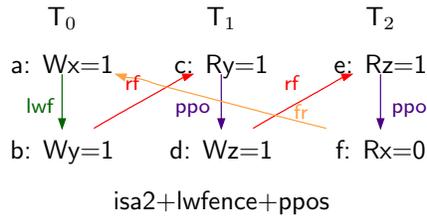}
\end{center}
\vspace*{-5mm}
\caption{The \DELETED{idiom}\NEW{pattern} \textsf{isa2} \NEW{with lighweight fence and ppo twice} \NEW{(forbidden)}\label{fig:isa2}}
\end{figure}

\DELETED{This shows that causality is preserved even if the two reads (from $y$
and $x$) do not belong to the same thread, if they are related by the
happens-before relation $\hb$.}

Once again TSO guarantees it without any help, but Power and ARM need a
lightweight fence between the writes, and (for example) an address or data
dependency between each pair of reads (\ie on both \myth{1} and \myth{2}).

\NEW{Thus on Power and ARM, the pattern \textsf{isa2+lwfence+ppos} illustrates
the B-cumulativity of the lightweight fence on \myth{0}, namely the
$\fences;\rstar{\hb}$ fragment of the definition of \cumul{} illustrated in
\myfig\ref{fig:cumul}.}
 
\subsection{\textsc{Propagation}} 

\textsc{propagation} constrains the order in which writes to memory are
propagated to the other threads, \DELETED{which cannot}\NEW{so that it does
not} contradict the coherence order, \ie $\acyclic(\co \cup \prop)$. 

\paragraph{On Power and ARM, \!\!\!\!} lightweight fences sometimes constrain
the propagation of writes, \NEW{as we have seen in the cases of \textsf{mp}
(see \myfig\ref{fig:mp}), \textsf{wrc} (see \myfig\ref{fig:wrc}) or
\textsf{isa2} (see \myfig\ref{fig:isa2})}. \DELETED{The \textsf{2+2w+lwfences}
idiom given in \myfig\ref{fig:2+2w} shows that a lightweight fence forces two
writes to propagate to the system in the order that they have been written by
the programmer.} \NEW{They can also, in combination with the coherence order,
create new ordering constraints.}
%
%

\NEW{The \textsf{2+2w+lwsync} pattern (given in \myfig\ref{fig:2+2w}) is for us
the archetypal illustration of coherence order and fences interacting to yield
new ordering constraints. It came as a surprise when we proposed it to our IBM
contact, as he wanted the lighweight fence to be as lightweight as
possible,\fixmeskip{jade@Luc: il faut qu'on verifie avec Derek si ca lui va
qu'on dise ca} for the sake of efficiency. However the pattern is acknowledged
to be forbidden. By contrast and as we shall see below, other patterns (such as
the \textsf{r} pattern in \myfig\ref{fig:r}) that mix the \co{} communication
with \fr{} require full fences to be forbidden.}

The \textsf{w+rw+2w+lwfences} \DELETED{idiom}\NEW{pattern} in
\myfig\ref{fig:w+rw+2w} distributes \textsf{2+2w+lwfences} over three threads.
\NEW{This pattern is to \textsf{2+2w+lwfences} what \textsf{wrc} is to
\textsf{mp}.} \NEW{Thus, j}ust as in the case of \textsf{mp} and \textsf{wrc},
the lightweight fence plays a\NEW{n} \NEW{A-}cumulative role, which ensures
that the two \DELETED{idioms}\NEW{patterns} \textsf{2+2w} and \textsf{w+rw+2w}
respond to the fence in the same way.

\begin{figure}[!h]
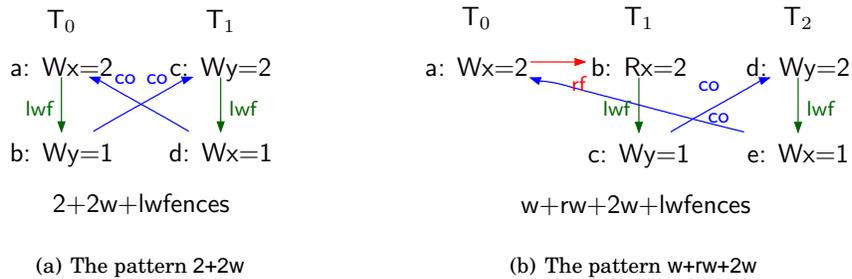

\centering
\subfigure[The \DELETED{idiom}\NEW{pattern} \textsf{2+2w}\label{fig:2+2w}]{
\input{img/2+2w+lwfences\bw.pstex_t}}
\hspace{6ex}
\subfigure[The \DELETED{idiom}\NEW{pattern} \textsf{w+rw+2w} \label{fig:w+rw+2w}]{
\input{img/wrw+2w+lwsyncs\bw.pstex_t}}
\caption{Two similar \DELETED{idioms}\NEW{patterns} \NEW{with lightweight fences} \NEW{(forbidden)}}
\end{figure}


\bigskip
\paragraph{On TSO, \!\!\!\!} every relation contributes to the propagation
order (\cf \myfig\ref{fig:sc+tso}), except the write-read pairs in program
order, which need to be fenced with a full fence (\mfence{} on TSO).  Consider
the store buffering (\textsf{sb+ffences}) \DELETED{idiom}\NEW{pattern} given in
\myfig\ref{fig:sb}. 

\begin{figure}[!h]
\begin{center}
\input{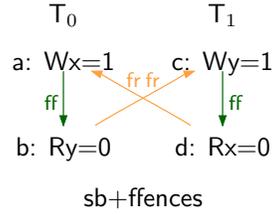}
\end{center}
\vspace*{-5mm}
\caption{The store buffering \DELETED{idiom}\NEW{pattern} \textsf{sb} \NEW{with full fences} \NEW{(forbidden)}\label{fig:sb}}
\end{figure}

We need a full fence on each side to prevent the reads $b$ and $d$ from reading
\DELETED{writes that are $\co$-before $a$ and $c$}\NEW{the initial state}.

\NEW{The pattern \textsf{sb} without fences being allowed, even on x86-TSO, is
perhaps one of the most well-known examples of a relaxed behaviour.  It can be
explained by writes being first placed into a thread-local store buffer, and
then carried over asynchronously to the memory system.  In that context, the
effect of a (full) fence can be described as flushing the store buffer. Of
course, on architectures more relaxed than TSO, the full fence has more work to
do, \eg cumulativity duties (as illustrated by the \textsf{iriw} example given
in \myfig\ref{fig:iriw}).}

On Power, the lightweight fence \lwsync{} does not order write-read pairs in
program order. Hence in particular it is not enough to prevent the \textsf{sb}
\DELETED{idiom}\NEW{pattern}; one needs to use a full fence on each side.
\NEW{The \textsf{sb} idiom, and the following \textsf{rwc} idiom, are instances
of the strong A-cumulativity of full fences.}

The read-to-write causality \DELETED{idiom}\NEW{pattern} \textsf{rwc+ffences}
(\cf \myfig\ref{fig:rwc})
distributes the \textsf{sb} \DELETED{idiom}\NEW{pattern} over three threads
with a read $b$ from $x$ between the write $a$ of $x$ and the read $c$ of $y$.
It is to \textsf{sb} what \textsf{wrc} is to \textsf{mp}, thus responds to
fences in the same way as \textsf{sb}. Hence it needs two full fences too.
\begin{figure}[!h]
\begin{center}
\input{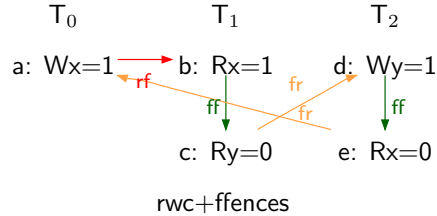}
\end{center}
\vspace*{-5mm}
\caption{The read-to-write causality \DELETED{idiom}\NEW{pattern} \textsf{rwc} \NEW{with full fences} \NEW{(forbidden)}\label{fig:rwc}}
\end{figure}
\begin{new}
Indeed, a full fence is required to order the write~$a$ and the read~$c$,
as lightweight fences do not apply from writes to reads.
The full fence on \myth{1} provides such an order, not only on \myth{1},
but also by (strong) A-cumulativity from the write~$a$ on \myth{1} to the
read~$c$ on~\myth{1}, as the read~$b$ on \myth{1} that reads from the write~$a$, \po-precedes the fence.
\end{new}

\bigskip
\bigskip
\bigskip
\bigskip

\DELETED{In the last \DELETED{idiom}\NEW{patterns} (\textsf{r+ffences}), \myth{0} writes to $x$ and $y$, whilst
\myth{1} writes to $y$ and reads from $x$. A full fence on each thread forces
the write $a$ to $x$ to propagate to \myth{1} before the write $b$ to $y$. Thus
if the write $b$ is $\co$-before the write $c$ on \myth{1}, the read $d$ of $x$
on \myth{1} cannot read from a write that is $\co$-before the write $a$\DELETED{:}\NEW{ (\cf \myfig\ref{fig:r}).}}
\NEW{The last two patterns, \textsf{r+ffences} and~\textsf{s+lwfence+ppo}
(\myfig\ref{fig:r}) illustrate the complexity of combining coherence order
and fences.
In both patterns, the thread \myth{0} writes to~$x$ (event~$a$)
and then to~$y$ (event~$b$).
In the former pattern \textsf{r+ffence},
\myth{1} writes to $y$ and reads from $x$. A full fence on each thread forces
the write $a$ to $x$ to propagate to \myth{1} before the write $b$ to $y$. Thus
if the write $b$ is $\co$-before the write $c$ on \myth{1}, the read $d$ of $x$
on \myth{1} cannot read from a write that is $\co$-before the write $a$.
By constrast, in the latter pattern~\textsf{s+lwfence+ppo},
\myth{1} reads from~$y$, reading the value stored by the write~$b$,
and then writes to~$x$. A lightweight fence on~\myth{0} forces the write~$a$
to~$x$ to propagate to \myth{1} before the write $b$ to~$y$.
Thus, as \myth{1} sees the write~$b$ by reading its value (read~$c$)
and as the write~$d$ is forced to occur by a dependency
(\ppo) after the read~$c$, that write~$d$ cannot \co-precede the write~$a$.}
\begin{figure}[!h]
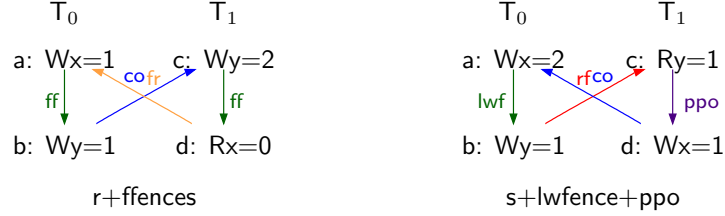

\begin{center}
\input{img/r+syncs\bw.pstex_t}\quad\quad\quad\quad\quad\input{img/s+lwsync+addr\bw.pstex_t}
\end{center}
\vspace*{-5mm}
\caption{The \DELETED{idiom}\NEW{patterns} \textsf{r} \NEW{with full fences and \textsf{s} with lightweight fence and ppo (both forbidden)}\label{fig:r}}
\end{figure}

\NEW{Following the architect's intent, inserting a lightweight fence \lwsync{}
between the writes $a$ and~$b$ does not suffice to forbid the \textsf{r}
pattern on Power.  It comes in sharp contrast with, for instance, the
\textsf{mp} pattern (\cf \myfig\ref{fig:mp}) and the \textsf{s} pattern, where
a lightweight fence suffices.  Thus the interaction between (lightweight) fence
order and coherence order is quite subtle, as it does forbid \textsf{2+2w} and
\textsf{s}, but not \textsf{r}. For completeness, we note that we did not
observe the \textsf{r+lwsync+sync} pattern on Power hardware.}

\subsection{Fences and propagation on Power and ARM}

Recall that architectures provides special instructions called fences to
prevent some weak behaviours. \DELETED{For example TSO provides \mfence{},
which prevents the reordering of write-read pairs in program order.}

\DELETED{Power provides the full fence \sync{}, the lightweight fence \lwsync{}
and the control fence \isync{}. ARM provides the \NEW{full} fences \dmb{} and
\dsb{}, and the control fence \isb{}.}

We summarise the fences and propagation order for Power and ARM in
\myfig\ref{fig:fences-power+arm} and~\ref{fig:prop-power+arm}. Note that the
control fences \NEW{(\isync{} for Power and \isb{} for ARM)} do not contribute
to the propagation order. \DELETED{It contributes}\NEW{They contribute} to the
definition of preserved program order, as explained in
\mysec\ref{sec:instr-sem} and \ref{sec:power}.

\begin{figure}[!h]
\begin{alignat*}{4}
  &\text{\textsf{Power}} &\qquad& \dfn{\ffence}{\sync} &\qquad& \dfn{\lwfence}{\lwsync \setminus \WR} &\qquad& \dfn{\cfence}{\isync} \\ 
  &\text{\textsf{ARM}} && \dfn{\ffence}{\dmb \cup \dsb} && \dfn{\lwfence}{\emptyset} &\qquad& \dfn{\cfence}{\isb}
\end{alignat*}
\caption{Fences for Power and ARM \label{fig:fences-power+arm}}
\end{figure}

\begin{figure}[!h]
\begin{mathpar}
\dfn{\hb}{\NEW{\ppo \cup \fences} \cup {\rfe}} \quad \DELETED{{\transc{(\ppo \cup \fences)}} \cup \rfe} \and
\dfn{\fences}{{\lwfence} \cup {\ffence}} \and
\NEW{\dfn{\Acumul}{\rfe;\fences}} \and
\DELETED{\dfn{\cumul}{\rfe;\fences \cup \fences;\rfe \cup
\rfe;\fences;\rfe}} \and
\NEW{\dfn{\propbase}{(\fences \cup \Acumul);\rstar{\hb}}} \and
\DELETED{\dfn{\propbase}{(\fences \cup \cumul);\rstar{\hb}}} \and
\dfn{\prop}{(\propbase \cap \WW) \cup (\rstar{\com}; \rstar{\propbase}; \ffence; \rstar{\hb})}
\end{mathpar}
\caption{Propagation order for Power and ARM\label{fig:prop-power+arm}}
\end{figure}

\paragraph{Fence\DELETED{s} placement} is the problem of automatically placing fences
between events to prevent undesired behaviours. For example, the message
passing \DELETED{idiom}\NEW{pattern} \textsf{mp} in \myfig\ref{fig:mp} can be prevented by placing
fences between events $a$ and $b$ in \myth{0} and events $c$ and $d$ in
\myth{1} in order to create a forbidden cycle.

This problem has already been studied in the past, but for TSO and its siblings
PSO and RMO (\cf\eg\cite{kvy10,bmm11,lnp12,bdm13}), or for previous versions of
the Power model (\cf\eg\cite{am11}). 

We emphasise how easy fence\DELETED{s} placement becomes, in the light of our new model.
We can read them off the definitions of the axioms and of the propagation
order.  Placing fences essentially amounts to counting the number of
communications (\ie the relations in $\com$) involved in the behaviour that we
want to forbid\DELETED{,}\NEW{.}

\fixmeskip{jade: more details for each of these paras}
To forbid a behaviour that involves only read-from ($\rf$) communications, or
only \emph{one} from-read ($\fr$) and otherwise $\rf$ communications, one can
resort to the \textsc{\DELETED{causality}\NEW{observation}} axiom, using the
$\propbase$ part of the propagation order.  Namely to forbid a behaviour of the
\textsf{mp} (\cf\myfig\ref{fig:mp}), \textsf{wrc} (\cf\myfig\ref{fig:wrc}) or
\textsf{isa2} (\cf\myfig\ref{fig:isa2}) type, one needs a lightweight fence on
the first thread, and some preserved program order mechanisms on the remaining
threads.

If coherence ($\co$) and read-from (\DELETED{$\fr$}\NEW{$\rf$}) are the only
communications involved, one can resort to the \textsc{propagation} axiom,
using the $\propbase$ part of the propagation order.  Namely to forbid a
behaviour of the \textsf{2+2w} (\cf\myfig\ref{fig:2+2w}) or \textsf{w+rw+2w}
(\cf\myfig\ref{fig:w+rw+2w}) type, one needs only lightweight fences on each
thread.

If more than one from-read ($\fr$) occurs, or if the behaviour involves both
coherence ($\co$) and from-read ($\fr$) communications, one needs to resort to
the part of the propagation order that involves the full fence (indeed that is
the only part of the propagation order that involves $\com$). Namely to forbid
behaviours of the type \textsf{sb} (\cf\myfig\ref{fig:sb}), \textsf{rwc}
(\cf\myfig\ref{fig:rwc}) or \textsf{r} (\cf\myfig\ref{fig:r}), one needs to use
full fences on each thread.

\paragraph{Power's~\eieio{} and ARM's~\dmbst{} and~\dsbst{}:} we make two
additional remarks, not reflected in the figures, about write-write barriers,
namely the \eieio~barrier on Power, and the \dmbst{} and~\dsbst{} barriers on
ARM.

On Power, the \eieio~barrier only maintains write-write pairs, as far as
ordinary (\emph{Memory Coherence Required}) memory is concerned~\cite[p.~702]{ppc:2.06}.  We demonstrate
(see~\url{http://diy.inria.fr/cats/power-eieio/}) that~\eieio{}
cannot be a full barrier, as this option is invalidated by hardware. For
example the following \textsf{w+rwc+eieio+addr+sync} \DELETED{idiom}\NEW{pattern}
(\cf\myfig\ref{fig:w+rwc}) would be forbidden if~\eieio{} was a full barrier,
yet is observed on Power~$6$ and Power~$7$ machines. Indeed this \DELETED{idiom}\NEW{pattern} involves
two from-reads ($\fr$), in which case our axioms require us to resort to the full
fence part of the propagation order. Thus we take~\eieio{} to be a lightweight
barrier that maintains only write-write pairs.
\begin{figure}[!h]
\begin{center}
\input{img/w+rwc+eieio+addr+sync\bw.pstex_t}
\end{center}
\caption{The \DELETED{idiom}\NEW{pattern} \textsf{w+rwc} \NEW{with \eieio{}, address dependency and
full fence} \NEW{(allowed)} \label{fig:w+rwc}}
\end{figure}

The ARM architecture features the \dmb{} and \dsb{} fences.  ARM
documentation~\cite{arm:cookbook} forbids \textsf{iriw+dmb}
(\cf\myfig\ref{fig:iriw}). Hence \dmb{} is a full fence, as this
behaviour involves two from-reads ($\fr$), and thus needs full fences to be
forbidden.
\begin{figure}[!h]
\begin{center}
\input{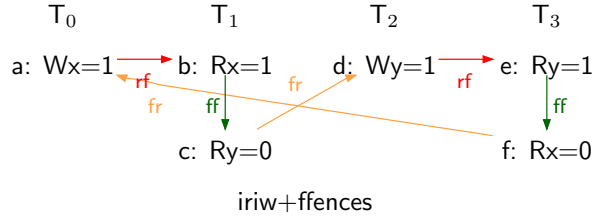}
\end{center}
\caption{The independent reads from independent writes \DELETED{idiom}\NEW{pattern} \textsf{iriw} \NEW{with full fences} \NEW{(forbidden)}\label{fig:iriw}}
\end{figure}

The ARM documentation also specifies \dsb{} to behave at least as \NEW{strongly
as} \dmb~\cite[p.~A3-49]{arm:arm08}: ``\emph{A \dsb{} behaves as a \dmb{} with
the same arguments, and also has [\ldots] additional properties [\ldots]}''.
Thus we take both \dmb{} and~\dsb{} to be full fences. \NEW{We remark that this
observation and our experiments (see \mysec\ref{sec:testing}) concern memory
litmus tests only; we do not know whether \dmb{} and~\dsb{} differ (or indeed
should differ) when out-of-memory communication (\eg{} interrupts) comes into
play.} 

Finally, the ARM architecture specifies that, when suffixed by \textsf{.st}, ARM
fences operate on write-to-write pairs only. It remains to be decided whether
the resulting \dmbst{} and~\dsbst{} fences are lightweight fences or not.  In
that aspect ARM documentation does not help much, nor do experiments on
hardware, as all our experiments are compatible with both alternatives (see
\url{http://diy.inria.fr/cats/arm-st-fences}).  Thus we choose
simplicity and consider that \textsf{.st} fences behave as their unsuffixed
counterparts, but limited to write-to-write pairs.
In other words, we assume
that the ARM architecture makes no provision for some kind of lightweight
fence, unlike Power which provides with \lwsync.

\NEW{Formally, to account for \textsf{.st} fences being full fences limited to
write-to-write pairs, we would proceed as follows. In
\myfig\ref{fig:fences-power+arm} for ARM, we would extend the definition of
$\ffence$ to $\dmb \cup \dsb \cup (\dmbst \cap \WW) \cup (\dsbst \cap \WW)$.
Should the option of \textsf{.st} fences being lightweight fences be prefered
(thereby allowing for instance the pattern \textsf{w+rwc+dmb.st+addr+dmb} of
\myfig\ref{fig:w+rwc}), one would instead define \lwfence{} as $(\dmbst \cap
\WW) \cup (\dsbst \cap \WW)$ in~\myfig\ref{fig:fences-power+arm} for ARM.}

\subsection{Some instances of our framework\label{sec:instances}}
We now explain how we instantiate our framework to produce the following
models: Lamport's Sequential Consistency~\cite{lam79} (SC), Total Store Order
(TSO), used in Sparc~\cite{sparc:94} and x86's~\cite{oss09} architectures, and
C++ R-A, \ie C++ restricted to the use of release-acquire
atomics~\cite{bos11,bdg13}.

\begin{figure}[!h]
\begin{mathpar}
\hspace*{-7mm} \textsf{SC: } \quad \dfn{\ppo}{\po} \and \dfn{\ffence}{\emptyset} \and
\dfn{\lwfence}{\emptyset} \and \dfn{\fences}{\ffence \cup \lwfence} \and \quad
\dfn{\prop}{\ppo \cup \fences \cup \rf \cup \fr} \\
\end{mathpar}
\vspace*{-6mm}
\hrule
\begin{mathpar}
\textsf{TSO: } \,\,\, \dfn{\ppo}{\po \setminus \WR} \and \dfn{\ffence}{\mfence} \and \dfn{\lwfence}{\emptyset} \and \dfn{\fences}{\ffence \cup \lwfence} \and \dfn{\prop}{\ppo \cup \fences \cup \rfe \cup \fr}
\end{mathpar}
\vspace*{-3mm}
\hrule
\begin{mathpar}
\textsf{C++ R-A $\approx$:} \quad \dfn{\ppo}{\syncbef} \quad \NEW{\text{(see \cite{bos11})}}
\and \dfn{\fences}{\emptyset} \and \dfn{\prop}{\hb}
\end{mathpar} 
\centerline{(we write $\approx$ to indicate that our definition
has a discrepancy with the standard)} 
\caption{Definitions of SC, TSO and C++ R-A\label{fig:sc+tso}} 
\end{figure}

\paragraph{SC and x86-TSO \!\!\!\!} are described in our terms on top
of~\myfig\ref{fig:sc+tso}, which gives their preserved program order, fences
and propagation order. 

We show that these instances of our model correspond\DELETED{s} to SC and TSO: 
\begin{lemma}Our model of SC as given in \myfig\ref{fig:sc+tso} is equivalent
to Lamport's SC~\cite{lam79}. Our model of TSO as given in
\myfig\ref{fig:sc+tso} is equivalent to Sparc TSO~\cite{sparc:94}.
\label{lem:sc+tso} \end{lemma}
\begin{proof} \noindent\small{\emph{ An execution $(\evts,\po,\co,\rf)$ is valid on
SC (resp.~TSO) iff $\acyclic(\po \cup \com)$ (resp.~$\acyclic(\ppo \cup \co
\cup \rfe \cup \fr \cup \fences)$) (all relations defined as in
\myfig\ref{fig:sc+tso}) by \cite[Def.~21,~p.~203]{alg12}
(resp.~\cite[Def.~23,~p.~204]{alg12}.)}} \qed \end{proof}

\paragraph{C++ R-A\!\!\!\!}, \ie C++ restricted to the use of
release-acquire atomics, appears at the bottom of~\myfig\ref{fig:sc+tso},
in a slightly stronger form than the current standard prescribes, as detailed
below. 

We take the \emph{\DELETED{synchronise}\NEW{sequenced} before} relation
$\syncbef$ of~\cite{bos11} to be the preserved program order, and the fences to
be empty. The happens-before relation $\dfn{\hb}{\syncbef \cup \rf}$ is the
only relation contributing to propagation, \ie $\prop=\transc{\hb}$. We take
the modification order $\modord$ of~\cite{bos11} to be our coherence order.  

The work of \cite{bdg13} shows that C++ R-A is defined by three axioms:
\textsc{Acyclicity}, \ie $\acyclic(\hb)$ (which immediately corresponds to our
\textsc{no thin air} axiom); \textsc{CoWR}, \ie $\forall r. \neg(\exists w_1,
w_2.$ $(w_1,w_2) \in \mo \wedge (w_1,r) \in \rf \wedge (w_2,r) \in
\transc{\hb})$ (which corresponds to our
\textsc{\DELETED{causality}\NEW{observation}} axiom by definition of $\fr$ and
since $\prop=\transc{\hb}$ here); and \textsc{HBvsMO}, \ie
$\irrefl(\transc{\hb};\modord)$. Our \DELETED{\textsc{uniproc}}\NEW{\textsc{sc
per location}} is implemented by \textsc{HBvsMO} for the \textsf{coWW} case,
and the eponymous axioms for the \textsf{coWR, coRW, coRR} cases.

Thus C++ R-A corresponds to our version, except for the \textsc{HBvsMO} axiom,
which requires the irreflexivity of $\transc{\hb};\modord$, whereas we require
its acyclicity via the axiom \textsc{propagation}. To adapt our framework to
C++ R-A, one simply needs to weaken the \textsc{propagation} axiom to
$\irrefl(\prop;\co)$.

\subsection{\NEW{A note on the genericity of our framework}\label{model:gen}}

\NEW{We remark that our framework is, as of today, not as generic as it could
be, for several reasons that we explain below.}

\paragraph{\NEW{Types of events}}
\NEW{For a given event (\eg read, write or barrier), we handle only one type of
this event. For example we can express C++ when all the reads are acquire and
all the writes are release, but nothing more. As of today, we could not express
a model where some writes can be relaxed writes (in the C++ sense), and others
can be release writes.  To embrace models with several types of accesses, \eg
C++ in its entirety, we would need to extend the expressivity of our
framework. We hope to investigate this extension in future work.}

\paragraph{\NEW{Choice of axioms}}
\NEW{We note that our axiom \textsc{sc per location} might be perceived as too
strong, as it forbids \emph{load-load hazard} behaviours (see \textsf{coRR} in
\myfig\ref{fig:co}). This pattern was indeed officially allowed by Sparc
RMO~\cite{sparc:94} and pre-Power 4 machines~\cite{tdf02}.}
\fixmeskip{jade@Luc: peux-tu faire une classification des tests qui sont
attrapes juste par acyclic(prop), et ceux qui sont attrapes par
acyclic(co;prop)?}

\NEW{Similarly, the \textsc{no thin air} axiom might be perceived as
controversial, as several software models, such as Java or C++, allow certain
\textsf{lb} patterns (see \myfig\ref{fig:lb}).}

\NEW{We also have already discussed, in the section just above, how the
\textsc{propagation} axiom needs to be weakened to reflect C++ R-A
accurately.}

\NEW{Yet we feel that this is much less of an issue than having only one type
of events. Indeed one can very simply disable the \textsc{no thin air} check,
or restrict the \DELETED{\textsc{uniproc}}\textsc{sc per location} check so
that it allows load-load hazard (see for example \cite[Sec. 5.1.2]{alg12}), or
weaken the \textsc{propagation} axiom (as we do above).}

\NEW{Rather than axioms set in stone, that we would declare as the absolute
truth, we present here some basic bricks from which one can build a model at
will.  Moreover, our new \prog{herd} simulator (see \mysec\ref{sec:herd})
allows the user to very simply and quickly modify the axioms of his model, and
re-run his tests immediately, without to have to dive into the code of the
simulator. This makes the cost of experimenting with several different variants
of a model, or fine-tuning a model, much less high. We give a flavour of such
experimentations and fine-tuning in our discussion about the ARM model (see for
example \mytab\ref{fig:summary-testing}.)}

\section{Instruction semantics}\label{sec:instr-sem}
\NEW{In this section, we specialise the discussion to Power, to make the
reading easier.} Before presenting the preserved program order for Power, given
in~\myfig\ref{fig:ppo}, we need to define \emph{dependencies}. We borrow the
names and notations of~\cite{ssa11} for more consistency.  \fixmeskip{jade:
formal dfn of addr, data, ctrl}

To define dependencies formally, we need to introduce some more possible
actions for our events. In addition to the read and write events relative to
memory locations, we can now have:
\begin{itemize}
\item read and write events relative to \emph{registers}, \eg for the register \textsf{r} we
can have a read of the value $0$, noted \textsf{Rr=0}, or a write of the value $1$,
noted \textsf{Wr=1};
\item branching events, which represent the branching decisions being made;
\item fence events, \eg \lwfence.
\end{itemize}

Note that in general a single instruction can involve several accesses, for
example, to registers.  Events coming from the same instruction can be related
by the relation $\iico$ (intra-instruction causality order). We give examples
of such a situation below.

\subsection{Semantics of instructions} We now give examples of the semantics
of instructions. We do not intend to be exhaustive, but rather to give the
reader enough understanding of the memory, register, branching and fence events
that an instruction can generate, so that we can define dependencies \wrt these
events in due course. We use Power assembly syntax~\cite{ppc:2.06}.

Here is the semantics of a load ``\textsf{lwz r2,0(r1)}'' with \textsf{r1}
holding the address \textsf{x}, assuming that \textsf{x} contains the value
$0$. The instruction reads the content of the register \textsf{r1}, and finds
there the memory address \textsf{x}; then (following $\iico$) it reads
\textsf{x}, and finds there the value $0$.  Finally, it writes this value into
register \textsf{r2}: 
\begin{center}
\scalebox{0.28}{\input{ins/load\bw.pstex_t}} 
\end{center}

Here is the semantics of a store~``\textsf{stw r1,0(r2)}'' with~\textsf{r1}
holding the value~$1$ and~\textsf{r2} holding the address~\textsf{x}. The
instruction reads the content of the register~\textsf{r2}, and finds there the
memory address~\textsf{x}. In parallel, it reads the content of the register
\textsf{r1}, and finds there the value~$1$. After (in $\iico$) these two
events, it writes the value $1$ into memory address~\textsf{x}:
\begin{center}
\scalebox{0.28}{\input{ins/store\bw.pstex_t}}
\end{center}

Here is the semantics of a ``\textsf{sync}'' fence, simply an eponymous event:
\begin{center}
\scalebox{0.28}{\input{ins/sync\bw.pstex_t}}
\end{center}

Here is the semantics for a branch ``\textsf{bne L0}'', which branches to the
label \textsf{L0} if the value of a special register (the Power ISA specifies
it to be register \textsf{CR0}~\cite[Chap.~2, p.~35]{ppc:2.06}) is not equal to
$0$ (``\textsf{bne}'' stands for ``branch if not equal''). Thus the instruction
reads the content of \textsf{CR0}, and emits a branching event. Note that it
emits the branching event regardless of the value of \textsf{CR0}, just to
signify that a branching decision has been made:
\begin{center}
\scalebox{0.28}{\input{ins/commit\bw.pstex_t}}
\end{center}
\vspace*{-1.5mm}

Here is the semantics for a ``\textsf{xor r9,r1,r1}'', which takes the bitwise
\textsf{xor} of the value in \textsf{r1} with itself and puts the result into
the register \textsf{r9}. The instruction reads (twice) the value in the
register \textsf{r1}, takes their \textsf{xor}, and writes the result
(necessarily $0$) in \textsf{r9}:
\begin{center}
\scalebox{0.28}{\input{ins/xor\bw.pstex_t}}
\end{center}
\vspace*{-1.5mm}

Here is the semantics for a comparison ``\textsf{cmpwi r1, 1}'', which compares
the value in register \textsf{r1} with $1$. The instruction writes the result
of the comparison ($2$~encodes equality) into the special register \textsf{CR0}
(the same that is used by branches to make branching decisions).  Thus the
instruction reads the content of \textsf{r1}, then writes to \textsf{CR0}
accordingly:
\begin{center}
\scalebox{0.28}{\input{ins/cmp\bw.pstex_t}}
\end{center}
\vspace*{-1.5mm}

Here is the semantics for an addition ``\textsf{add r9,r1,r1}'', which reads
the value in register \textsf{r1} (twice) and writes the sum into register
\textsf{r9}:
\begin{center} 
\scalebox{0.28}{\input{ins/add\bw.pstex_t}}
\end{center}
\vspace*{-1.5mm}

\subsection{Dependencies}
We can now define dependencies formally, in terms of the events generated by
the instructions involved in implementing a dependency. We borrow the textual
explanations from~\cite{ssa11}.

\NEW{In \myfig\ref{fig:dep}, we give the definitions of address
($\addr$), data ($\data$), control ($\ctrl$) and control+$\cfence$
($\ctrlcfence$) dependencies. Below we detail and illustrate what they mean.}

\begin{figure}[!h]
\[
\begin{tabular}{r@{~}c@{~}>{$}l<{$}p{.5\linewidth}}
\ddreg & = & \transc{(\rfreg \cup \iico)} & Data dependency over registers \\
\addr & = & \ddreg \cap \RM & The last \rfreg{} is to the address entry port of the target instruction.\\
\data & = & \ddreg \cap \RW & The last \rfreg{} is to the value entry port of the target store instruction.\\
\ctrl & = & (\ddreg \cap \RB);\po
& On Power or ARM, control dependencies targetting a read do not belong to~\ppo.\\
\ctrlcfence & = & (\ddreg \cap \RB);\cfence & On Power or ARM, branches followed
by a control fence (\isync{} on Power, \isb{} on ARM) targetting a read
belong to~\ppo.
\end{tabular}
\]
\caption{\NEW{Definitions of dependency relations}\label{fig:dep}}
\end{figure}

\NEW{In \myfig\ref{fig:dep}, we use the notations that we have defined before
(see \mysec\ref{sec:model} for sets of events. We write ``\textsf{M}'' for the
set of all memory events, so that for example \textsf{RM} is the set of pairs
$(r,e)$ where $r$ is a read and $e$ any memory event (\ie a write or a read).
We write ``\textsf{B}'' for the set of branch events (coming from a branch
instruction); thus \textsf{RB} is the set of pairs $(r,b)$ where $r$ is a read
and $b$ a branch event.}

\NEW{Each definition uses the relation $\ddreg$, defined as $\transc{(\rfreg
\cup \iico)}$.  This relation $\ddreg$ builds chains of read-from through
registers (\rfreg) and intra-instruction causality order (\iico). Thus it is a
special kind of data dependency, over register accesses only.}

\NEW{We find that the formal definitions in \myfig\ref{fig:dep} make quite
evident that all these dependencies ($\addr, \data, \ctrl$ and $\ctrlcfence$)
all correspond to data dependencies over registers ($\ddreg$), starting with a
read. The key difference between each of these dependencies is the kind of
events that they target: whilst $\addr$ targets any kind of memory event,
$\data$ only targets writes. A $\ctrl$ dependency only targets branch events; a
$\ctrlcfence$ also targets only branch events, but requires a control fence
\cfence{}  to immediately follow the branch event.}

\begin{figure}[!h]
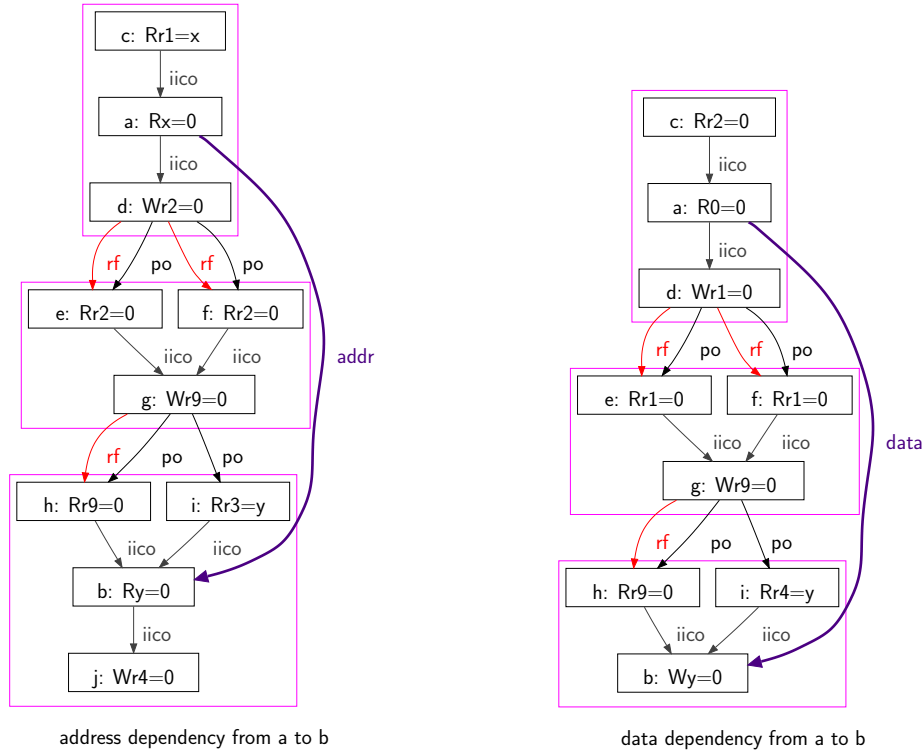

\begin{center}
\scalebox{0.28}{\input{ins/addr\bw.pstex_t}}\quad\quad\quad\quad\scalebox{0.28}{\input{ins/data\bw.pstex_t}}
\end{center}
\caption{\NEW{Data-flow dependencies}\label{fig:addr:data}}
\end{figure}

\subsubsection{Address dependencies\label{sec:addr-dep}} Address dependencies
are gathered in the~$\addr$ relation.  There is an address dependency from a
memory read~$r$ to a~$\po$-subsequent memory event~$e$ (either read or write)
if there is a data flow path \NEW{(\ie a \ddreg{} relation)} from~$r$ to the
address of~$e$ through registers and arithmetic or logical operations (but not
through memory accesses). Note that this notion also includes
\DELETED{so-called} \emph{false dependencies}, \eg when \textsf{xor}'ing a
value with itself and using the result of the \textsf{xor} in an address
calculation. For example, in Power (on the left) or ARM assembly (on the
right), the following excerpt

\begin{tabular}{m{.5\linewidth} m{.5\linewidth}}
{\sf (1) lwz r2,\DELETED{0,r1}\NEW{0(r1)}}  & \sf ldr r2,[r1] \\
{\sf (2) xor r9,r2,r2} & \sf eor r9,\DELETED{r1,r1}\NEW{r2,r2} \\
{\sf (3) lwzx r4,r9,r3} & \sf ldr r4,[r9,r3]
\end{tabular}
ensures that the load at line~{\sf (3)} cannot be reordered with the load at
line~{\sf (1)}, despite the result of the~{\sf xor} at line~{\sf (2)} being
always~$0$.

Graphically \NEW{(\cf the left diagram of \myfig\ref{fig:addr:data})}, the read $a$ from address $x$ is related by $\addr$ to the
read $b$ from address $y$ because there is a path of $\rf$ and $\iico$ (through
register events) between them.
\NEW{Notice that the last $\rf$ is to the index register (here \textsf{r9})
of the load from~$y$ instruction.}

\subsubsection{Data dependencies} Data dependencies are gathered in the~$\data$
relation.  There is a data dependency from a memory read~$r$ to
a~$\po$-subsequent memory write~$w$ if there is a data flow path \NEW{(\ie a
\ddreg{} relation)} from~$r$ to the value written by~$w$ through registers and
arithmetic or logical operations (but not through memory accesses). This also
includes false dependencies, as described above.  For example

\begin{tabular}{m{.5\linewidth} m{.5\linewidth}}
{\sf (1) lwz r2,\DELETED{0,r1}\NEW{0(r1)}}  & \sf ldr r2,[r1] \\
{\sf (2) xor r9,r2,r2} & \sf eor r9,r2,r2\\
{\sf (\DELETED{4}\NEW{3}) \DELETED{stwx r9,0,r4}\NEW{stw r9,0(r4)}} & \sf str r9,[r4]
\end{tabular}
ensures that the store at line~{\sf (\DELETED{4}\NEW{3})} cannot be reordered with the load at
line~{\sf (1)}, despite the result of the~{\sf xor} at line~{\sf (2)} being
always~$0$.

Graphically \NEW{(see the right diagram of \myfig\ref{fig:addr:data})}, the read $a$ from address $x$ is related by $\data$ to the
write $b$ to address $y$ because there is a path of $\rf$ and $\iico$
(through registers) between them\NEW{, the last $\rf$ being to the value entry
port of the store.}

\NEW{\paragraph{Remark:} Our semantics does not account for conditional
execution in the ARM sense (see \cite[\mysec A8.3.8 ``Conditional
execution'']{arm:arm08} and \cite[\mysec 6.2.1.2]{arm:cookbook}).  Informally,
most instructions can be executed or not, depending on condition flags.  It is
unclear how to handle them in full generality, both as target of dependencies
(conditional load and conditional store instructions); or in the middle of a
dependency chain (\eg conditional move).  In the target case, a dependency
reaching a conditional memory instruction through its condition flag could act
as a control dependency.  In the middle case, the conditional move could
contribute to the data flow path that defines address and data dependencies. We
emphasise that we have not tested these hypotheses.}

\begin{figure}[!h]
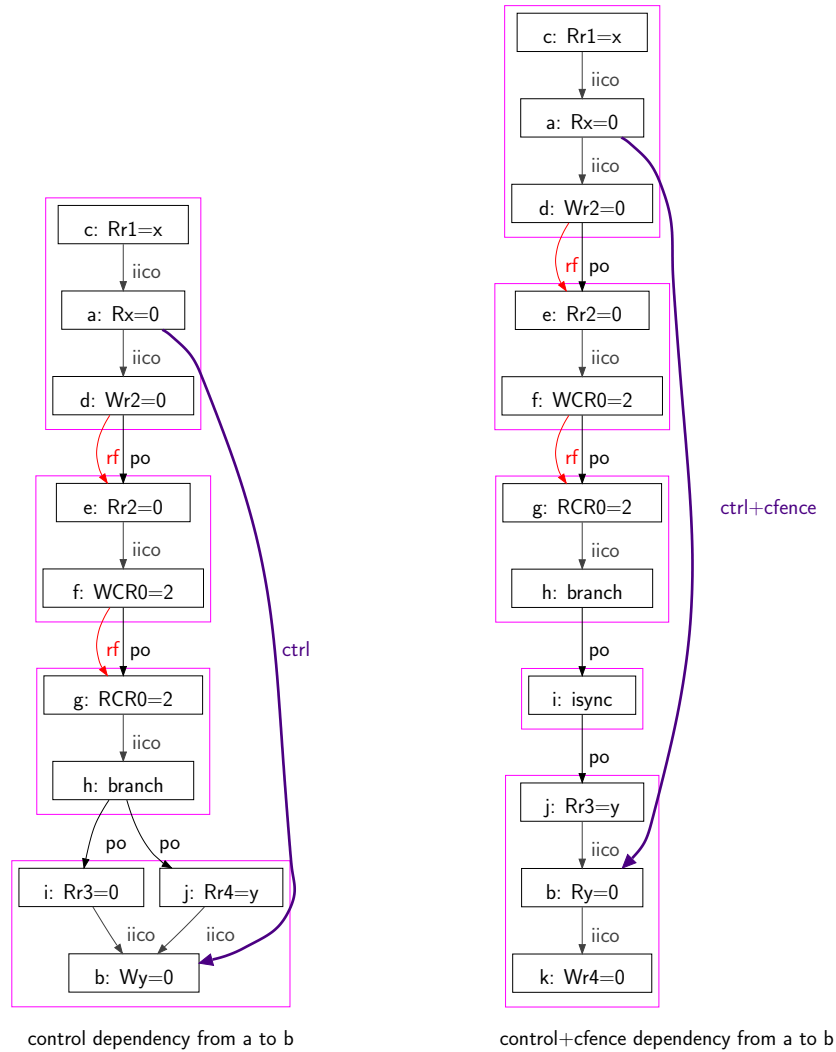

\begin{center}
\begin{tabular}{b{.4\linewidth}b{.4\linewidth}}
\scalebox{0.28}{\input{ins/ctrl\bw.pstex_t}} & \scalebox{0.28}{\input{ins/ctrlcfence\bw.pstex_t}}
\end{tabular}
\end{center}
\caption{\NEW{Control-flow dependencies}\label{fig:ctrl}}
\end{figure}

\subsubsection{Control dependencies} Control dependencies are gathered in
the~$\ctrl$ relation.  There is a control dependency from a memory read~$r$ to
a~$\po$-subsequent memory write~$w$ if there is a data flow path \NEW{(\ie a
\ddreg{} relation)} from~$r$ to the test of a conditional branch that
precedes~$w$ in~$\po$. For example

\begin{tabular}{m{.5\linewidth} m{.5\linewidth}}
{\sf (1) lwz r2,\DELETED{0,r1}\NEW{0(r1)}}  & \sf ldr r2,[r1]\\
{\sf (2) cmpwi r2,0} &  \sf cmp r2,\#0\\
{\sf (3) bne L0} & \sf bne L0 \\
{\sf (4) \DELETED{stwx r3,0,r4}\NEW{stw r3,0(r4)}} & \sf str r3,[r4]\\
{\sf (5) L0:} & \sf L0:
\end{tabular}
ensures that the store at line~{\sf (4)} cannot be reordered with the load at
line~{\sf (1)}.
\NEW{We note that there will still be a control dependency from the load to the store, even if the label immediately follows the branch, \ie the label \textsf{L0} is placed between the conditional branch instruction at line \textsf{(3)}
and the store.}

Graphically\NEW{ (\cf the left diagram\footnote{The diagram depicts the situation for Power; ARM status
flags are handled differently.} of \myfig\ref{fig:ctrl})}, the read $a$ from
address $x$ is related by $\ctrl$ to the
write $b$ to address $y$ because there is a path of $\rf$ and $\iico$
(through registers) between $a$ and a branch event ($h$ in that case)
$\po$-before $b$ (some $\po$ edges are omitted for clarity).

Such a data flow path between two memory reads is not enough to order them in
general.  To order two memory reads, one needs to place a \emph{control fence}
$\cfence$ ($\isync$ on Power, $\isb$ on ARM, as shown on top
of~\myfig\ref{fig:ppo}) after the branch, as described below.

\subsubsection{Control+$\cfence$ dependencies} Control+$\cfence$ dependencies
are gathered in the~$\ctrlcfence$ relation. There is such a dependency from a
memory read~$r_1$ to a~$\po$-subsequent memory read~$r_2$ if there is a data
flow path \NEW{(\ie a \ddreg{} relation)} from~$r_1$ to the test of a
conditional branch that $\po$-precedes a control fence, the fence itself
preceding $r_2$ in~$\po$. For example

\begin{tabular}{m{.5\linewidth} m{.5\linewidth}}
{\sf (1) lwz r2,\DELETED{0,r1}\NEW{0(r1)}}  & \sf ldr r2,[r1]\\
{\sf (2) cmpwi r2,0} &  \sf cmp r2,\#0\\
{\sf (3) bne L0} & \sf bne L0 \\
{\sf (4) isync} & \sf isb \\
{\sf (5) lwz r4,\DELETED{0,r3}\NEW{0(r3)}} & \sf ldr r4,[r3]\\
{\sf (6) L0:} & \sf L0:
\end{tabular}
ensures, thanks to the control fence at line {\sf (4)}, that the load at
line~{\sf (5)} cannot be reordered with the load at line~{\sf (1)}.

\addtocounter{footnote}{-1}
Graphically \NEW{(\cf the right diagram\footnotemark{}
of \myfig\ref{fig:ctrl})},
\addtocounter{footnote}{1}%
the read $a$ from address $x$ is related by $\ctrlcfence$
to the read $b$ from address $y$ because there is a path of $\rf$ and
$\iico$ (through registers) between $a$ and a branch event ($h$ here)
$\po$-before a $\cfence$ ($i:\isync$ here)
$\po$-before $b$.

\section{Preserved program order for Power}
\label{sec:power}

We can now present how to compute the preserved program order for Power, which
we give in~\myfig\ref{fig:ppo}. Some readers might find it easier to read the
equivalent specification given in \myfig\ref{fig:herd-ppc-model}.  ARM is
broadly similar; we detail it in the next section, in the light of our
experiments.
 
To define the preserved program order, we first need to distinguish two parts
for each memory event.  To name these parts, we again borrow the terminology of
the models of~\cite{ssa11,mms12} for more consistency. We give a table of
correspondence between the names of~\cite{ssa11,mms12} and the present paper in
\mytab\ref{fig:term}.

\begin{table}[!h] 
\begin{center} 
\begin{tabular}{c|c|c} 
present paper & \cite{ssa11} & \cite{mms12} \\\hline 
init read $\init(r)$ & satisfy read & satisfy read \\ 
commit read $\commit(r)$ & commit read & commit read \\ 
init write $\init(w)$ & n/a & init write \\ 
commit write $\commit(w)$ & commit write & commit write 
\end{tabular} 
\end{center} 
\caption{Terminology correspondence\label{fig:term}}
\end{table} 

\bigskip
A memory read $r$ consists of an \emph{init} $\init(r)$, where it
\DELETED{decides}\NEW{reads} its value, and a \emph{commit} part $\commit(r)$,
where it becomes irrevocable. A memory write $w$ consists of an \emph{init}
part $\init(w)$, where its value becomes available locally, and a \emph{commit}
part $\commit(w)$, where the write is ready to be sent out to \DELETED{the
system}\NEW{to other threads}.  

\begin{figure}[!h]
\begin{mathpar}
\dfn{\dep}{\addr \cup \data} \and
\dfn{\rdw}{{\poloc} \cap {(\efr;\rfe)}} \and
\dfn{\detour}{{\poloc} \cap {(\ews ; \rfe)}} \and

\dfn{\iiz}{\dep \cup \rdw \cup \rfi} \and
\dfn{\ciz}{(\ctrlcfence) \cup \detour} \\
\dfn{\icz}{\emptyset} \and
\dfn{\ccz}{{\dep} \cup {\poloc} \cup {\ctrl} \cup {(\addr;\po)}} \\

\dfn{\ii}{\iiz \cup \ci \cup (\ic;\ci) \cup (\ii;\ii)} \and 
\dfn{\ci}{\ciz \cup (\ci;\ii) \cup (\cc;\ci)} \and
\dfn{\ic}{\icz \cup \ii \cup \cc \cup (\ic;\cc) \cup (\ii;\ic)} \and
\dfn{\cc}{\ccz \cup \ci \cup (\ci;\ic) \cup (\cc;\cc)} \\ \\
\dfn{\ppo}{(\ii \cap \RR) \cup (\ic \cap \RW)} 
\end{mathpar}
\caption{Preserved program order for Power\label{fig:ppo}}
\end{figure}

We now \DELETED{expose}\NEW{describe} how the parts of events relate to one
another. We do a case disjunction over the part of the events we are concerned
with (init or commit).

Thus we define four relations (\cf~\myfig\ref{fig:ppo}): $\ii$~relates the init
parts of events; $\ic$~relates the init part of a first event to the commit
part of a second event; $\ci$~relates the commit part of a first event to the
init part of a second event; $\cc$~relates the commit parts of events. We
define these four relations recursively, with a least fixpoint semantics; we
write \textsf{r}$_0$ for the base case of the recursive definition of a
relation \textsf{r}.

Note that the two parts of an event $e$ are ordered: its init~$\init(e)$
precedes its commit~$\commit(e)$. Thus for two events~$e_1$ and~$e_2$, if for
example the commit of~$e_1$ is before the commit of~$e_2$ (\ie~$(e_1,e_2) \in
\cc$), then the init of~$\e_1$ is before the commit of~$e_2$ (\ie~$(e_1,e_2)
\in \ic$). Therefore we have the following inclusions (see
also~\myfig\ref{fig:ppo} and \NEW{\ref{fig:ic}}): $\ii$~contains $\ci$;
$\ic$~contains $\ii$ and $\cc$; $\cc$ contains $\ci$.

\begin{figure}[!h]
\begin{center}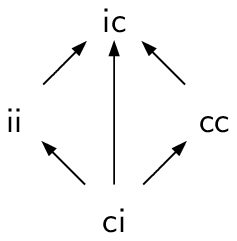\end{center}
\vspace*{-4mm}
\caption{\NEW{Inclusions between subevents relations}\label{fig:ic}}
\end{figure}

Moreover, these orderings hold transitively: for three events~$e_1,e_2$
and~$e_3$, if the init of~$e_1$ is before the commit of~$e_2$ (\ie~$(e_1,e_2)
\in \ic$), which is itself before the init of~$e_3$ (\ie $(e_2,e_3) \in \ci$),
then the init of~$e_1$ is before the init of~$e_3$ (\ie~$(e_1,e_3) \in \ii$).
Therefore we have the following inclusions (see also~\myfig\ref{fig:ppo}):
$\ii$~contains $(\ic;\ci)$ and $(\ii;\ii)$; $\ic$~contains $(\ic;\cc)$ and
$(\ii;\ic)$; $\ci$~contains $(\ci;\ii)$ and $(\cc;\ci)$; $\cc$~contains
$(\ci;\ic)$ and $(\cc;\cc)$.

We now describe the base case for each of our four relations, \ie
$\iiz,\icz,\ciz$ and $\ccz$. We do so by a case disjunction over whether the
concerned events are reads or writes. We omit the cases where there is no
ordering, \eg between two init writes.
 
\paragraph{Init reads \!\!\!} ($\iiz \cap \RR$ in~\myfig\ref{fig:ppo}) are
ordered by \textsf{(i)} the $\dep$ relation (which stands for
``dependencies''), which gathers address ($\addr$) and data ($\data$)
dependencies as defined above; \textsf{(ii)} the $\rdw$ relation (``read
different writes''), defined as ${\poloc} \cap {(\efr;\rfe)}$. 

The \textsf{(i)} case means that if two reads are separated (in program order)
by \DELETED{either }an address \DELETED{or a data dependency}, then their init
parts are ordered.  \NEW{The micro-architectural intuition is here quite clear:
we simply lift to the model level the constraints induced by data-flow paths in
the core. The vendors documentations~\cite[Book~II, \mysec1.7.1]{ppc:2.06}
and~\cite[\mysec6.2.1.2]{arm:cookbook} support our intuition, as the following
quotes show. }

\NEW{Quoting Power's documentation~\cite[Book~II, \mysec1.7.1]{ppc:2.06}:
``[i]f a load instruction depends on the value returned by a preceding load
instruction (because the value is used to compute the effective address
specified by the second load), the corresponding storage accesses are performed
in program order with respect to any processor or mechanism [\dots].'' We
interpret this bit as a justification for address dependencies (\addr)
contributing to the $\ppo$.}

\NEW{Furthermore, Power's documentation adds: ``[t]his applies even if the
dependency has no effect on program logic (e.g., the value returned by the
first load is ANDed with zero and then added to the effective address specified
by the second load).'' We interpret this as a justification for ``false
dependencies'', as described in \mysec\ref{sec:addr-dep}.}

\NEW{Quoting ARM's documentation~\cite[\mysec6.2.1.2]{arm:cookbook}: ``[if] the
value returned by a read is used to compute the virtual address of a subsequent
read or write (this is known as an address dependency), then these two memory
accesses will be observed in program order. An address dependency exists even
if the value read by the first read has no effect in changing the virtual
address (as might be the case if the value returned is masked off before it is
used, or if it had no effect on changing a predicted address value).'' We
interpret this ARM rule as we do for the Power equivalent.}

\NEW{We note however that the Alpha architecture (see our discussion of
\textsf{mp+fence+addr} being allowed on Alpha on page~\pageref{alpha:addr})
demonstrate that sophisticated hardware may invalidate the lifting of core
constraint to the complete system.}

The \textsf{(ii)} case means that the init parts of two reads $b$ and $c$ are
ordered when the instructions are in program order, reading from the same
location, and the first read $b$ reads from an external write which is
$\co$-before the write $a$ from which the second read $c$ reads, as shown in
\myfig\ref{fig:rdw}.
\begin{figure}[!h]
\begin{center}
\input{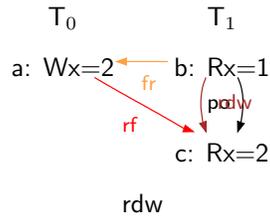}
\end{center}
\vspace*{-5mm}
\caption{The read different writes \textsf{rdw} \DELETED{idiom}\NEW{relation}\label{fig:rdw}}
\end{figure}

\NEW{We view this as a side-effect of the write propagation model: as the
read~$b$ reads from some write that is \co-before the write~$a$, $b$ is
satisfied before the write~$a$ is propagated to \myth{1}; furthermore as the
read~$c$ reads from the write~$a$, it is satisfied after the write~$a$ is
propagated to~\myth{1}. As a result, the read~$b$ is satisfied before the
read~$c$ is.}

\paragraph{Init read and init write \!\!\!\!} ($\iiz \cap \RW$) relate by
address and data dependencies, \ie~$\dep$. 
\NEW{This bit is similar to the ordering between two init reads (see $\iiz \cap
\RR$), but here both address and data dependencies are included.  Vendors
documentations \cite[Book~II, \mysec 1.7.1]{ppc:2.06}
and~\cite[\mysec A3.8.2]{arm:arm08} document address and data dependencies from
read to write.}

\NEW{Quoting Power's documentation \cite[Book~II, \mysec 1.7.1]{ppc:2.06}:
``[b]ecause stores cannot be performed ``out-of-order'' (see Book III), if a
store instruction depends on the value returned by a preceding load instruction
(because the value returned by the load is used to compute either the effective
address specified by the store or the value to be stored), the corresponding
storage accesses are performed in program order.'' We interpret this bit as a
justification for lifting both address and data dependencies to the $\ppo$.}

\NEW{Quoting ARM's documentation~\cite[\mysec A3.8.2]{arm:arm08}:
``[i]f the value returned by a read access is used as data written by a
subsequent write access, then the two memory accesses are observed in program
order.'' We interpret this ARM rule as we did for Power above.}

\emph{Init write and init read} ($\iiz \cap \WR$) relate by the internal
read-from~$\rfi$.
\NEW{This ordering constraint stems directly from our interpretation of init
subevents (\cf \mytab\ref{fig:term}): init for a write is the point in time
when the value stored by the write becomes available locally, while init for a
read is the point in time when the read is satisfied. Thus a read can be
satified from a local write only once the write in question has made its value
available locally.}

\emph{Commit read and init read \NEW{or write}} ($\ciz \cap \NEW{\RM}$
\DELETED{$\RR$}) relate by $\ctrlcfence$ dependencies.  \NEW{The
$\ctrlcfence$ relation models the situation where a first read controls the
execution of a branch which contains a control fence that \po-precedes the
second memory access.  An implementation of the control fence could refetch the
instructions that \po-follows the fence (see for example the quote of ARM's
documentation~\cite[Sec.~A3.8.3]{arm:arm08} that we give below), and prevent
any speculative execution of the control fence.  As a result, instructions that
follow the fence may start only once the branching decision is irrevocable, \ie
once the controling read is irrevocable.} 

\NEW{Quoting Power's documentation~\cite[Sec~1.7.1]{ppc:2.06}: ``[b]ecause an
isync instruction prevents the execution of instructions following the \isync{}
until instructions preceding the isync have completed, if an \isync{} follows a
conditional branch instruction that depends on the value returned by a
preceding load instruction, the load on which the branch depends is performed
before any loads caused by instructions following the \isync{}.'' We interpret
this bit as saying that a branch followed by an \isync{} orders read-read pairs
on Power (read-write pairs only need a branch, without an \isync{}).
Furthermore the documentation adds: ''[t]his applies even if the effects of the
``dependency'' are independent of the value loaded (e.g., the value is compared
to itself and the branch tests the EQ bit in the selected CR field), and even
if the branch target is the sequentially next instruction.'' This means that
the dependency induced by the branch and the \isync{} can be a ``false
dependency''.}

\NEW{Quoting ARM's documentation~\cite[Sec. A3.8.3]{arm:arm08}: ``[a]n \isb{}
instruction flushes the pipeline in the processor, so that all instructions
that come after the \isb{} instruction in program order are fetched from cache
or memory only after the \isb{} instruction has completed.'' We interpret this
bit as saying that a branch followed by an \isb{} orders read-read and
read-write pairs on ARM.} 

\emph{Commit write and init read} ($\ciz \cap \WR$)
relate by the relation $\dfn{\detour}{{\poloc} \cap {(\ews;\rfe)}}$. This means
that the commit of a write $b$ to memory location $x$ precedes the init of a
read $c$ from $x$ when $c$ reads from an external write $a$ which follows $b$
in coherence, as shown in \myfig\ref{fig:detour}.  \begin{figure}[!h]
\begin{center}
\input{img/detour\bw.pstex_t} \end{center} \vspace*{-5mm} \caption{The \textsf{detour}
\DELETED{idiom}\NEW{relation} \label{fig:detour}} \end{figure} 

\NEW{This effect is similar to the \textsf{rdw} effect (\cf
\myfig\ref{fig:rdw}) but applied to write-read pairs instead of read-read pairs
in the case of \textsf{rdw}.  As a matter of fact, in the write propagation
model of~\cite{ssa11,mms12}, the propagation of a write~$w_2$ to a
thread~\textsf{T} before \textsf{T} makes a write~$w_1$ to the same address
available to the memory system implies that $w_2$ $\co$-precedes $w_1$. Thus,
if the local $w_1$ $\co$-precedes the external $w_2$, it must be that $w_2$
propagates to~\textsf{T} after $w_1$ is committed.  In turn this implies that
any read that reads from the external~$w_2$ is satisfied after the local~$w_1$
is committed.}

\paragraph{Commits \!\!\!\!} relate by program order if they access the
same location, \ie $\poloc$. Thus in~\myfig\ref{fig:ppo}, $\ccz$ contains
$\poloc$. 

\NEW{We inherit this contraint from~\cite{mms12}.  From the implementation
standpoint, the core has to perform some actions so as to enforce the
\textsc{sc per location} check, \ie the five patterns of \myfig\ref{fig:co}. One way could be to perform these actions at commit time, which this bit of the $\ppo$ represents.}

\NEW{However note that our model performs  the \textsc{sc per location} check
independently of the definition of the $\ppo$. Thus, the present
commit-to-commit ordering constraint is not required to enforce this particular
axiom of the model. Nevertheless, if this bit of the $\ppo$ definition is
adopted, it will yield specific ordering constraint, such as $\poloc \in \ccz$.
As we shall see in \mysec\ref{sec:arm-testing} this very constraint needs to be
relaxed to account for some behaviours observed on Qualcomm systems.}

In addition, \emph{commit read\DELETED{s}} \NEW{and commit read or write}
($\ccz \cap \NEW{\RM}\DELETED{\RR}$) relate by address, data
and control dependencies, \ie $\dep$ \NEW{and $\ctrl$}.
\DELETED{\emph{Commit read and commit write} ($\ccz \cap \RW$)
relate by control dependencies, \ie $\ctrl$.}

\NEW{Here and below (see \myfig\ref{fig:lb+addrs+ww}) we express ``chains of
irreversibility'', when some memory access depends, by whatever means, on a
\po-preceding read it can become irrevocable only when the read it depends upon
has.} \fixmeskip{jade@Luc: je ne comprends pas pourquoi}

Finally, a read $r$ must be
committed before the commit of any access $e$ that is program order after any
access $e'$ which is $\addr$-after $r$, as in the
\textsf{lb+addrs+ww}\footnote{\NEW{We note that this pattern does not quite
follow the naming convention that we have outlined in
\myfig\ref{fig:gloss-litmus}, but we keep this name for more consistency with
previous works, \eg~\cite{ssa11,smo12,mms12}.}}\DELETED{idiom}\NEW{pattern} in
\myfig\ref{fig:lb+addrs+ww}.

\begin{figure}[h]
\begin{center}
\input{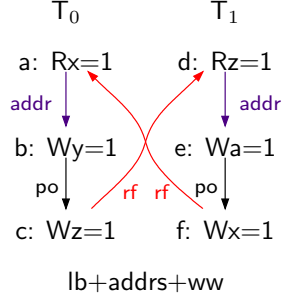}
\end{center}
\caption{\NEW{A variant of the load buffering pattern \textsf{lb} (see also \myfig\ref{fig:lb})}\label{fig:lb+addrs+ww}}
\end{figure}

In \NEW{the thread} \myth{0} \DELETED{above}\NEW{of
\myfig\ref{fig:lb+addrs+ww}}, the write to~$z$ cannot be read by another thread
before the write $b$ knows its own address, which in turn requires the read $a$
to be committed because of the address dependency between $a$ and $b$. Indeed
if the address of $b$ was $z$, the two writes $b$ and $c$ to $z$ could violate
the \textsf{coWW} \DELETED{idiom}\NEW{pattern}, an instance of our \textsc{sc
per location} axiom.  Hence the commit of the write $c$ has to be delayed (at
least) until after the read $a$ is committed.
 
Note that the same pattern with data instead of address dependencies is allowed
and observed (\cf \url{http://diy.inria.fr/cats/data-addr}).

\fixmeskip{add remarks on what difference it makes if we put (for example) rfi in
cc0, etc (which tests does it break)?}

\section{Operational models}
\label{sec:pldi}

We introduce here an operational equivalent of our model, our
\emph{intermediate machine}, given in \myfig\ref{fig:mach}, which we then
compare to a previous model, the \emph{PLDI machine} of~\cite{ssa11}. 

%
%





\subsection{Intermediate machine~\label{sec:int-mach}}
Our intermediate machine is simply a reformulation of our axiomatic model as a
transition system. We give its formal definition in~\myfig\ref{fig:mach}, which
we explain below. In this section, we write $\udr(\textsf{r})$ for the union of
the domain and the range of the relation \textsf{r}. \NEW{We write
$\textsf{r}{\small\textsf{++}}[e]$ to indicate that we append the element $e$
at the end of a total order $\textsf{r}$.}

\begin{figure}[!h]
\begin{mathpar}
\textsf{Input data: $(\ppo,\fences,\prop)$, $(\evts,\po)$ and $\textsf{p}$} \\
\textsf{Derived from \textsf{p}: $\dfn{\co(\evts,\textsf{p})}{\{(w_1,w_2) \mid}$ $\loc(w_1) = \loc(w_2) \wedge (\Flushw(w_1), \Flushw(w_2)) \in \textsf{p}\}$}
\bigskip

\inferrule[Commit write]{
\vbox{\hbox to 12.5cm {$
\begin{array}{c cccc c}
\textsc{(cw\label{cw:coww}: } \textsc{sc per location/} \textsf{coWW}) & & & & & 
\textsc{(cw\label{cw:prop}: } \textsc{propagation)} \\ 
\neg(\exists w' \in \NEW{\buff} \DELETED{\udr(\buff)} \text{ s.t. } (w,w') \in \poloc) & & & & &
\neg(\exists w' \in \NEW{\buff} \DELETED{\udr(\buff)} \text{ s.t. } (w,w') \in \prop) \\
\multicolumn{6}{c}{$\cwciwr$} \\
\multicolumn{6}{c}{\neg(\exists r \in \queue \text{ s.t. } (w,r) \in {\fences})} \\
\end{array}
$}}}
{s \xrightarrow{\Delw(w)} (\NEW{\buff \cup \{w\}} \DELETED{\buff{\small\textsf{++}}[w]},\rcp,\queue,\cread)}

\bigskip
\bigskip

\inferrule[Write reaches coherence point]{
\vbox{\hbox to 12.5cm {$\begin{array}{c cccc c}
\multicolumn{6}{c}{\textsc{(cpw: write is committed)}}  \\
\multicolumn{6}{c}{\NEW{w \in \buff} \DELETED{\udr(\buff)}} \\
\textsc{(cpw\label{cpw:accord}: } \NEW{\poloc}\DELETED{\buff} \textsc{and} \rcp \textsc{are in accord)} & & & & & 
\textsc{(cpw\label{cpw:prop}: propagation)} \\
\neg(\exists w' \in \udr(\rcp) \text{ s.t. } {\NEW{(w,w') \in \poloc} \DELETED{\buff}}) & & & & & 
\neg(\exists w' \in \udr(\rcp) \text{ s.t. } {(w,w') \in \NEW{\prop} \DELETED{(\co;\prop)}})\\
\end{array}$}}}
{s \xrightarrow{\Flushw(w)} (\buff,\NEW{\rcp{\small\textsf{++}}[w]} \DELETED{\cup \{w\}},\queue,\cread)}

\bigskip
\bigskip

\inferrule[Satisfy read]{
\vbox{\hbox to 13.5cm {$
\begin{array}{c cccc c}
\multicolumn{6}{c}{\textsc{(sr: write is either local or committed)}} \\ 
\multicolumn{6}{c}{{(w,r) \in \poloc} \vee {\NEW{w \in \buff} \DELETED{\udr(\buff)}}} \\ 
\textsc{(sr\label{sr:iiwr}: } \textsc{ppo/}\iiz \cap \RR) & & & & & 
\textsc{(sr\label{sr:caus}: } \textsc{\NEW{observation}\DELETED{causality})} \\
\neg(\exists r' \in \queue \text{ s.t. } (r,r') \in \ppo \cup \fences) & & & & & 
\neg(\exists w' \text{ s.t. } (w,w') \in \NEW{\co} \DELETED{\buff} \wedge (w',r) \in \prop;\rstar{\hb})
\end{array}$}}}
{s \xrightarrow{\Delr(w,r)} (\buff,\rcp,\queue \cup \{r\},\cread)}

\bigskip
\bigskip

\inferrule[Commit read]{
\vbox{\hbox to 14cm {$\begin{array}{c cccc c}
\textsc{(cr: read is satisfied)} & & & & & \textsc{(cr: sc per location/} \textsf{coWR, coRW\{1,2\}, coRR)} \\
r \in \queue & & & & & \visible(w,r) \\
\textsc{(cr: ppo/} \ccz \cap \RW) & & & & & \textsc{(cr: ppo/} (\ciz \cup \ccz) \cap \RR)  \\
\neg(\exists w' \in \buff \text{ s.t. } (r,w') \in \ppo \cup \fences) & & & & & 
\neg(\exists r' \in \queue \text{ s.t. } (r,r') \in \ppo \cup \fences) \\
\end{array}$}}}
{s \xrightarrow{\Flushr(w,r)} (\buff,\rcp,\queue,\cread \cup \{r\})}
\end{mathpar}
\caption{Intermediate machine\label{fig:mach}}
\end{figure}

Like the axiomatic model, our machine takes as input the events~$\evts$, the
program order~$\po$ and an architecture~$(\ppo, \fences, \prop)$.  

It also needs a \emph{path of labels}, \ie a total order over labels; a label
triggers a transition of the machine. We borrow the names of the labels
from~\cite{ssa11,smo12} for consistency.

We build the labels from the events~$\evts$ as follows: a write $w$ corresponds
to a \emph{commit write} label $\Delw(w)$ and a \emph{write reaching coherence
point} label $\Flushw(w)$; a read $r$ first guesses from which write $w$ it
might read, in an angelic manner; the pair $(w,r)$ then yields two labels:
\emph{satisfy read} $\Delr(w,r)$ and \emph{commit read} $\Flushr(w,r)$.  

For the architecture's functions $\ppo, \fences$ and $\prop$ to make sense, we need
to build a coherence order and a read-from map. We define them from the events
$\evts$ and the path $\textsf{p}$ as follows:
\begin{itemize}
\item $\co(\evts,\textsf{p})$ gathers the writes to the same memory location in the
order that their corresponding coherence point labels have in $\textsf{p}$: 
$\dfn{\co(\evts,\textsf{p})}{\{(w_1,w_2) \mid}$ $\loc(w_1) = \loc(w_2) \wedge (\Flushw(w_1), \Flushw(w_2)) \in \textsf{p}\}$;
\item $\rf(\evts,\textsf{p})$ gathers the write-read pairs with same location 
and value which have a commit label in $\textsf{p}$: 
$\dfn{\rf(\evts,\textsf{p})}{\{(w,r) \mid}$ $\loc(w) = \loc(r) \wedge \val(w) = \val(r) \wedge \Flushr(w,r) \in \udr(\textsf{p})\}$.
\end{itemize}
In the definition of our intermediate machine, we consider the relations
defined \wrt $\co$ and $\rf$ in the axiomatic model (\eg happens-before $\hb$,
propagation $\prop$) to be defined \wrt the coherence and read-from above.

Now, our machine operates over a state $(\buff,\rcp,\queue,\cread)$ composed of
\begin{itemize}
\item a \DELETED{total order} \NEW{set} $\buff$ (``committed writes'')
\DELETED{over}\NEW{of} writes \DELETED{to the same memory location} that have
been committed;
\item a \DELETED{set}\NEW{relation} $\rcp$ \DELETED{of}\NEW{over} writes having
reached coherence point, \NEW{which is a total order per location};
\item a set $\queue$ (``satisfied reads'') of reads having been satisfied;
\item a set $\cread$ (``committed reads'') of reads having been committed.
\end{itemize}

\subsubsection{Write transitions}
The order \NEW{in which writes enter the set} $\buff$ for a given location
corresponds to the coherence order for that location. Thus a write $w$ cannot
enter $\buff$ if it contradicts the \DELETED{\textsc{uniproc}}\NEW{\textsc{sc
per location}} and \textsc{propagation} axioms.  Formally, a commit write
$\Delw(w)$ yields a commit write transition, which appends $w$ at the end of
$\buff$ for its location if 
\begin{itemize}
\item \cwcoww{} there is no $\poloc$-subsequent write $w'$ which is already
committed, which forbids the \textsf{coWW} case of the
\DELETED{\textsc{uniproc}}\NEW{\textsc{sc per location}} axiom, and
\item \cwprop{} there is no $\prop$-subsequent write $w'$ which is
already committed, which ensures that the \textsc{propagation} axiom holds, and
\item \cwciwr{} there is no ${\fences}$-subsequent read $r$ which has
already been satisfied, which contributes to the semantics of the full fence.
\end{itemize}

A write can reach coherence point (\ie \DELETED{enter the $\rcp$ set}\NEW{take
its place at the end of the $\rcp$ order}) if:
\begin{itemize}
\item \cpwcom{} the write has already been committed, and
\item \cpwaccord{} all the \DELETED{previously committed writes (\ie
$\buff$-previous)} \NEW{$\poloc$-previous writes} have reached coherence point,
and
\item \cpwprop{} all the \DELETED{$(\co;\prop)$}\NEW{$\prop$}-previous writes have reached coherence point.
\end{itemize}

This ensures that the order in which writes reach coherence point is compatible
with \DELETED{the commit order,} the coherence order and the propagation order. 

\subsubsection{Read transitions}
The read set $\queue$ is ruled by the happens-before relation $\hb$. The way
reads enter $\queue$ ensures \textsc{no thin air} and
\textsc{\DELETED{causality}\NEW{observation}}, whilst the way they enter
$\cread$ ensures parts of \DELETED{\textsc{uniproc}}\NEW{\textsc{sc per
location}} and of the preserved program order. 

\paragraph{Satisfy read} The satisfy read transition $\Delr(w,r)$ places
the read $r$ in $\queue$ if:
\begin{itemize}
\item \srokw{} the write $w$ from which $r$ reads is either local (\ie $w$ is $\poloc$-before 
$r$), or has been committed already (\ie $w \in \NEW{\buff}\DELETED{\udr(\buff)}$), and
\item \sriirr{} there is no $(\ppo \cup \fences)$-subsequent read $r'$ that has
already been satisfied, which implements the $\iiz \cap \RR$ part of the
preserved program order, and
\item \srcaus{} there is no write $w'$ \DELETED{committed }\NEW{$\co$-}after
$w$ was\footnote{\NEW{Recall that in the context of our intermediate machine, $(w,w') \in \co$ means that $w$ and $w'$ are relative to the same
address and $(\Flushw(w),\Flushw(w')) \in \textsf{p}$.}} s.t.  $(w',r) \in
\prop;\rstar{\hb}$, which ensures
\textsc{\DELETED{causality}\NEW{observation}}.  
\end{itemize}

\paragraph{Commit read}

\addtocounter{footnote}{-1}
To define the commit read transition, we need a preliminary notion. We define a
write $w$ to be \emph{visible} to a read $r$ when
\begin{itemize}
\item $w$ and $r$ share the same location~$\lo$;
\item $w$ is equal to, or \DELETED{committed after
($\buff$-after)}\NEW{$\co$-}after\footnotemark{}, the last write $w_b$ to $\lo$
that is $\poloc$-before $r$, and
\addtocounter{footnote}{1}
\item $w$ is $\poloc$-before $r$, or \DELETED{committed before
($\buff$-before)}\NEW{$\co$-}before the first write $w_a$ to
$\lo$ that is $\poloc$-after $r$.
\end{itemize} 

We give here an illustration of the case where $w$ is
\DELETED{committed}\NEW{$\co$-}after $w_b$ and before $w_a$:
\begin{center}
\resizebox{.25\linewidth}{!}{\input{visible\bw.pstex_t}}
\end{center}

Now, recall that our read labels contain both a read $r$ and a write $w$ that it might
read. The commit read transition $\Flushr(w,r)$ records $r$ in $\cread$ when:
\begin{itemize}
\item \crsat{} $r$ has been satisfied (\ie is in $\queue$), and
\item \cruni{} $w$ is \emph{visible} to $r$, which prevents the \textsf{coWR},
\textsf{coRW1} and \textsf{coRW2} cases of
\DELETED{\textsc{uniproc}}\NEW{\textsc{sc per location}}, and
\item \crccrw{} there is no $(\ppo \cup \fences)$-subsequent write $w'$ that
has already been committed, which implements the ${\ccz} \cap \NEW{\RW}$
\DELETED{${\ccz} \cap \WR$} part of the preserved program order, and
\item \crci{} there is no $(\ppo \cup \fences)$-subsequent read $r'$ that has
already been satisfied, which implements the $\ciz \cap \RR$ and $\ccz \cap
\RR$ parts of the preserved program order.
\end{itemize}

To forbid the \textsf{coRR} case of \DELETED{\textsc{uniproc}}\NEW{\textsc{sc
per location}}, one needs to \textsf{(i)} make the read set $\cread$ record the
write from which a read takes its value, so that $\cread$ is a set of pairs
$(w,r)$ and no longer just a set of reads; and \textsf{(ii)} augment the
definition of $\visible(w,r)$ to require that there is no $(w',r')$ s.t. $r'$
is $\poloc$-before $r$ yet $w'$ is \DELETED{committed after} \NEW{$\co$-after}
$w$ \DELETED{(\ie $\buff$-after $w$)}.  We chose to present the simpler version
of the machine to ease the reading.

\subsection{Equivalence of axiomatic model and intermediate machine (proof of~\mythm\ref{thm:equiv})}

We can now state our equivalence result:
\begin{theorem}\label{thm:equiv}
All behaviours allowed by the axiomatic model are allowed by the intermediate
machine and conversely.  
\end{theorem}

We prove \mythm\ref{thm:equiv} in two steps. We first show that given a set of
events $\evts$, a program order $\po$ over these, and a path $\textsf{p}$ over
the corresponding labels such that $(\evts,\po,\textsf{p})$ is accepted by our
intermediate machine, the axiomatic execution
$(\evts,\po,\co(\evts,\textsf{p}),\rf(\evts,\textsf{p}))$ is valid in our
axiomatic model:
\begin{lemma}\label{lem:interm-ax}
All behaviours allowed by the intermediate machine are allowed by the axiomatic model.
\end{lemma}

Conversely, from a valid axiomatic execution $(\evts,\po,\co,\rf)$, we build a
path accepted by the intermediate machine:
\begin{lemma}\label{lem:ax-interm}
All behaviours allowed by the axiomatic model are allowed by the intermediate machine.
\end{lemma}

We give below the main arguments for proving our results. For more confidence,
we have implemented the equivalence proof between our axiomatic model and our
intermediate machine in the Coq proof assistant~\cite{coqart}. We give our
proof scripts online: \url{http://www0.cs.ucl.ac.uk/staff/j.alglave/cats}.

\subsubsection{From intermediate machine to axiomatic model (proof of \mylem\ref{lem:interm-ax})}

We show here that a path of labels~$\textsf{p}$ relative to a set of
events~$\evts$ and a program order~$\po$ accepted by our intermediate machine
leads to a valid axiomatic execution. To do so, we show that the
execution~$(\evts,\po,\co(\evts,\textsf{p}),\rf(\evts,\textsf{p}))$ is valid in
our axiomatic model.

\paragraph{Well formedness of $\co(\evts,\textsf{p})$ \!\!\!\!} (\ie $\co$ is a
total order on writes to the same location) follows from \textsf{p} being a
total order.

\paragraph{Well formedness of $\rf(\evts,\textsf{p})$ \!\!\!\!} (\ie $\rf$
relates a read to a unique write to the same location with same value) follows
from the fact that we require that each read $r$ in $\evts$ has a unique
corresponding write $w$ in $\evts$, s.t. $r$ and $w$ have same location and
value, and $\Flushr(w,r)$ is a label of \textsf{p}.

\paragraph{The \DELETED{\textsc{uniproc}}\NEW{\textsc{sc per location}} axiom
holds: \!\!\!} to prove this, we show that \textsf{coWW}, \textsf{coRW1},
\textsf{coRW2}, \textsf{coWR} and \textsf{coRR} are forbidden. 

\paragraph{\textsf{coWW} is forbidden: \!\!\!\!} suppose as a contradiction two
writes $e_1$ and $e_2$ to the same location s.t. $(e_1,e_2) \in \po$ and
$(e_2,e_1) \in \co(\evts,\textsf{p})$. The first hypothesis entails that
$\NEW{(\Flushw(e_1),\Flushw(e_2))}\DELETED{(\Delw(e_1),\Delw(e_2))} \in
\textsf{p}$, otherwise we would contradict the premise \DELETED{\cwcoww{} of
the \textsc{Commit write} rule}\NEW{\cpwaccord{} of the \textsc{Write reaches
coherence point} rule}.  The second hypothesis means by definition that
$(\Flushw(e_2),\Flushw(e_1)) \in \textsf{p}$. \DELETED{, which entails that
$(\Delw(e_2),\Delw(e_1)) \in \textsf{p}$ (otherwise we would contradict the
\cpwaccord{} premise of the \textsc{Write reaches coherence point} rule).} This
contradicts the acyclicity of \textsf{p}.

\paragraph{\textsf{coRW1} is forbidden: \!\!\!\!} suppose as a contradiction a
read $r$ and a write $w$ relative to the same location, s.t.  $(r,w) \in \po$
and $(w,r) \in \rf(\evts,\textsf{p})$. Thus $w$ cannot be visible to $r$ as it
is $\po$-after $r$. This contradicts the premise \cruni{} of the
\textsc{Commit read} rule.

\paragraph{\textsf{coRW2} is forbidden: \!\!\!\!} suppose as a contradiction a
read $r$ and two writes $w_1$ and $w_2$ relative to the same location, s.t.
$(r,w_2) \in \po$ and $(w_2,w_1) \in \co(\evts,\textsf{p})$ and $(w_1,r) \in
\rf(\evts,\textsf{p})$. Thus $w_1$ cannot be visible to $r$ as it is
$\co$-after $w_2$, $w_2$ itself being either equal or $\co$-after the first
write $w_a$ in $\poloc$ after $r$. This contradicts the premise \cruni{}
of the \textsc{Commit read} rule.

\paragraph{\textsf{coWR} is forbidden: \!\!\!\!} suppose as a contradiction two
writes $w_0$ and $w_1$ and a read $r$ relative to the same location, s.t.
$(w_1,r) \in \po$, $(w_0,r) \in \rf(\evts,\textsf{p})$ and $(w_0,w_1) \in
\co(\evts,\textsf{p})$. Thus $w_0$ cannot be visible to $r$ as it is
$\co$-before $w_1$, $w_1$ being itself either equal or $\co$-before the last
write $w_b$ in $\poloc$ before $r$. This contradicts the premise
\cruni{} of the \textsc{Commit read} rule.

\paragraph{\textsf{coRR} is forbidden: \!\!\!} (note that this holds only for
the modified version of the machine outlined at the end of
\mysec\ref{sec:int-mach}) suppose as a contradiction two writes $w_1$ and $w_1$
and two reads $r_1$ and $r_2$ relative to the same location, s.t. $(r_1,r_2)
\in \po$, $(w_1,r_1) \in \rf$, $(w_2,r_2) \in \rf$ and $(w_2,w_1) \in \co$.
Thus $w_2$ cannot be visible to $r_2$ as it is \DELETED{$\buff$}\NEW{$\co$}-before $w_1$ (following their order in $\co$), and $r_1$ is $\poloc$-before
$r_2$.  This contradicts the premise \cruni{} of the \textsc{Commit read} rule.

\paragraph{The \textsc{no thin air} axiom holds: \!\!\!\!} suppose as a
contradiction that there is a cycle in $\transc{\hb}$, \ie there is an event
$x$ s.t. $(x,x) \in \transc{(\mo{(\ppo \cup \fences)};\mo{\rfe})}$. Thus there
exists $y$ s.t. \textsf{(i)} $(x,y) \in \mo{(\ppo} \cup$ $\mo{\fences)};
\mo{\rfe}$ and \textsf{(ii)} $(y,x) \in \transc{(\mo{(\ppo \cup \fences)};
\mo{\rfe})}$. Note that $x$ and $y$ are reads, since the target of $\rf$ is
always a read.

We now show that \textsf{(iii)} for two reads $r_1$ and $r_2$, having
$(r_1,r_2) \in \transc{(\mo{(\ppo \cup \fences)};\mo{\rfe})}$ implies
$(\Delr(r_1),\Delr(r_2)) \in \textsf{p}$ --- we are abusing our notations here,
writing $\Delr(r)$ instead of $\Delr(w,r)$ where $w$ is the write from which
$r$ reads. From the fact \textsf{(iii)} and the hypotheses \textsf{(i)} and
\textsf{(ii)}, we derive a cycle in $\textsf{p}$, a contradiction since
\textsf{p} is an order.

Proof of \textsf{(iii)}: let us show the base case; the inductive case follows
immediately. Let us have two reads $r_1$ and $r_2$ s.t. $(r_1,r_2) \in
\mo{(\ppo \cup \fences)};\mo{\rfe}$. Thus there is $w_2$ s.t.  $(r_1,w_2) \in
\ppo \cup \fences$ and $(w_2,r_2) \in \rfe$. Now, note that when we are about
to take the \textsc{Satisfied read} transition triggered by the label
$\Delr(w_2,r_2)$, we know that the premise \srokw{} holds. Thus we know that
either $w_2$ and $r_2$ belong to the same thread, which immediately contradicts
the fact that they are in $\rfe$, or that $w_2$ has been committed. Therefore
we have \textsf{(iv)} $(\Delw(w_2),\Delr(r_2)) \in \textsf{p}$. Moreover, we
can show that \textsf{(v)} $(\Delr(r_1),\Delw(w_2)) \in \textsf{p}$ by the fact
that $(r_1,w_2) \in \ppo \cup \fences$. Thus by \textsf{(iv)} and \textsf{(v)}
we have our result.

Proof of \textsf{(v)}: take a read $r$ and a write $w$ s.t. $(r,w) \in \ppo
\cup \fences$. We show below that \textsf{(vi)} $(\Flushr(r),\Delw(w)) \in
\textsf{p}$. Since it is always the case that $(\Delr(r),\Flushr(r)) \in
\textsf{p}$ (thanks to the fact that a read is satisfied before it is
committed, \cf premise \crsat{} of the \textsc{Commit read} rule), we can
conclude. Now for \textsf{(vi)}: since \textsf{p} is total, we have either our
result or $(\Delw(w),\Flushr(r)) \NEW{\in \textsf{p}}$. Suppose the latter:
then when we are about to take the \textsc{Commit read} transition triggered by
$\Flushr(r)$, we contradict the premise \crccrw. Indeed we have $w \in \buff$
by $\Delw(w)$ preceding $\Flushr(r)$ in \textsf{p}, and $(r,w) \in \ppo \cup
\fences$ by hypothesis.

\paragraph{The \textsc{\DELETED{causality}\NEW{observation}} axiom holds:
\!\!\!\!} suppose as a contradiction that the relation $\efr;\prop;\rstar{\hb}$
is not irreflexive, \ie there are $w$ and $r$ s.t. $(r,w_2) \in \efr$ and
$(w_2,r) \in \prop;\rstar{\hb}$. Note that $(r,w_2) \in \fr$ implies the
existence of a write $w_1$ s.t. $(w_1,w_2) \in \co$ and $(w_1,r) \in \rf$.
Observe that this entails that $r$, $w_1$ and $w_2$ are relative to the same
location.

Thus we take two writes $w_1, w_2$ and a read $r$ relative to the same location
s.t.  $(w_1,w_2) \in \co$, $(w_1,r) \in \rf$ and $(w_2,r) \in
\prop;\rstar{\hb}$ as above. This contradicts the \srcaus{} hypothesis. Indeed
when we are about to process the transition triggered by the label $\Delr(r)$,
we have $(w_2,r) \in \prop;\rstar{\hb}$ by hypothesis, and
\DELETED{$(\Delw(w_1),\Delw(w_2)) \in \textsf{p}$ (entailed by $(w_1,w_2) \in
\co$)}\NEW{$(w_1,w_2) \in \co$ by hypothesis}. 

\paragraph{The \textsc{propagation} axiom holds: \!\!\!\!} suppose as a
contradiction that there is a cycle in $\transc{(\co \cup \prop)}$, \ie there
is an event $x$ s.t. $(x,x) \in \transc{(\co \cup \prop)}$. In other terms
there is $y$ s.t. \textsf{(i)} $(x,y) \in \co;\prop$ and \textsf{(ii)} $(y,x)
\in \transc{(\co; \prop)}$.  Note that $x$ and $y$ are writes, since the source
of $\co$ is always a write.

We now show that \textsf{(iii)} for two writes $w_1$ and $w_2$, having
$(w_1,w_2) \in  \transc{(\co;\prop)}$ implies $(\Flushw(w_1),\Flushw(w_2)) \in
\transc{\textsf{p}}$.  From the fact \textsf{(iii)} and the hypotheses
\textsf{(i)} and \textsf{(ii)}, we derive a cycle in \textsf{p}, a
contradiction since \textsf{p} is an order.

Proof of \textsf{(iii)}:  let us show the base case; the inductive case follows
immediately. Let us have two writes $w_1$ and $w_2$ s.t. $(w_1,w_2) \in
\co;\prop$; thus there is a write $w$ s.t. $(w_1,w) \in \co$ and $(w,w_2) \in
\prop$. Since $\textsf{p}$ is total, we have either the result or
$(\Flushw(w_2),\Flushw(w_1)) \in \textsf{p}$. Suppose the latter. \DELETED{; observe that
this implies $(\Delw(w_2),\Delw(w_1)) \in \textsf{p}$ since otherwise we would
contradict the \cpwaccord{} premise of the \textsc{Write reaches coherence
point} rule.} Thus we contradict the \DELETED{\cwprop{}}\NEW{\cpwprop{}}
hypothesis.  Indeed when we are about to take the \DELETED{\textsc{Commit
write}}\NEW{\textsc{Write reaches coherence point}} transition triggered by the
label \DELETED{$\Delw(w)$}\NEW{$\Flushw(w)$}, we have $(w,w_2) \in \prop$ by
hypothesis, and $w_2$ already in \DELETED{$\buff$}\NEW{$\rcp$}: the hypothesis
$(w_1,w) \in \co$ entails that $\NEW{(\Flushw(w_1),\Flushw(w))} \in \textsf{p}
\DELETED{(\Delw(w_1),\Delw(w))}$, and we also have
$\NEW{(\Flushw(w_2),\Flushw(w_1))} \in \textsf{p} \DELETED{
(\Delw(w_2),\Delw(w_1))}$ by hypothesis. Therefore when we are about to
process \DELETED{$\Delw(w)$}\NEW{$\Flushw(w)$} we have placed $w_2$ in
\DELETED{$\buff$}\NEW{$\rcp$} by taking the transition
\DELETED{$\Delw(w_2)$}\NEW{$\Flushw(w_2)$}.

\subsubsection{From axiomatic model to intermediate machine (proof of \mylem\ref{lem:ax-interm})}

We show here that an axiomatic execution $(\evts,\po,\co,\rf)$ leads to a valid
path $\textsf{p}$ of the intermediate machine. To do so, we show that the
intermediate machine accepts \DELETED{any}\NEW{certain}
path\NEW{s}\footnote{\label{foot:fifo}\NEW{The path has to linearise \textsf{r}
so that for all writes $w_1$ and $w_2$, if $(\Flushw(w_1),\Flushw(w_2) \in
\textsf{p}$ then $(\Delw(w_1),\Delw(w_2) \in \textsf{p}$. We refer to this
property as ``\textsf{p} being fifo'.'In other words, the linearisation must be
such that coherence point labels and commit labels are in accord. Note that
this does not affect the generality of \mylem\ref{lem:ax-interm}, as to prove
this lemma, we only need to find one valid intermediate machine path for a
given axiomatic execution; our path happens to be so that coherence point and
commit labels are in accord. }} that linearise\DELETED{s} the transitive
closure of the relation $\textsf{r}$ defined inductively as follows (we abuse
our notations here, writing \eg $\Flushr(r)$ instead of $\Flushr(w,r)$ where
$w$ is the write from which $r$ reads): 

\begin{itemize}
\item for all $r \in \evts$, $(\Delr(r), \Flushr(r)) \in \textsf{r}$, \ie we
satisfy a read before committing it; 
\item for all $w \in \evts$, $(\Delw(w),\Flushw(w)) \in \textsf{r}$, \ie we
commit a write before it can reach its coherence point;
\item forall $w$ and $r$ separated by a fence in program order,
$(\Delw(w),\Delr(r)) \in \textsf{r}$, \ie we commit the write $w$ before we satisfy the read
$r$;
\item for all $(w,r) \in \rfe$, $(\Delw(w),\Delr(w,r)) \in \textsf{r}$, \ie we
commit a write before reading externally from it;
\item for all $(w_1,w_2) \in \co$, $(\Flushw(w_1),\Flushw(w_2)) \in \textsf{r}$,
\ie $\rcp$ and the coherence order are in accord;
\item for all $(r,e) \in \ppo \cup \fences$, we commit a read $r$
before processing any other event $e$ (\ie satisfying $e$ if $e$ is
a read, or committing $e$ if $e$ is a write), if $r$ and $e$ are
separated \eg by a dependency or a barrier; 
\item for all $(w_1,w_2) \in \transc{\prop}$, $(\Flushw(w_1),\Flushw(w_2)) \in
\textsf{r}$, \ie $\rcp$ and propagation order are in accord. 
\end{itemize}
\fixmeskip{jade: in write reaches coh point transition, could we put w,w' in (co U prop)+ instead of the sequence co;prop as it is now?}

Since we build \textsf{p} as a linearisation of the relation \textsf{r} defined
above, we first need to show that we are allowed to linearise \textsf{r}, \ie
that \textsf{r} is acyclic. 

\paragraph{Linearisation of \textsf{r}: \!\!\!} suppose as a contradiction that
there is a cycle in \textsf{r}, \ie there is a label $l$ s.t. $(l,l) \in
\transc{\textsf{r}}$. Let us write $S_1$ for the set of commit writes, satisfy
reads and commit reads, and $S_2$ for the set of writes reaching coherence
points. We show by induction that for all pair of labels $(l_1,l_2) \in
\transc{\textsf{r}}$, either:

\begin{itemize}
\item $l_1$ and $l_2$ are both in $S_1$, and their corresponding events $e_1$
and $e_2$ are ordered by happens-before, \ie $(e_1,e_2) \in \transc{\hb}$, or
\item $l_1$ and $l_2$ are both in $S_2$ and their corresponding events $e_1$
and $e_2$ are ordered by $\transc{(\co \cup \prop)}$, or 
\item $l_1$ is in $S_1$, $l_2$ in $S_2$, and their corresponding events $e_1$
and $e_2$ are ordered by happens-before, or
\item $l_1$ is in $S_1$, $l_2$ in $S_2$, and their corresponding events $e_1$
and $e_2$ are ordered by $\transc{(\co \cup \prop)}$, or  
\item $l_1$ is in $S_1$, $l_2$ in $S_2$, and their corresponding events $e_1$
and $e_2$ are ordered by $\transc{\hb};\transc{(\co \cup \prop)}$, or
\item $l_1$ is a satisfy read and $l_2$ the corresponding commit read.
\item $l_1$ is a commit write and $l_2$ the write reaching coherence point.
\end{itemize}

Each of these items contradicts the fact that $l_1=l_2$: the first two resort
to the axioms of our model prescribing the acyclicity of $\hb$ on the one hand
(\textsc{no thin air}), and ${\co} \cup {\prop}$ on the second hand
(\textsc{propagation}); all the rest resorts to the transitions being
different. 
 
We now show that none of the transitions of the machine given in
\myfig\ref{fig:mach} can block.

\paragraph{\textsc{Commit write} does not block: \!\!\!} suppose as a
contradiction a label $\Delw(w)$ s.t. the transition of the intermediate machine
triggered by $\Delw(w)$ blocks. This means that one of the premises of the
\textsc{Commit write} rule is not satisfied.

First case: the premise \cwcoww{} is not satisfied, \ie  there exists $w'$ in
$\buff$ s.t. $(w,w') \in \poloc$. Since $(w,w') \in \poloc$ we have $(w,w') \in
\ws$ by \DELETED{\textsc{uniproc}}\NEW{\textsc{sc per location}}.  \NEW{By
construction of \textsf{p}, we deduce $(\Flushw(w),\Flushw(w')) \in
\textsf{p}$. By $\textsf{p}$ being fifo (see footnote on
page~\pageref{foot:fifo}), we deduce \textsf{(i)} $(\Delw(w),\Delw(w')) \in
\textsf{p}$.} Moreover if $w'$ is in $\buff$ when we are about to process
$\Delw(w)$, then $w'$ has been committed before $w$, hence \textsf{(ii)}
\DELETED{$(w',w) \in \ws$}\NEW{$(\Delw(w'),\Delw(w))\in \textsf{p}$}. By
\textsf{(i)} and \textsf{(ii)}, we \DELETED{contradict the \textsc{uniproc}
axiom}\NEW{derive a cycle in \textsf{p}, a contradiction since \textsf{p} is an
order (since we build it as a linearisation of a relation)}. 

Second case: the premise \cwprop{} is not satisfied, \ie  there exists $w'$ in
$\buff$ s.t. $(w,w') \in \prop$. Since $w'$ is in $\buff$ when we are about to
process the label $\Delw(w)$, we have \NEW{\textsf{(i)}} $(\Delw(w'),\Delw(w))
\in \textsf{p}$\DELETED{, which entails \textsf{(i)} $(\Flushw(w'),\Flushw(w))
\in \textsf{p}$}.  Since $(w,w') \in \prop$, we have \textsf{(ii)}
$(\Flushw(w),\Flushw(w')) \in \textsf{p}$ (recall that we build \textsf{p}
inductively; in particular the order of the $\Flushw(w)$ transitions in
\textsf{p} respects the order of the corresponding events in $\prop$).
\NEW{Since we build \textsf{p} so that it is fifo (see footnote on
page~\pageref{foot:fifo}), we deduce from \textsf{(i)} the fact \textsf{(iii)}
$(\Delw(w),\Delw(w')) \in \textsf{p}$.}  From \DELETED{\textsf{(i)} and}
\textsf{(ii)} \NEW{and \textsf{(iii)}}, we derive a cycle in \textsf{p}, a
contradiction since \textsf{p} is an order (since we build it as a
linearisation of a relation). 

Third case: the premise \cwciwr{} is not satisfied, \ie there exists $r$ in
$\queue$ s.t. $(w,r) \in {\fences}$. From $r \in \queue$ we deduce
$(\Delr(r),\Delw(w)) \in \textsf{p}$. From $(w,r) \in {\fences}$ we deduce (by
construction of \textsf{p}) $(\Delw(w),\Delr(r)) \in \textsf{p}$, which creates
a cycle in \textsf{p}. 

 
\paragraph{\textsc{Write reaches coherence point} does not block: \!\!\!}
suppose as a contradiction  a label $\Flushw(w)$ s.t. the transition of the
intermediate machine triggered by $\Flushw(w)$ blocks. This means that one of
the premises of the \textsc{Write reaches coherence point} rule is not
satisfied.

First case: the premise \cpwcom{} is not satisfied, \ie $w$ has not been
committed. This is impossible since $(\Delw(w),\Flushw(w)) \in \textsf{p}$ by
construction of \textsf{p}.

Second case: the premise \cpwaccord{} is not satisfied, \ie there is a write
$w'$ that has reached coherence point s.t. $(w,w') \in \NEW{\poloc}\DELETED{
\buff}$. From $(w,w') \in \NEW{\poloc}\DELETED{\buff}$, we know
\NEW{by \textsc{sc per location}} that $(w,w') \in \co$. \NEW{Thus by
construction of \textsf{p}, we know \textsf{(i)} $(\Flushw(w),\Flushw(w')) \in
\textsf{p}$.} From $w'$ having reached coherence point before $w$, we know
\textsf{(ii)} $(\Flushw(w'),\Flushw(w)) \in \textsf{p}$. \DELETED{, hence by
the construction of \textsf{p} we know that \textsf{(ii)} $(w',w) \in
\transc{(\co \cup \prop)}$.} By \textsf{(i)} and \textsf{(ii)}, we
\DELETED{contradict the \textsc{propagation} axiom.}\NEW{derive a cycle in
\textsf{p}, a contradiction since \textsf{p} is an order (since we build it as
a linearisation of a relation)}.

Third case: the premise \cpwprop{} is not satisfied, \ie there is a write $w'$
that has reached coherence point s.t. $(w,w') \in \NEW{\prop}\DELETED{
\co;\prop}$.  \NEW{By construction of \textsf{p}, we deduce \textsf{(i)}
$(\Flushw(w),\Flushw(w') \in \textsf{p})$.} From $w'$ having reached coherence
point before $w$, we know \textsf{(ii)} $(\Flushw(w'),\Flushw(w)) \in
\textsf{p}$. \DELETED{, hence by the construction of \textsf{p} we know that
\textsf{(ii)} $(w',w) \in \transc{(\co \cup \prop)}$.} By \textsf{(i)} and
\textsf{(ii)}, we \DELETED{contradict the \textsc{propagation}
axiom}\NEW{derive a cycle in \textsf{p}, a contradiction since \textsf{p} is an
order (since we build it as a linearisation of a relation)}.

\paragraph{\textsc{Satisfy read} does not block: \!\!\!} suppose as a
contradiction a label $\Delr(w,r)$ s.t. the transition of the intermediate
machine triggered by $\Delr(w,r)$ blocks. This means that one of the premises
of the \textsc{Satisfy read} rule is not satisfied. Note that since
$\Delr(w,r)$ is a label of \textsf{p}, we have \textsf{(i)} $(w,r) \in \rf$.

First case: the premise~\srokw{} is not satisfied, \ie $w$ is neither local nor
committed. Suppose $w$ not local (otherwise we contradict our hypothesis); let
us show that it has to be committed. Suppose it is not, therefore we have
$(\Delr(r),\Delw(w)) \in \textsf{p}$. Since $w$ is not local, we have $(w,r) \in
\rfe$, from which we deduce (by construction of \textsf{p}) that
$(\Delw(w),\Delr(r)) \in \textsf{p}$; this leads to a cycle in \textsf{p}.

Second case: the premise~\sriirr{} is not satisfied, \ie there is a satisfied
read $r'$ s.t.  $(r,r') \in \ppo \cup \fences$. From $r'$ being satisfied we
deduce $(\Delr(r'),\Delr(r)) \in \textsf{p}$. From $(r,r') \in \ppo \cup
\fences$ and by construction of \textsf{p}, we deduce $(\Flushr(r),\Delr(r'))
\in \textsf{p}$. Since $\Delr(r)$ precedes $\Flushr(r)$ in \textsf{p} by
construction of \textsf{p}, we derive a cycle in \textsf{p}, a contradiction.

Third case: the premise~\srcaus{} is not satisfied, \ie there is a write~$w'$
s.t.  $(w,w') \in \NEW{\co}\DELETED{\buff}$ and $(w',r) \in
\prop;\rstar{\hb}$.  Since \DELETED{$(w,w') \in \buff$, we have} $(w,w') \in
\co$, by \textsf{(i)} $(r,w') \in \fr$.  Therefore we contradict the
\textsc{\DELETED{causality}\NEW{observation}} axiom. 

\paragraph{\textsc{Commit read} does not block: \!\!\!} suppose as a
contradiction a label $\Flushr(w,r)$ s.t. the transition of the intermediate
machine triggered by $\Flushr(w,r)$ blocks. This means that one of the premises
of the \textsc{Commit read} rule is not satisfied.

First case: the premise~\crsat{} is not satisfied, \ie $r$ is not in $\queue$.
This is impossible since we impose $(\Delr(r),\Flushr(r)) \in \textsf{p}$ when
building \textsf{p}.

Second case: the premise~\cruni{} is not satisfied, \ie~$w$ is not visible
to~$r$.  This contradicts the~\DELETED{\textsc{uniproc}}\NEW{\textsc{sc per
location}} axiom, as follows.  Recall that the visibility definition
introduces~$w_a$ as the first write to the same location as~$r$ which
is~$\poloc$-after $r$; and~$w_b$ as the last write to the same location as $r$
which is $\poloc$-before~$r$. Now, if~$w$ is not visible to $r$ we have either
\textsf{(i)} $w$ is \DELETED{$\buff$}\NEW{$\co$}-before~$w_b$, or \textsf{(ii)} equal or \DELETED{$\buff$}\NEW{$\co$}-after $w_a$. 

Suppose \textsf{(i)}, we have $(w,w_b) \in \NEW{\co}\DELETED{\buff}$.
\DELETED{; hence $(\Flushw(w),\Flushw(w_b)) \in \textsf{p}$ because commits and
coherence points are in accord. Recall the linearisation lemma, we then know
that \textsf{(i)} $(w,w_b) \in \transc{(\co \cup \prop)}$. This entails that
\textsf{(ii)} $(w,w_b) \in \co$, since $\co$ is total \NEW{per location} (thus
$w$ and $w_b$ must appear in one order or the other in $\co$), and $\co \cup
\prop$ is acyclic by the \textsc{(propagation)} axiom, thus their order in
$\co$ cannot contradict their order in $\transc{(\co \cup \prop)}$ as given by
\textsf{(i)}.} Hence we have $(w,w_b) \in \co$\DELETED{(by \textsf{(ii)})},
$(w_b,r) \in \poloc$ by definition of $w_b$, and $(w,r) \in \rf$ by definition
of $\Delr(w,r)$ being in \textsf{p}.  Thus we contradict the \textsf{coWR} case
of the \DELETED{\textsc{uniproc}}\NEW{\textsc{sc per location}} axiom.  The
\textsf{(ii)} case is similar, and contradicts \textsf{(coRW1)} if $w=w_a$ or
\textsf{(coRW2)} if $(w_a,w) \in \buff$.

Third case: the premise~\crccrw{} is not satisfied, \ie there is a write $w'$ in
$\buff$ s.t. $(r,w') \in \ppo \cup \fences$. From $w' \in \buff$ we deduce
$(\Delw(w'),\Flushr(r)) \in \textsf{p}$. From $(r,w') \in \ppo \cup \fences$ we
deduce $(\Flushr(r),\Delw(w')) \in \textsf{p}$ by construction of \textsf{p}.
This leads to a cycle in~\textsf{p}, a contradiction.
 

Fourth case: the premise~\crci{} is not satisfied, \ie there is a read 
$r'$ in $\queue$ s.t. $(r,r') \in \ppo \cup \fences$. From $r' \in
\queue$ we deduce $(\Delr(r'),\Flushr(r)) \in \textsf{p}$. From $(r,r') \in
\ppo \cup \fences$ we deduce $(\Flushr(r),\Delr(r')) \in \textsf{p}$
by construction of \textsf{p}. This leads to a cycle in~\textsf{p}, a
contradiction.


\subsection{Comparing our model and the PLDI machine}

The PLDI machine is an operational model, which we describe here briefly
(see~\cite{ssa11} for details). This machine maintains a coherence order (a
strict partial order over the writes to the same memory location), and, per
thread, a list of the writes and fences that have been propagated to that
thread.

A load instruction yields two transitions of this machine (amongst others): a
\emph{satisfy read} transition, where the read takes its value, and a
\emph{commit read} transition, where the read becomes irrevocable.  A store
instruction yields a \emph{commit write} transition, where the write becomes
available to be read, several \emph{propagate write} transitions, where the
write is sent out to different threads of the system, and a \emph{reaching
coherence point} transition, where the write definitely takes place in the
coherence order. We summarise the effect of a PLDI transition on a PLDI state
in the course of this section. 

We show that \DELETED{a certain class of valid paths}\NEW{a valid path} of the
PLDI machine leads to valid path\DELETED{s} of our intermediate machine.
\DELETED{That is, we assume that, for two writes $w$ and $w'$, if their commit
transitions and their coherence point transitions are not in the same order
in~\textsf{p}, then there is another path~\textsf{p'} from $s_0$ to $s_n$ where
the commit and coherence point transitions are in accord. In the next section,
we explain how we observed experimentally that, for each of our litmus tests,
all executions allowed by the PLDI machine are also allowed by our model.}

\medskip

First, we show how to relate the two machines. 

\subsubsection{Mapping PLDI Objects (labels and states) to intermediate objects}

We write $\pltol(l)$ to map a PLDI $l$ to a label of the intermediate machine.
For technical convenience we assume a special \noop intermediate label such
that, from any state $s$ of the intermediate machine, we may perform a
transition from $s$ to $s$ via \noop.

We can then define $\pltol(l)$ as being the eponymous label in the intermediate
machine if it exists (\ie for commit write, write reaches coherence point,
satisfy read and commit read), and \noop otherwise. We derive from this mapping
the set $\mathbb{L}_i$ of intermediate labels composing our intermediate path.





We build a state of our intermediate machine from a PLDI state~$s$ and an
accepting PLDI path~$\textsf{p}$; we write $\pstos(\textsf{p},s) = (\buff,
\rcp, \queue, \cread)$ for the intermediate state built as follows: 
\begin{itemize}
\item for a given location, \DELETED{we take} $\buff$ \DELETED{to be all
the}\NEW{is simply the set of} writes to this location that have been committed
in $s$\DELETED{, ordered \wrt \textsf{p}};
\item \NEW{we take} $\rcp$ \DELETED{is simply the set of} \NEW{to be all the}
writes having reached coherence point in $s$\NEW{, ordered \wrt \textsf{p}};
\item the set $\queue$ gathers the reads that have been satisfied in the state
$s$: we simply add a read to $\queue$ if the corresponding satisfy transition
appears in \textsf{p} before the transition leading to $s$;
\item the set $\cread$ gathers the reads that have been committed in the state
$s$.
\end{itemize}
 



\subsubsection{Building a path of the intermediate machine from a PLDI path\label{sec:path}} 

A given path \textsf{p} of the PLDI machine entails a run $s_0 \rln{l_1} s_1
\rln{l_2} \cdots \rln{l_n} s_n$ such that $(l,l') \in \textsf{p}$ if and only if
there exist $i$ and $j$ such that $i < j$ and $l = l_i$ and $l' = l_j$. 

We show that $\pstos(s_0) \rln{\pltol(l_1)} \pstos(s_1) \rln{\pltol(l_2)}
\cdots \rln{\pltol(l_n)} \pstos(s_n)$ is a path of our intermediate machine.
We proceed by induction for $0 \leq m \leq n$. The base case $m=0$ is
immediately satisfied by the single-state path $\pstos(s_0)$.
 
Now, inductively assume that $\pstos(s_0) \rln{\pltol(l_1)} \pstos(s_1)
\rln{\pltol(l_2)} \cdots \rln{\pltol(l_m)} \pstos(s_m)$ is a path of the
intermediate machine.  We prove the case for $m+1$.  Take the transition $s_m
\rln{l_{m+1}} s_{m+1}$ of the PLDI machine.  We prove $\pstos(s_m)
\rln{\pltol(l_{m+1})} \pstos(s_{m+1})$ to complete the induction. There are
several cases.

When $\pltol(l_{m+1}) = \noop$ we have that $\pstos(s_m) = \pstos(s_{m+1})$
simply because the PLDI labels that have $\noop$ as an image by $\pltol$ do not
affect the components $\buff,\rcp,\queue$ and $\cread$ of our state. 

Only the PLDI transitions that have an eponymous transition in our machine
affect our intermediate state. Thus we list below the corresponding four cases.

\paragraph{Commit write:} in the PLDI machine, a commit transition of a
write $w$ makes this write~$\co$-after all the writes to the same location that
have been propagated to its thread. The PLDI transition guarantees
that~\textsf{(i)} $w$ had not been committed in~$s_1$. 

Observe that $w$ takes its place in \NEW{the set} $\buff$ \DELETED{after all
the writes to the same location that have been committed before $w$}, ensuring
that we modify the state as prescribed by our \textsc{commit write} rule. 

Now, we check that we do not contradict the premises of our \textsc{Commit
write} rule. 

First case: contradict the premise~\cwcoww, \ie take a write $w'$ in~$\buff$
s.t. $(w,w') \in \poloc$.  In that case, we contradict the fact that the commit
order respects $\poloc$ (\cf~\cite[p.~7, item 4 of \S~Commit in-flight
instruction]{ssa11}). 

Second case: contradict the premise~\cwprop, \ie take a write~$w'$ in~$\buff$
s.t.~$(w,w') \in \prop$. In that case, $(w,w') \in \prop$ guarantees that~$w$
was propagated to the thread of~$w'$ in~$s_1$. Therefore (\cf~\cite[p.~6,
premise of \S~Propagate write to another thread]{ssa11}), $w$ was seen in~$s_1$.  For
the write to be seen, it needs to have been sent in a write request~\cite[p.~6,
item 1 of \S Accept write request]{ssa11}; for the write request to be sent,
the write must have been committed~\cite[p.~8, action 4 of \S Commit in-flight
instruction]{ssa11}. Thus we contradict \textsf{(i)}.

Third case: contradict the premise~\cwciwr, \ie take a read $r$ in $\queue$
s.t. $(w,r) \in {\fences}$. Since $r$ is in $\queue$, $r$ is satisfied. Note
that the write from which $r$ reads can be either local (\cf\cite[p.~8,
\S~Satisfy memory read by forwarding an in-flight write directly to reading
instruction]{ssa11}) or committed (\cf\cite[p.~8, \S~Satisfy memory read from
storage subsystem]{ssa11}).

In both cases, the fence between $w$ and $r$ must have been
committed (\cf~\cite[p.~8, item 2 of \S~Satisfy memory read by forwarding
an in-flight write directly to reading instruction and item 2 of \S~Satisfy
memory read from storage subsystem]{ssa11}). Thus by~\cite[p.~7, item 6 of
\S~Commit in-flight instruction]{ssa11}, $w$ has been committed, a contradiction of
\textsf{(i)}.


\paragraph{Write reaches coherence point:} in the PLDI machine, write
reaching coherence point transitions order writes following a linearisation of
~$\transc{(\co \cup \prop)}$. Our $\buff$ implements that per location, then we
make the writes reach coherence point following $\buff$ and $\transc{(\co \cup
\prop)}$. 

Observe that a write $w$ reaching coherence point takes its place \DELETED{in
the set of}\NEW{after all the} writes having \NEW{already} reached coherence
point, ensuring that we modify the intermediate state as prescribed by our
\textsc{write reaches coherence point} rule.

Now, we check that we do not contradict the premises of our \textsc{Write
reaches coherence point} rule.

First case: contradict the premise~\cpwcom, \ie suppose that $w$ is not
committed. This is impossible as the PLDI machine requires a write to have been
seen by the storage subsystem for it to reach coherence point~\cite[p.~3, \S
Write reaches its coherence point]{smo12}. For the write to be seen, it needs
to have been sent in a write request~\cite[p.~6, item 1 of \S Accept write
request]{ssa11}; for the write request to be sent, the write must have been
committed~\cite[p.~7, action 4 of \S Commit in-flight instruction]{ssa11}. 

Second case: contradict the premise~\cpwaccord, \ie take a write $w'$ in $\rcp$
s.t. $(w,w') \in \NEW{\poloc}\DELETED{\buff}$. This means that
\textsf{(i)} $w'$ has reached coherence point before $w$, despite $w$ preceding
$w'$ in \DELETED{the commit order $\buff$}\NEW{$\poloc$}. \DELETED{Recall that
\DELETED{we} our path \textsf{p} guarantees that commit and coherence point
transitions are in the same order in \textsf{p}, thus an immediate
contradiction.}\NEW{This is a contradiction, as we now explain. If $(w,w') \in
\poloc$, then $w$ is propagated to its own thread before $w'$~\cite[p.~7,
action 4 of \S Commit in-flight instruction]{ssa11}.  Since $w$ and $w'$ access
the same address, when $w'$ is propagated to its own thread, $w'$ is recorded
as being $\co$-after all the writes to the same location already propagated to
its thread~\cite[p.~6, item 3 of \S Accept write request]{ssa11}, in particular
$w$. Thus we have $(w,w') \in \co$. Now, when $w'$ reaches coherence point, all
its coherence predecessors must have reached theirs~\cite[p. 4, item 2 of \S
Write reaches its coherence point]{smo12}, in particular $w$. Thus $w$ should
reach its coherence point before $w'$, which contradicts \textsf{(i)}.}

Third case: contradict the premise~\cpwprop, \ie take a write $w'$ in $\rcp$
s.t. $(w,w') \in \NEW{\prop}\DELETED{\co;\prop}$. This contradicts the
fact that writes cannot reach coherence point in an order that contradicts
\DELETED{coherence nor} propagation (\cf \cite[p.~4, item\DELETED{s 2 and} 3 of
\S Write reaches coherence point]{smo12}).   

\paragraph{Satisfy read:} in the PLDI machine, a satisfy read transition
does not modify the state (\ie $s_1=s_2$). In the intermediate state, we
augment $\queue$ with the read that was satisfied. 

Now, we check that we do not contradict the premises of our \textsc{Satisfy
read} rule.

First case: contradict the \srokw{} premise, \ie suppose that $w$ is neither
local nor committed. Then we contradict the fact that a read can read either
from a local $\poloc$-previous write (\cf \cite[p.~8, item 1 of \S Satisfy
memory read by forwarding an in-flight write directly to reading
instruction]{ssa11}), or from a write from the storage subsytem --- which
therefore must have been committed \fixmeskip{jade: explain why}
(\cf~\cite[p.~8, \S~Satisfy memory read from storage subsystem]{ssa11}).

Second case: contradict the \sriirr{} premise, \ie take a satisfied read $r'$
s.t.  $(r,r') \in \ppo \cup \fences$. Then we contradict the fact that read
satisfaction follows the preserved program order and the fences (\cf\cite[p.~8,
all items of both \S Satisfy memory read by forwarding an in-flight write
directly to reading instruction and \S Satisfy memory read from storage
subsystem]{ssa11}). \fixmeskip{jade: pas sure pour local write}

Third case: contradict the \srcaus{} premise, \ie take a write~$w'$ in
\DELETED{$\buff$}\NEW{$\co$} after~$w$ s.t.~$(w',r) \in \prop;\rstar{\hb}$.
Since $w$ and $w'$ are related by \DELETED{$\buff$}\NEW{$\co$}, they have the
same location. The PLDI transition guarantees that \textsf{(i)} $w$ is the most
recent write to~$\loc(r)$ propagated the thread of~$r$ (\cf\cite[p.~6, \S Send
a read response to a thread]{ssa11}).  Moreover, $(w',r) \in \prop;\rstar{\hb}$
ensures that \textsf{(ii)} $w'$ has been propagated to the thread of $r$, or
there exists a write $e$ such that $(w',e) \in \co$ and $e$ is propagated to
the thread of $r$. Therefore, we have $(w',w) \in \co$ by \cite[p.~6, item 2 of
\S Propagate write to another thread]{ssa11}.

\DELETED{Now, recall that we assume commit and coherence point transitions to
be in accord.}  Therefore by $(w,w') \in \NEW{\co}\DELETED{\buff}$, we
know that $w$ reaches its coherence point before $w'$.  Yet $w'$ is a
$\co$-predecessor of $w$, which contradicts \cite[p.~4, item 2 of \S Write
reaches coherence point]{smo12}.

Proof of \textsf{(ii)}:\fixmeskip{jade: bien relire tout ca} we take $(w',r)
\in \prop;\rstar{\hb}$ as above. This gives us a write $w''$ s.t. $(w',w'') \in
\prop$ and $(w'',r) \in \rstar{\hb}$.  Note that $(w',w'') \in \prop$ requires
the presence of a barrier between $w'$ and $w''$.  

We first remind the reader of a notion from~\cite{ssa11}, the \emph{group A of
a barrier}: the group A of a barrier is the set of all the writes that have
been propagated to the thread holding the barrier when the barrier is sent to
the system (\cf~\cite[p.~5,\S Barriers (sync and lwsync) and cumulativity by
fiat]{ssa11}). When a barrier is sent to a thread, all the writes in its group
A must have been propagated to that thread (\cf~\cite[p.~6 item 2 of \S
Propagate barrier to another thread]{ssa11}).  Thus if we show that
\textsf{(a)} the barrier between $w'$ and $w''$ is propagated to the thread of
$r$ and \textsf{(b)} $w'$ is in the group A of this barrier, we have our
result.

Let us now do a case disjunction over $(w',w'') \in \prop$.

When $(w',w'') \in \propbase$, we have a barrier $b$ such that \textsf{(i)}
$(w',b) \in \fences \cup (\rfe;\fences)$ and \textsf{(ii)} $(b,r) \in
\rstar{\hb}$. Note \textsf{(i)} immediately entails that $w'$ is in the group A
of $b$. For \textsf{(ii)}, we reason by induction over $(b,r) \in \rstar{\hb}$,
the base case being immediate. In the inductive case, we have a\DELETED{n}
write $e$ such that $b$ is propagated to the thread of $e$ before $e$ and $e$
is propagated to the thread of $r$ before $r$. Thus, by \cite{ssa11}[p.~6, item
3 of \S Propagate write to another thread], $b$ is propagated to $r$.

When $(w',w'') \in \rstar{\com}; \rstar{\propbase}; \ffence; \rstar{\hb}$, we
have a barrier $b$ (which is a full fence) such that $(w',b) \in \rstar{\com};
\rstar{\propbase}$ and $(b,r) \in \rstar{\hb}$.  We now proceed by reasoning
over $\rstar{\com}$. 

If there is no $\com$ step, then we have $(w',b) \in \rstar{\propbase}$, thus
$w'$ is propagated to the thread of $b$ before $b$ by the $\propbase$ case
above. Note that this entails that $w'$ is in the group A of $b$. Since $b$ is
a full fence, $b$ propagates to all threads (\cf\cite[premise of \S
Acknowledge sync barrier]{ssa11}), in particular to the thread of $r$. Thus by
\cite[item 2 of \S Propagate barrier to another thread]{ssa11}, $w'$ is
propagated to $r$. 

In the $\transc{\com}$ case (\ie $(w',b) \in \rstar{\com}; \rstar{\propbase}$),
we remark that $\transc{\com} = \com \cup \ws;\rf \cup \fr;\rf$. Thus since $w'$
is a write, only the cases $\rf, \ws$ and $\ws;\rf$ apply. In the $\rf$ case,
we have $(w',b) \in \rstar{\propbase}$, which leads us back to the base case
(no $\com$ step) above. In the $\ws$ and $\ws;\rf$ cases, we have $(w',b) \in
\ws;\rstar{\propbase}$, \ie there exists a write $e$ such that $(w',e) \in \co$
and $(e,b) \in \rstar{\propbase}$, \ie our result.  
 
\paragraph{Commit read: \!\!\!} in the PLDI machine, a commit read transition
does not modify the state. In the intermediate state, we augment $\cread$ with
the read that was committed.  We now check that we do not contradict the
premises of our \textsc{Commit read} rule.

First case: contradict the premise~\crsat, \ie suppose that the read that we
want to commit is not in $\queue$; this means that this read has not been
satisfied.  This is impossible since a read must be satisfied before it is
committed (\cf~\cite[p.~7, item 1 of \S Commit in-flight instruction]{ssa11}).

Second case: contradict the \cruni{} premise. This is impossible since the PLDI
machine prevents~\textsf{coWR}, \textsf{coRW1} and~\textsf{coRW2})
(\cf~\cite[p.~3, \S Coherence]{ssa11}). 

Third case: contradict the premise~\crccrw, \ie take a committed write $w'$
s.t. $(r,w') \in \ppo \cup \fences$. Since $r$ must have been committed before
$w'$ (by~\cite[p.~7, items 2, 3, 4, 5, 7 of \S Commit in-flight
instruction]{ssa11}), we get a contradiction. 


Fourth case: contradict the premise~\crci, \ie take a satisfied read $r'$ s.t.
$(r,r') \in \ppo \cup \fences$. Since $r$ must have been committed before $r'$
was satisfied (by~\cite[p.~8, item 3 of \S Satisfy memory read from
storage subsystem and item 3 of \S Satisfy memory read by forwarding an
in-flight write directly to reading instruction]{ssa11}), we get a contradiction.

\paragraph{\DELETED{Remark}} \DELETED{We prove above that for all valid PLDI paths where commit and
coherence points transitions are in accord, we have a valid intermediate path.
By \mythm\ref{thm:equiv}, all such PLDI paths correspond to a valid execution
in our axiomatic model. }

\DELETED{As we explain in the next section, we observe experimentally that all
valid PLDI paths (without the assumption on commit and coherence point
transitions) corresponds to a valid axiomatic execution.}

\section{Testing and simulation}\label{sec:testing}

As usual in this line of work, we developed our model in tandem with extensive
experiments on hardware. We report here on our experimental results on Power
and ARM hardware. Additionally, we experimentally compared our model to the
ones of~\cite{ssa11} and \cite{mms12}. Moreover, we developed a new simulation tool
called \prog{herd}\footnote{We acknowledge that we reused some code written by
colleagues, in particular Susmit Sarkar, in an earlier version of the
tool.}. Finally, we adapted the \prog{CBMC} tool~\cite{akt13} to our new
models.  

\subsection{Hardware testing}

We performed our testing on several platforms using the \prog{diy} tool
suite~\cite{ams10,ams11,ams12}. This tool generates \emph{litmus tests}, \ie
very small programs in x86, Power or ARM assembly code, with specified initial
and final states. It then runs these tests on hardware and collects the memory
and register states that it observed during the runs.  Most of the time, litmus
tests violate SC: if one can observe their final state on a given machine, then
this machine exhibits features beyond SC. 

We generated \npower{} tests for Power and~\narm{} tests for ARM,
illustrating various features of the hardware, \eg
\textsf{lb}, \textsf{mp}, \textsf{sb}, and their variations with dependencies
and barriers, \eg \textsf{lb+addrs}, \textsf{mp+lwsync+addr},
\textsf{sb+syncs}. 

We tested the model described in \myfig\ref{fig:model},
\ref{fig:prop-power+arm}, and \ref{fig:ppo} on Power and ARM machines, to check
experimentally if this model was suitable for these two architectures. \NEW{In
the following, we write ``Power model'' for this model instantiated for Power,
and ``Power-ARM model'' for this model instantiated for ARM.} We give a summary
of our experiments in \mytab\ref{fig:summary-testing}. 
\begin{table}
\begin{center}
\begin{tabular}{c*{2}{|>{$}r<{$}}}
         &
\multicolumn{1}{|c}{Power} &
\multicolumn{1}{|c}{ARM} \\\hline
\rule{0ex}{2ex}\# tests & \npower & \narm \\
invalid & \nmodelpowerinvalid & \nmodelarminvalid \\
unseen  & \nmodelpowerunseen  & \nmodelarmunseen 
\end{tabular}
\end{center}
\caption{Summary of our experiments on Power and ARM h/w\label{fig:summary-testing}}
\end{table}

For each architecture, the row ``unseen'' gives the number of tests that our
model allows but that the hardware does not exhibit. This can be the case
because our model is too coarse (\ie fails to reflect the architectural
intent in forbidding some behaviours), or because the behaviour is intended to
be allowed by the architect, but is not implemented yet.

The column ``invalid'' gives the number of tests that our model forbids but that
the hardware does exhibit. This can be because our model is too strict and
forbids behaviours that are actually implemented, or because the behaviour is a
hardware bug.

\subsubsection{On Power} We tested three generations of machines:
Power G5, 6 and 7. The complete logs of our experiments can be found at
\url{http://diy.inria.fr/cats/model-power}. 

Our \NEW{Power} model is not invalidated by \NEW{Power} hardware (there is no
test in the ``invalid'' columns on Power in \mytab\ref{fig:summary-testing}).
In particular it allows \textsf{mp+lwsync+addr-po-detour}, which~\cite{ssa11}
wrongly forbids, as this behaviour is observed on hardware (\cf
\url{http://diy.inria.fr/cats/pldi-power/\#lessvs}). 

Our \NEW{Power} model allows some behaviours (see the ``unseen'' columns on
Power), \eg\textsf{lb}, that are not observed on \NEW{Power} hardware. This is
to be expected as the \textsf{lb} \DELETED{idiom}\NEW{pattern} is not yet
implemented on Power hardware, despite being clearly architecturally
allowed~\cite{ssa11}.

\subsubsection{On ARM\label{sec:arm-testing}} We tested several system
configurations: \NEW{NVDIA} Tegra 2 and 3, \NEW{Qualcomm} APQ8060 and APQ8064,
\NEW{Apple} A5X and A6X, and \NEW{Samsung} Exynos 4412, 5250 \NEW{and 5410}.
The complete logs of our experiments can be found at
\url{http://diy.inria.fr/cats/model-arm}.  \NEW{This section about ARM testing
is articulated as follows:} \begin{itemize} \item \NEW{we first explain how
o}ur \NEW{Power-ARM} model is invalidated by \NEW{ARM} hardware (see the
``invalid'' column on ARM) by $\nmodelarminvalid$ tests \NEW{(see \S ``Our
Power-ARM model is invalidated by ARM hardware'')}.  We detail and document the
discussion below at \url{http://diy.inria.fr/cats/arm-anomalies}.  \item
\NEW{we then propose a model for ARM (see \S ``Our proposed ARM model'').}
\item \NEW{we then explain how we tested this model on ARM machines, and the
anomalies that we have found whilst testing (see \S ``Testing our model'').}
\end{itemize} 

\paragraph{\NEW{Our Power-ARM model is invalidated by ARM hardware}} Amongst
the tests we have ran on ARM hardware, some unveiled a \DELETED{so-called}
\emph{load-load hazard} bug in the coherence mechanism of all machines\DELETED{
except A6X and Exynos 5250}. This bug is a violation of the \textsf{coRR}
\DELETED{idiom}\NEW{pattern} shown in \mysec\ref{sec:model}, and was later
acknowledged as such by ARM\NEW{, in the context of Cortex-A9 cores, in the
note~\cite{arm:rar}.}\footnote{\NEW{We note that violating \textsf{coRR}
invalidates the implementation of C++ modification order \textsf{mo} (\eg{} the
implementation of {\tt memory\_order\_relaxed}), which explicitely requires the
five coherence patterns of \myfig\ref{fig:co} to be forbidden~\cite[p.  6, col.
1]{bos11}.}}

\NEW{Amongst the machines that we have tested, this note applies directly
to Tegra 2 and~3,
A5X, Exynos 4412. Additionally Qualcomm's APQ8060 is supposed to have many
architectural similarities with the ARM Cortex-A9, thus we believe that the
note might apply to APQ8060 as well. Morever we have observed
load-load hazards anomalies on cortex-A15 based systems (Exynos~5250 and~5410),
on the cortex-A15 compatible Apple ``Swift'' (A6X) and
on the Krait-based APQ8064, although much less often than on cortex-A9 based
systems.}
Note that we observed \DELETED{such a violation}\NEW{the violation of
\textsf{coRR} itself} quite frequently,
as illustrated by the first line of
the table in~\mytab\ref{fig:arm-runs}.
\begin{figure}
\begin{new}
\begin{center}\input{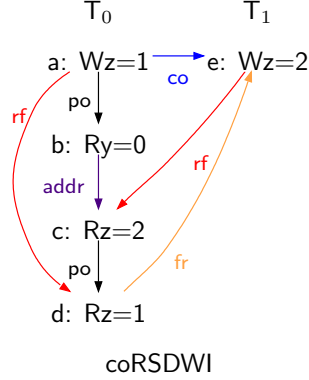}\end{center}
\caption{An observed behaviour that features a \textsf{coRR} violation\label{fig:coRSDWI}}
\end{new}
\end{figure}
\NEW{The second line of \mytab\ref{fig:arm-runs}
refers to a more sophisticated test,
\textsf{coRSDWI} (see \myfig\ref{fig:coRSDWI}), whose executions reveal violations of the \textsf{coRR} pattern on the location~$z$.
Both tests considered, we observed the load-load hazard bug
on all tested machines}.

\begin{table}[!h]
\begin{center}
\newcommand{\handletest}[1]{\rule{0ex}{1.8ex}\textsf{#1}}
\def\color#1{}
\begin{tabular}{l|r|r}
& \multicolumn{1}{|c}{model}& \multicolumn{1}{|c}{machines} \\ \hline

\handletest{coRR}& Forbid                    & Ok, 10M/95G              \\

\handletest{coRSDWI}& Forbid                    & Ok, 409k/18G             \\

\handletest{mp+dmb+fri-rfi-ctrlisb}& Forbid                    & Ok, 153k/178G            \\

\handletest{lb+data+fri-rfi-ctrl}& Forbid                    & Ok, 19k/11G              \\

\handletest{moredetour0052}& Forbid                    & Ok, 9/17G                \\

\handletest{mp+dmb+pos-ctrlisb+bis}& Forbid                    & Ok, 81/32G               \\
\end{tabular}

\end{center}

\caption{\label{fig:arm-runs}Some counts of invalid observations on ARM machines}
\end{table}

Others, such as the \textsf{mp+dmb+fri-rfi-ctrlisb} behaviour of
\myfig\ref{fig:arm-feature}, were claimed to be desirable behaviours by the
designers that we talked to.  This behaviour is a variant of the message
passing example, with some more accesses to the flag variable $y$ before the
read of the message variable $x$.  We observed this behaviour quite frequently
(\cf the \DELETED{second}\NEW{third} line of \mytab\ref{fig:arm-runs}) albeit
on one machine (of type APQ8060) only. 

\NEW{Additionally, we observed similar behaviours on APQ8064 (see also}
\url{http://moscova.inria.fr/~maranget/cats/model-qualcomm/compare.html#apq8064-invalid}).
\NEW{We give three examples of these behaviours in
\myfig\ref{fig:arm-put-features}. The first two are variants of the load
buffering example of \myfig\ref{fig:lb}. The last is a variant of the
``\textsf{s}'' idiom of \myfig\ref{fig:s}. We can only assume that these are as
desirable as the behaviour in \myfig\ref{fig:arm-feature}}

\NEW{For reasons explained in the next paragraph, we gather all such behaviours
under the name ``early commit behaviours'' (see reason \textsf{(ii)} in \S
``Our proposed ARM model'').} 

\begin{figure}[!h]
\begin{center}
\input{img/mp+dmb+fri-rfi-ctrlisb\bw.pstex_t} 
\end{center}
\vspace*{-4mm}
\caption{A \DELETED{putative}feature of some ARM machines\label{fig:arm-feature}}
\end{figure}

\begin{figure}[!h]
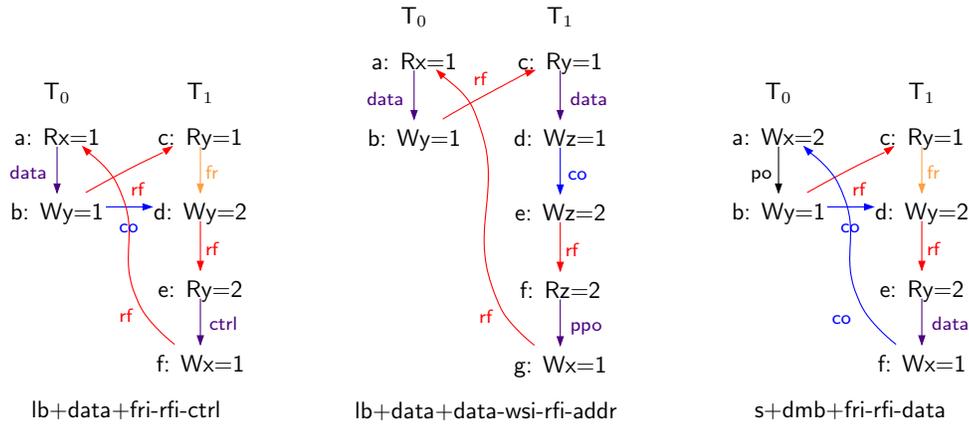

\begin{center}\def\tst#1{\input{img/#1\bw.pstex_t}}%
\scalebox{.9}{
\begin{tabular}{c@{}c@{}c}
\tst{lb+data+fri-rfi-ctrl} &
\tst{lb+data+data-wsi-rfi-addr} &
\tst{s+dmb+fri-rfi-data}
\end{tabular}}
\end{center}
\vspace*{-4mm}
\caption{\NEW{Putative features of some ARM machines}\label{fig:arm-put-features}}
\end{figure}

\paragraph{\NEW{Our proposed ARM model}}
Our \NEW{Power-ARM} model rejects \DELETED{this}\NEW{the
\textsf{mp+dmb+fri-rfi-ctrlisb}} behaviour via the
\textsc{\DELETED{causality}\NEW{observation}} axiom, as the event~$c$ is
$\ppo$-before the event~$f$.  More precisely, from our description of preserved
program order (see \myfig~\ref{fig:ppo}), the order from $c$ to~$f$ derives
from three reasons: \textsf{(i)} reads are satisfied before they are committed
($\init(r)$ precedes $\commit(r)$), \textsf{(ii)} instructions that touch the
same location commit in order ($\poloc$ is in $\ccz$), and~\textsf{(iii-a)}
instructions in a branch that are \po-after a control fence (here \isb) do not
start before the \isb{} executes, and \isb{} does not execute before the branch
is settled, which in turn requires the read ($e$ in the diagram) that controls
the branch to be committed ($\ctrlcfence{}$ is in $\ciz$).

\NEW{Our Power-ARM model rejects the \textsf{s+dmb+fri-rfi-data} behaviour via the \textsc{propagation} axiom for the same reason: the event $c$ is $\ppo$-before the event $f$.}

\begin{new}Similarily, our Power-ARM model rejects the
\textsf{lb+data+fri-rfi-ctrl} behaviour via the \textsc{no thin air} axiom, as
the event~$c$ is $\ppo$-before the event~$f$. Here, still from our description
of preserved program order (see \myfig~\ref{fig:ppo}) the order from $c$ to~$f$
derives from three reasons: \textsf{(i)} reads are satisfied before they commit
($\init(r)$ precedes $\commit(r)$), \textsf{(ii)} instructions that touch the
same location commit in order ($\poloc$ is in $\ccz$), and~\textsf{(iii-b)}
instructions (in particular store instructions) in a branch do not commit
before the branch is settled, which in turn requires the read ($e$ in the
diagram) that controls the branch to be committed ($\ctrl{}$ is~in~$\ccz$).
\end{new}

\NEW{Finally, our Power-ARM model rejects the
\textsf{lb+data+data-wsi-rfi-addr} via the \textsc{no thin air} axiom, because
because the event $c$ is $\ppo$-before the event $g$. Again, from our
description of preserved program order (see \myfig~\ref{fig:ppo}) the order
from $c$ to~$f$ derives from three reasons: \textsf{(i)} reads are satisfied
before they commit ($\init(r)$ precedes $\commit(r)$), \textsf{(ii)}
instructions that touch the same location commit in order ($\poloc$ is in
$\ccz$), and~\textsf{(iii-c)} instructions (in particular store instructions)
do not commit before a read they depend on (\eg the read $f$) is satisfied
($\addr{}$ is~in~$\icz$ because it is in $\iiz$ and $\iiz$ is included in
$\icz$).  }

The reasons~\textsf{(i)} \DELETED{and~}\NEW{, }\textsf{(iii\NEW{-a})},
\NEW{\textsf{(iii-b)} and \textsf{(iii-c)}} seem uncontroversial. In particular
for~\textsf{(iii\NEW{-a})}, if \ctrlcfence{} is not included in $\ppo$, then
neither the compilation scheme from C++ nor the entry barriers of locks would
work~\cite{smo12}.  \begin{new}For \textsf{(iii-b)},  if \ctrl{} to a write
event is not included in $\ppo$, then some instances of the pattern
\textsf{lb+ppos} (\cf \myfig\ref{fig:lb}) such as \textsf{lb+ctrls} would be
allowed. From the architecture standpoint, this could be explained by value
speculation, an advanced feature that current commodity ARM processors do not
implement. \NEW{A similar argument applies for \textsf{(iii-c)}, considering
this time that \textsf{lb+addrs} should not be allowed by a hardware model.  In
any case, allowing such simple instances of the pattern \textsf{lb+ppos} would
certainly contradict (low-level) programmer intuition.} \end{new} 

As for \textsf{(ii)} however, one could argue that this could be explained by
an ``early commit'' feature.  For example, looking at
\textsf{mp+dmb+fri-rfi-ctrlisb} in \myfig\ref{fig:arm-feature} and
\textsf{lb+data+fri-rfi-ctrl} in \myfig\ref{fig:arm-put-features}, the read~$e$
(which is satisfied by forwarding the local write~$d$) could commit without
waiting for the satisfying write $d$ to commit\NEW{, nor for any other write to
the same location that is \po-before~$d$}.  \begin{new}Moreover, the read~$e$
could commit without waiting for the commit of any read from the same location
that is \po-before its satisfying write. We believe that this might be
plausible from a micro-architecture standpoint: the value read by~$e$ cannot be
changed by any later observation of incomming writes performed by load
instructions that are \po-before the satisfying write~$d$; thus the value read
by $e$ is irrevocable.  \end{new}

\DELETED{Additionally, even after relaxing the commit order between a write and
a read to the same location, we need to allow reads from the same location
(here $c$ and~$e$) not to commit in program order, at least when the second
read is satisfied by store forwarding.}

In conclusion, to allow the \DELETED{\textsf{mp+fri-rfi-ctrlisb} }behaviour\NEW{s} of \myfig\ref{fig:arm-feature}, we need to weaken the definition of preserved
program order of~\myfig~\ref{fig:ppo}. For the sake of simplicity, we chose to
remove $\poloc$ altogether from the $\ccz$ relation, which is the relation
ordering the commit parts of events.  This means that two accesses relative to
the same location and in program order do not have to commit in this
order.\footnote{This is the most radical option; one could choose to remove
only $\poloc \cap \WR$ and $\poloc \cap \RR$
\NEW{, as that would be
enough to explain }\DELETED{\textsf{mp+dmb+fri-rfi-ctrlisb}}\NEW{the behaviours of}
\DELETED{(}\myfig\ref{fig:arm-feature}\DELETED{)}\NEW{ and similar others}.
We detail our experiments with alternative
formulations for $\ccz$ at
\url{http://diy.inria.fr/cats/arm-anomalies/index.html\#alternative}.
Ultimately we chose to adopt the weakest model since, as we explain in this
section, it still exhibits hardware anomalies.}

\NEW{Thus we propose the following model for ARM, which has so far not been
invalidated on hardware (barring the load-load hazard behaviours, acknowledged
as bugs by ARM~\cite{arm:rar} and the other anomalies presented in \S ``Testing
our model'', that we take to be undesirable). We go back to the soundness of
our ARM model at the end of this section, in \S ``Remark on our proposed ARM
model''.}

\NEW{The general skeleton of our ARM model should be the four axioms given in
\myfig\ref{fig:model}, and the propagation order should be as given in
\myfig\ref{fig:prop-power+arm}. For the preserved program order, we take it to
be as the Power one given in \myfig\ref{fig:ppo}, except for $\ccz$ which now
excludes $\poloc$ entirely, to account for the early commit behaviours, \ie
$\ccz$ should now be $\dep \cup \ctrl \cup (\addr;\po)$.
\mytab\ref{fig:arm-summary} give a summary of the various ARM models that we
consider in this section.}

\begin{table}
\begin{tabular}{c|p{.1\linewidth}|p{.2\linewidth}|p{.4\linewidth}}
         & Power-ARM & ARM & ARM llh \\
skeleton & \myfig\ref{fig:model} & \myfig\ref{fig:model} & \myfig\ref{fig:model}s.t. \textsc{sc} \textsc{per} \textsc{location} \. becomes $\acyclic(\polocllh \cup \com)$, with $\dfn{\polocllh}{\poloc \setminus \RR}$   \\
propagation & \myfig\ref{fig:prop-power+arm} & \myfig\ref{fig:prop-power+arm} & \myfig\ref{fig:prop-power+arm} \\   
$\ppo$ & \myfig\ref{fig:ppo} & \myfig\ref{fig:ppo} s.t. $\ccz$ becomes $\dep
\cup \ctrl \cup (\addr;\po)$ & \myfig\ref{fig:ppo}  s.t. $\ccz$ becomes $\dep
\cup \ctrl \cup (\addr;\po)$\end{tabular}
\caption{\NEW{Summary of ARM models}\label{fig:arm-summary}} \end{table}

\paragraph{\NEW{Testing our model}}
For the purpose of these experiments only, because our machines suffered from
the load-load hazard bug, we removed the read-read pairs from the
\DELETED{\textsc{uniproc}}\NEW{\textsc{sc per location}} check as well. This
allowed us to have results that were not cluttered by this bug.

We call the resulting model ``ARM llh'' (ARM load-load hazard), \ie the model
presented in~\myfig\ref{fig:model}, \ref{fig:prop-power+arm} and \ref{fig:ppo},
where we remove read-read pairs from $\poloc$ in the
\DELETED{\textsc{uniproc}}\NEW{\textsc{sc per location}} axiom, thus allowing
load-load hazard, and where $\ccz$ is ${\dep \cup \ctrl \cup (\addr;\po)}$.
\NEW{\mytab\ref{fig:arm-summary} gives a summary of this model, as well as
Power-ARM and our proposed ARM model.}

We then compared our original \NEW{Power-}ARM model (\ie taking literally the
definitions of~\myfig\ref{fig:model}, \ref{fig:prop-power+arm}
and~\ref{fig:ppo}) and the ARM llh~model with hardware:

\begin{table}[!h]
\begin{center}\newcommand{\fmtmdl}[1]{\multicolumn{1}{|c}{\textsc{#1}}}
\begin{tabular}{l*{10}{|>{$}r<{$}}}
 & \fmtmdl{All} & \fmtmdl{S} & \fmtmdl{T} & \fmtmdl{P} & \fmtmdl{ST} & \fmtmdl{SO} & \fmtmdl{SP} &
\fmtmdl{OP} & \fmtmdl{STO} & \fmtmdl{SOP} \\\hline
\rule{0ex}{2ex}Power-ARM & \acemodelarminvalid & \acemodelarmSinvalid & \acemodelarmTinvalid & \acemodelarmPinvalid & \acemodelarmSTinvalid & \acemodelarmSCinvalid & \acemodelarmSPinvalid & \acemodelarmCPinvalid & \acemodelarmSTCinvalid & \acemodelarmSCPinvalid \\
ARM~llh & \acerelaxedinvalid & \acerelaxedSinvalid & \acerelaxedTinvalid & \acerelaxedPinvalid & \acerelaxedSTinvalid & \acerelaxedSCinvalid & \acerelaxedSPinvalid & \acerelaxedCPinvalid & \acerelaxedSTCinvalid & \acerelaxedSCPinvalid 
\end{tabular}
\end{center}
\caption{Classification of anomalies observed on ARM hardware\label{fig:arm-bugs}}
\end{table}

More precisely, we classify the executions of both models: for each model we
count the number of invalid executions (in the sense of
\mysec\ref{sec:prelim}).  By invalid we mean that an execution is forbidden by
the model yet observed on hardware.  A given test can have several executions,
which explains why the numbers in \mytab\ref{fig:arm-bugs} are much higher than
the number of tests.

The table in \mytab\ref{fig:arm-bugs} is organised by sets of
axioms of our model (note that these sets are pairwise disjoint):
\DELETED{``U''}\NEW{``S''} is for \DELETED{\textsc{uniproc}}\NEW{\textsc{sc per
location}}, ``T'' for \textsc{no thin air}, \DELETED{``C''}\NEW{``O''} for
\textsc{\DELETED{causality}\NEW{observation}}, and ``P'' is for
\textsc{propagation}.  For each set of axioms (column) and model (row), we
write the number of executions forbidden by said axiom(s) of said model, yet
have been observed on ARM hardware. We omit a column (namely O, TO, TP, STP,
TOP, STOP) if the counts are $0$ for both models. 

For example, an execution is counted in the column \DELETED{``U''}\NEW{``S''}
of the row ``\NEW{Power-}ARM'' if it is forbidden by the
\DELETED{\textsc{uniproc}}\NEW{\textsc{sc per location}} check
of~\myfig\ref{fig:model} (and allowed by other checks).  The column
\DELETED{``U''}\NEW{``S''} of the row ``ARM llh'' is
\DELETED{\textsc{uniproc}}\NEW{\textsc{sc per location}} minus the read-read
pairs.  An execution is counted in the column ``OP'' of the row
``\NEW{Power-}ARM'' if it is forbidden by both
\textsc{\DELETED{causality}\NEW{observation}} and \textsc{propagation} as
defined in \myfig\ref{fig:model} (and allowed by other checks); for the row
``ARM llh'', one needs to take into account the modification to $\ccz$
mentioned above in the definition of the $\ppo$.

While our original \NEW{Power-ARM} model featured $\nmodelarminvalid$ tests that
exhibited~$\acemodelarminvalid$ invalid executions forbidden by the model yet
observed on ARM machines, those numbers drop to $\nrelaxedinvalid$ tests and
$\acerelaxedinvalid$ invalid executions for the ARM~llh model (see column ``ARM
llh'', row ``\textsc{All}''; see also
\url{http://diy.inria.fr/cats/relaxed-classify/index.html}). 

An inspection of each of these anomalies revealed what we believe to be more
bugs.  We consider the violations of \DELETED{\textsc{uniproc}}\NEW{\textsc{sc
per location}} to be particularly severe (see all rows mentioning
\DELETED{U}\NEW{S}). By contrast, the load-load hazard behaviour could be
argued to be desirable, or at least not harmful, and was indeed officially
allowed by Sparc RMO~\cite{sparc:94} and pre-Power 4
machines~\cite{tdf02}\NEW{, as we already say in \mysec\ref{model:gen}}.
 
\begin{figure}[!h] \vspace*{-3mm} \begin{center}
\NEW{\input{img/moredetour0052\bw.pstex_t}} \end{center} \caption{\DELETED{Two striking
violations}\NEW{A violation} of \DELETED{\textsc{uniproc}}\NEW{\textsc{sc per
location}} observed on ARM hardware\label{fig:arm-corw-bugs}} \end{figure}

\DELETED{These behaviours are once again variants of the message passing \DELETED{idiom}\NEW{pattern}, like
the~\textsf{mp+fri-rfi-ctrlisb} of~\myfig\ref{fig:arm-feature}.
Unlike~\textsf{mp+fri-rfi-ctrlisb}, they cannot be claimed to be a feature, in
a way that we believe uncontroversial.  Indeed they exhibit a violation of
the~\textsf{coRW1} \DELETED{idiom}\NEW{pattern} (see \mysec\ref{sec:model}). This basically allows a
load to read from the future: for instance in the left diagram, the read~$c$
from location~$y$ reads from the~$\po$-successor write~$d$. }

\myfig\ref{fig:arm-corw-bugs} shows \DELETED{two striking examples}\NEW{a
violation of~\textsc{sc per location}}.  \NEW{Despite the apparent complexity
of the picture, the violation is quite simple.  The violation occurs on
\myth{1} wich loads the value $4$ from the location~$y$  (event~$f$), before
writing the value~$3$ to same location~$y$ (event~$g$).  However, $4$ is the
final value of the location~$y$, as the test harness has observed once the test
has completed. As a consequence, the event $e$ $\co$-precedes the event $f$ and
we witness a cycle $g \stackrel{\co}{\rightarrow} e \stackrel{\rf}{\rightarrow}
f \stackrel{\poloc}{\rightarrow}  g$. That is we witness a violation of the
~\textsf{coRW2} pattern  (see \mysec\ref{sec:model})}.

Note that we observed \DELETED{these}\NEW{this} behaviour\DELETED{s} rather infrequently, as shown in the
\DELETED{third and fourth}\NEW{fith} line\DELETED{s} of \mytab\ref{fig:arm-runs}. \NEW{However, the behaviour is observed on two machines of type
Tegra3 and~Exynos4412.}


In addition to the violation\DELETED{s} of
\DELETED{\textsc{uniproc}}\NEW{\textsc{sc per location}} shown in
\myfig\ref{fig:arm-corw-bugs}, we observed the two behaviours given
in\myfig\ref{fig:arm-causality-violation} (rather infrequently, as shown on the last line of
\mytab\ref{fig:arm-runs}, and on one machine only, of type Tegra3), which
violate \textsc{\DELETED{causality}\NEW{observation}}.

In addition to the violations of \DELETED{\textsc{uniproc}}\NEW{\textsc{sc per
location}} shown in \myfig\ref{fig:arm-corw-bugs}, we observed the\DELETED{
following} two behaviours\NEW{~of~\myfig\ref{fig:arm-causality-violation}}
(rather infrequently, as shown on the last line of \mytab\ref{fig:arm-runs},
and on one machine only, of type Tegra3), which violate
\textsc{\DELETED{causality}\NEW{observation}}.
 
\begin{figure}[!h]
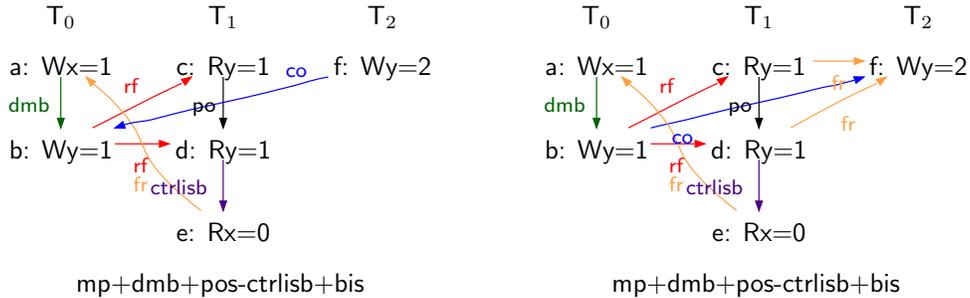

\begin{center}
\input{img/mp-bug2\bw.pstex_t}\quad\quad\input{img/mp-bug3\bw.pstex_t} 
\end{center}
\caption{\NEW{Two violations of \textsc{observation}} observed on ARM hardware\label{fig:arm-causality-violation}}
\end{figure}

The test \textsf{mp+dmb+pos-ctrlisb+bis} includes the simpler test
\textsf{mp+dmb+ctrlisb} plus one extra read ($c$ on \myth{1}) and one extra
write ($f$ on~\myth{2}) of the flag variable~$y$.  The  depicted behaviours are
violations of the \textsf{mp+dmb+ctrlisb} \DELETED{idiom}\NEW{pattern}, which
must uncontroversially be forbidden. Indeed the only way to allow
\textsf{mp+dmb+ctrlisb} is to remove \ctrlcfence{} from the preserved program
order $\ppo$. We have argued above that this would for example break the
compilation scheme from C++ to Power (\cf~\cite{smo12}). 

\begin{new}It is worth noting that we have observed other violations of
\textsc{\DELETED{causality}\NEW{observation}} on Tegra3, as one can see at
\url{http://diy.inria.fr/cats/relaxed-classify/OP.html}.\end{new} \NEW{For
example we have observed \textsf{mp+dmb+ctrlisb}, \textsf{mp+dmb+addr},
\textsf{mp+dmb.st+addr}, which should be uncontroversially forbidden.  We tend
to classify such observations as bugs of the tested chip. However, since the
tested chip exhibits the acknowledged read-after-read hazard bug, the blame can
also be put on the impact of this acknowledged bug on our testing
infrastructure. Yet this would mean that this impact on our testing
infrastructure would reveal on Tegra3 only.}

\DELETED{More generally} \NEW{In any case,} the interplay between having
several consecutive accesses relative to the same location on one thread (\eg
$c,d$ and~$e$ on~\myth{1} in \textsf{mp+dmb+fri-rfi-ctrlisb} ---
\cf\myfig\ref{fig:arm-feature}), in particular two reads ($c$ and~$e$), and the
message passing \DELETED{idiom}\NEW{pattern} \textsf{mp}, seems to pose
implementation difficulties (see the violations of
\textsc{\DELETED{causality}\NEW{observation}} listed in
~\mytab\ref{fig:arm-bugs}, in the columns containing
\DELETED{``C''}\NEW{``O''}, and the two \DELETED{following} examples in
\myfig\ref{fig:arm-causality-violation}). 

\paragraph{\NEW{Remarks on our proposed ARM model}}  

Given the state of affairs for ARM, we do not claim our model \NEW{(see model
``ARM'' in \mytab\ref{fig:summary-testing})} to be definitive. \DELETED{
However, we express serious doubts \wrt the~\textsf{dmb+fri-rfi-ctrlisb}
behaviour of \myfig\ref{fig:arm-feature} being as desirable as the designers
claimed it to be.Indeed, this behaviour for now appears only on machines which
suffer from other quite serious flaws (see the violations of \textsc{uniproc}
listed in ~\mytab\ref{fig:arm-bugs}, in the columns starting with ``U'', and
our two striking examples in \myfig\ref{fig:arm-corw-bugs}).}

\DELETED{We thus wonder if the behaviour \textsf{mp+dmb+fri-rfi-ctrlisb} of
\myfig\ref{fig:arm-feature} can only be implemented on a machine with load-load
hazards, which ARM acknowledged to be a flaw (\cf~\cite{arm:rar}).  Indeed, the
new generation of machines that we have tested (Apple A6X and Samsung Exynos
5250) do not invalidate our original model (see
\texttt{http://diy.inria.fr/cats/model-newgeneration}), and are the only ARM
implementations that we have tested that do not exhibit the load-load hazard
bug.  Coincidentally, we have not observed \textsf{mp+dmb+fri-fri-ctrlisb} on
these machines either.}

\NEW{In particular, we wonder if the behaviour \textsf{mp+dmb+fri-rfi-ctrlisb}
of \myfig\ref{fig:arm-feature} can only be implemented on a machine with
load-load hazards, which ARM acknowledged to be a flaw (\cf~\cite{arm:rar}), as
it involves two reads from the same address. }

\NEW{Nevertheless, our ARM contacts were fairly positive that they would like
this behaviour to be allowed. Thus we think a good ARM model should account for
it. As to the similar early commit behaviours given in
\myfig\ref{fig:arm-put-features}, we can only assume that they should be
allowed as well.}

\NEW{Hence our ARM model allows such behaviours, by excluding $\poloc$ from the
commit order $\ccz$ (see \myfig\ref{fig:ppo} and \ref{fig:summary-testing}).
We have performed experiments to compare our ARM model and ARM hardware. To do
so, we have excluded the load-load hazard related
behaviours.}\footnote{\NEW{More precisely, we have built the model that only
allows load-load hazard behaviours. In herd parlance, this is a model that only
has one check: $\reflexive(\poloc;\fr;\rf)$. We thus filtered the behaviours
observed on hardware by including only the behaviours that are not allowed by
this load-load hazard model (\ie{} all but load-load hazard behaviours). We
then compared these filtered hardware behaviours with the ones allowed by our
ARM model.}}

\NEW{We give the full comparison table at
\url{http://diy.inria.fr/cats/proposed-arm/}.  As one can see, we still
have~$31$ behaviours that our model forbids yet are observed on hardware (on
Tegra 2, Tegra 3 and Exynos 4412).}

\NEW{All of them seem to present anomalies, such as the behaviours that we show
in \myfig\ref{fig:arm-corw-bugs} and \ref{fig:arm-causality-violation}. We will
consult with our ARM contacts for confirmation.}

\subsection{\NEW{Experimental comparisons of models}}

\paragraph{\DELETED{Experimental comparisons of models} \!\!\!\!} \DELETED{was
done using the same $\npower$ and $\narm$ tests that we used to exercise Power
and ARM machines. }

\DELETED{We compare our model to the one of~\cite{ssa11}: our experimental data
can be found at \texttt{http://diy.inria.fr/cats/pldi-model}.
Experimentally, our model allows all the behaviours that are allowed
by the one of~\cite{ssa11}. The only exception to that claim is a test which
involves the \eieio{} barrier, which we take to be a bug in the \prog{ppcmem}
simulator of~\cite{ssa11}. }

\DELETED{We also observe experimentally that our model and the one
of~\cite{ssa11} differ \emph{only} on the behaviours that~\cite{ssa11} wrongly
forbids (\cf \texttt{http://diy.inria.fr/cats/pldi-power/\#lessvs}).}

\DELETED{Finally, we observe experimentally that our model and the one
of~\cite{mms12} are equivalent on our set of tests: our experimental data can
be found at \texttt{http://diy.inria.fr/cats/cav-model}.}

\NEW{Using the same $\npower$ and $\narm$ tests that we used to exercise Power
and ARM machines, we have experimentally compared our model to the one
of~\cite{ssa11} and the one of~\cite{mms12}.}

\paragraph{\NEW{Comparison with the model of} \!\!\!\!}\NEW{~\cite{ssa11}: our
experimental data can be found at \url{http://diy.inria.fr/cats/pldi-model}.
Experimentally, our Power model allows all the behaviours that are allowed by
the one of~\cite{ssa11}.} 

\NEW{We also observe experimentally that our Power model and the one
of~\cite{ssa11} differ \emph{only} on the behaviours that~\cite{ssa11} wrongly
forbids (\cf \url{http://diy.inria.fr/cats/pldi-power/\#lessvs}). We give one
such example in \myfig\ref{fig:mp-detour}: the behaviour
\textsf{mp+lwsync+addr-po-detour} is observed on hardware yet forbidden by the
model of~\cite{ssa11}.}

\begin{figure}[!h]
\begin{center}\input{img/mp+lwsync+addr-po-detour\bw.pstex_t}\end{center}
\caption{\NEW{A behaviour forbidden by the model of Sarkar et al. but observed on Power hardware}\label{fig:mp-detour}}
\end{figure}

\NEW{We note that some work is ongoing to adapt the model of~\cite{ssa11} to
allow these tests (\cf \url{http://diy.inria.fr/cats/op-power}). This variant
of the model of~\cite{ssa11} has so far not been invalidated by the hardware.}

\NEW{Finally, the model of~\cite{ssa11} forbids the ARM ``fri-rfi'' behaviours such
as the ones given in \myfig\ref{fig:arm-feature}. Some work is ongoing to adapt
the model of~\cite{ssa11} to allow these tests.}

\paragraph{\NEW{Comparison with the model of} \!\!\!\!}\NEW{~\cite{mms12}: our
experimental data can be found at \url{http://diy.inria.fr/cats/cav-model}.}
\NEW{Our Power model and the one of~\cite{mms12} are experimentally equivalent
on our set of tests, except for a few tests of similar structure.
Our model allows them, whereas the
model of~\cite{mms12} forbids them, and they are not observed on hardware. We
give the simplest such test in \myfig\ref{fig:bigdetour}.
The test is a refinement of the \textsf{mp+lwsync+ppo} pattern (\cf \myfig\ref{fig:mp}). The difference of acceptance between the two models can be explained
as follows: the model of~\cite{mms12}
does preserve the program order from \myth{1} initial read~$c$ to
\myth{1} final read~$f$, while our model does not.
More precisely, the issue reduces to reads~$d$ and~$e$ (on~\myth{1})
being ordered or not.
And, indeed, the propagation model for writes of~\cite{mms12}
enforces the order, while our definition of \ppo{} does~not.}
\begin{new}\newsavebox{\savecite}\sbox{\savecite}{\cite{mms12}}
\begin{figure}[!h]
\begin{new}
\begin{center}\input{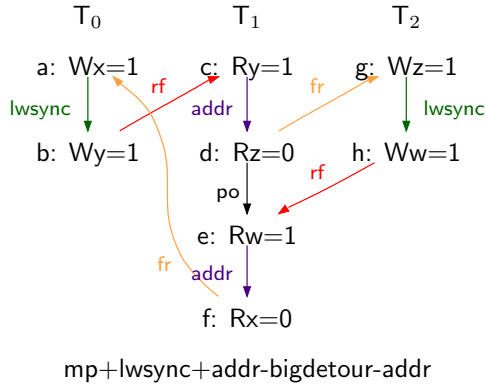}\end{center}
\caption{A behaviour allowed by our model while forbidden by~\usebox{\savecite}\label{fig:bigdetour}}
\end{new}
\end{figure}
\end{new}

\NEW{If such a test is intentionally forbidden by the architect, it seems to
suggest that one could make the preserved program order of Power (see
\myfig\ref{fig:ppo}) stronger. Indeed one could take into account the
effect of barriers (such as the one between the two writes $g$ and $h$ on \myth{2}
in the figure above) within the preserved program order. }

\NEW{Yet, we think that one should tend towards more simplicity in the
definition of the preserved program order. It feels slightly at odds with our
intuition that the preserved program order should take into account dynamic
notions such as the propagation order of the writes g and h. By dynamic
notions, we here mean notions whose definitions require execution relations
such as $\rf$ or $\prop$.}

\NEW{As a related side note, although we did include the dynamic relations
\textsf{rdw} and \textsf{detour} into the definition of the preserved program
order in \myfig\ref{fig:ppo}, we would rather prescribe not to include them.
This would lead to a weaker notion of preserved program order, but more
stand-alone. By this we mean that the preserved program order would just
contain per-thread information (\eg the presence of a control fence, or a
dependency between two accesses), as opposed to external communications such as
\rfe.} 

\NEW{We experimented with a weaker, more static, version of the preserved
program order for Power and ARM, where we excluded \rdw{} from $\iiz$ and
\detour{} from $\ciz$ (see \myfig\ref{fig:ppo}). We give the full experiment
report at \url{http://diy.inria.fr/cats/nodetour-model/} On our set of tests,
this leads to only \nnodetourpower{} supplementary behaviours allowed on Power
and \nnodetourarm{} on ARM. We believe that this suggests that it might not be
worth complicating the $\ppo$ for the sake of only a few behaviours being
forbidden.  Yet it remains to be seen whether these patterns are so common that
it is important to determine their precise status \wrt a given model.}


\subsection{Model-level simulation\label{sec:herd}}

Simulation was done using our new \prog{herd} tool:
given a model specified in the terms of~\mysec\ref{sec:model} and a litmus
test, \prog{herd} computes all the executions allowed by the model. We distribute our
tool, its sources and documentation at \url{http://diy.inria.fr/herd}.

Our tool \prog{herd} understands \DELETED{any} model\NEW{s} specified
\DELETED{as in}\NEW{in the style of} \mysec\ref{sec:model}, \ie defined in
terms of relations over events, and irreflexivity or acyclicity of these
relations. For example, \myfig\ref{fig:herd-ppc-model} gives the \prog{herd}
model corresponding to our Power model (\cf \mysec\ref{sec:model} and
\ref{sec:power}).

We emphasise the concision of~\myfig\ref{fig:herd-ppc-model}, which contains
the \emph{entirety} of our Power model. 

\begin{figure*}[!t]
\begin{verbatim}
(* sc per location *) acyclic po-loc|rf|fr|co

(* ppo *)
let dp = addr|data 
let rdw = po-loc & (fre;rfe)
let detour = po-loc & (coe;rfe)

let ii0 = dp|rdw|rfi
let ic0 = 0
let ci0 = (ctrl+isync)|detour
let cc0 = dp|po-loc|ctrl|(addr;po)

let rec ii = ii0|ci|(ic;ci)|(ii;ii)
and ic = ic0|ii|cc|(ic;cc)|(ii;ic)
and ci = ci0|(ci;ii)|(cc;ci)
and cc = cc0|ci|(ci;ic)|(cc;cc)
let ppo = RR(ii)|RW(ic)

(* fences *)
let fence = RM(lwsync)|WW(lwsync)|sync

(* no thin air *) 
let hb = ppo|fence|rfe
acyclic hb

(* prop *) 
let prop-base = (fence|(rfe;fence));hb*
let prop = WW(prop-base)|(com*;prop-base*;sync;hb*)

(* observation *) irreflexive fre;prop;hb*
(* propagation *) acyclic co|prop
\end{verbatim}
\caption{\prog{herd} definition of our Power model\label{fig:herd-ppc-model}}
\end{figure*}

\paragraph{Language description: \!\!\!\!} we build definitions with {\tt let,
let rec} and {\tt let rec ... and ...} operators. We build unions,
intersections and sequences of relations with ``{\tt |}'', ``{\tt \&}'' and
``{\tt ;}'' respectively; transitive closure with ``{\tt +}'', and transitive
and reflexive closure with ``{\tt *}''. The empty relation is ``{\tt 0}''.

We have some built-in relations, \eg $\poloc, \rf, \fr, \co$, $\addr, \data$,
and operators to specify whether the source and target events are reads or
writes. For example {\tt RR(r)} gives the relation {\tt r} restricted to both
the source and target being reads.

To some extent, the language that \prog{herd} takes as input shares some
similarities with the much broader Lem project~\cite{obz11}. However, we merely
intend to have a concise way of defining a variety of memory models, whereas
Lem aims at (citing the webpage:
\url{http://www.cs.kent.ac.uk/people/staff/sao/lem/}) \emph{``large scale
semantic definitions. It is also intended as an intermediate language for
generating definitions from domain-specific tools, and for porting definitions
between interactive theorem proving systems.''}

\NEW{The \prog{alloy} tool~\cite{jac02} (see
also~\url{http://alloy.mit.edu/alloy}) is closer to \prog{herd} than Lem. Both
\prog{alloy} and \prog{herd} allow a concise relational definition of a given
system. But while \prog{alloy} is very general, \prog{herd} is only targeted at
memory models definitions.}

\NEW{Thus one could see \prog{herd} as a potential front-end to \prog{alloy}.
For example, \prog{herd} provides some built-in objects (\eg{} program order,
read-from), that spare the user the effort of defining these objects;
\prog{alloy} would need the user to make these definitions explicit. }

\NEW{More precisely, to specify a memory model in \prog{alloy}, one would need
to explicitly define an object ``memory event'', for example a record with an
identifier, a direction, \ie{} write or read, a location and a value, much like
we do in our Coq development
(see~\url{http://www0.cs.ucl.ac.uk/staff/j.alglave/cats}).}

\NEW{One would also need to handcraft relations over events (\eg{} the program
order \po), as well as the well-formedness conditions of these relations (\eg{}
the program order is total order per thread), using first order logic. Our tool
\prog{herd} provides all these basic bricks (events, relations and their
well-formedness conditions) to the user.}

\NEW{Finally, \prog{alloy} uses a SAT solver as a backend, whereas \prog{herd}
uses a custom solver optimised for the limited constraints hat \prog{herd}
supports (namely acyclicity and irreflexivity of relations).}

\paragraph{Efficiency of simulation: \!\!\!}  our axiomatic description
underpins \prog{herd}, which allows for a greater efficiency in the simulation.
By contrast, simulation tools based on operational models (\eg
\prog{ppcmem}~\cite{ssa11}, or the tool of~\cite{bps12}\footnote{All the
results relative to the tool of~\cite{bps12} are courtesy of Arthur Guillon,
who exercised the simulator of~\cite{bps12} on a subset of the tests that we
used for exercising the other tools. }) are not able to process all tests
within the memory bound of $40$~GB for~\cite{ssa11} and $6$~GB
for~\cite{bps12}: \prog{ppcmem} processes $\npldipower$ tests out of $\npower$;
the tool of~\cite{bps12} processes $396$ tests out of~$518$.

Tools based on \NEW{multi-event} axiomatic models \DELETED{but with more than
one event per instruction} (\eg our reimplementation of~\cite{mms12}
inside~\prog{herd}) are able to process all $\npower$ tests, but require more
than eight times the time that our single-event axiomatic model needs.

\mytab\ref{fig:simulation-tools-comparison} gives a summary of the number of
tests that each tool can process, and the time needed to do so.
\begin{table}[!h] \scalebox{0.9}{ \begin{tabular}{c|c|c|c|c} tool & model &
style & \# of tests & (user) time in $s$ \\\hline \prog{ppcmem} & \cite{ssa11}
&  operational & $\npldipower$ & $\timepldipower$ \\ \prog{herd} & \cite{mms12}
& multi-event axiomatic & $\ncavpower$ & $\timecavpower$ \\ --- & \cite{bps12}
& operational & $396$ &
$53100$ \\ \prog{herd} & this model & single-event axiomatic & $\nmodelpower$ &
$\timemodelpower$ \end{tabular}} \caption{Comparison of simulation tools (on
Power)\label{fig:simulation-tools-comparison}} \end{table}

As we have implemented the model of~\cite{mms12}\footnote{The implementations
tested in~\cite{mms12} were much less efficient.} and the present model inside
\prog{herd} using the same techniques, we claim that the important gain in
runtime efficiency originates from reducing the number of events. On a reduced
number of events, classical graph algorithms such as acyclicity test and, more
significantly, transitive closure and other fixed point calculations run much
faster. 

We note that simulation based on axiomatic models outperforms simulation based
on operational models.  This is mostly due to a state explosion issue, which is
aggravated by the fact that Power and ARM are very relaxed architectures. Thus
in any given state of the operational machine, there are numerous operational
transitions enabled. 

We note that \prog{ppcmem} is not coded as efficiently as it could be.
Better implementations are called for, but the distance to \prog{herd} is
considerable: \prog{herd} is about 45000 times faster than \prog{ppcmem}, and
\prog{ppcmem} fails to process about half of the tests.  

\NEW{We remark that our single-event axiomatic model also needs several
subevents to describe a given instruction (see for example our definition of
the preserved program order for Power, in \myfig\ref{fig:ppo}). Yet the
opposition between multi-event and single-event axiomatic models lies in the
number of events needed to describe the propagation of writes to the system. In
multi-event models, there is roughly one propagation event per thread,
mimicking the transitions of an operational machine. In single-event models,
there is only one event to describe the propagation to several different
threads; the complexity of the propagation mechanism is captured through our
use of the relations (\eg $\rf, \co, \fr$ and $\prop$).}

We note that \NEW{single-event} axiomatic simulators also suffer from
combinatorial explosion. The initial phase computes executions (in the sense
of~\mysec~\ref{sec:prelim}) and thus enumerates all possible $\rf$ and $\co$
relations.  However, as clearly shown
in~\mytab\ref{fig:simulation-tools-comparison}, the situation is less severe,
and we can still process litmus tests of up to four or five threads.

\subsection{Verification of C programs\label{sec:verif}}

While assembly-level litmus tests enable detailed study of correctness of the
model, the suitability of our model for the verification of high-level programs
remains to be proven. To this effect, we experimented with a modified version
of \prog{CBMC}~\cite{ckl04}, which is a bounded model-checker for C programs.
Recent work~\cite{akt13} has implemented the framework of~\cite{ams10,ams12} in
\prog{CBMC}, and observed speedups of an order of magnitude \wrt other
verification tools. \prog{CBMC} thus features several models, ranging from SC
to Power.

In addition, the work of~\cite{akn13} proposes an instrumentation technique,
which transforms a concurrent program so that it can be processed by an SC
verification tool, \eg \prog{CBMC} in SC mode. This relies on an operational
model equivalent to the one of~\cite{ams12}; we refer to it
in~\mytab\ref{fig:verif-axiomatic-operational} under the name
``\prog{goto-instrument+tool}''.  The advantage of supporting existing tools in
SC mode comes at the price of a considerably slower verification time when
compared to the implementation of the equivalent axiomatic model within the
verification tool, as \mytab\ref{fig:verif-axiomatic-operational} shows.

\begin{table}[!h]
\begin{center}
\begin{tabular}{c|c|c|c}
tool & model & \# of tests & time in $s$ \\\hline
\prog{goto-instrument+CBMC (SC)} & \cite{ams12} & $555$ & $2511.6$ \\
\prog{CBMC (Power)} & \cite{ams12} & $555$ & $14.3$ \\
\end{tabular}
\end{center}
\caption{Comparison of operational vs.~axiomatic model implementation\label{fig:verif-axiomatic-operational}}
\end{table}

We adapted the encoding of~\cite{akt13} to our present framework, and recorded
the time needed to verify the reachability of the final state of more than
$4000$ litmus tests (translated to C). As a comparison point, we also
implemented the model of~\cite{mms12} in \prog{CBMC}, and compared the
verification times, given in~\mytab\ref{fig:verif-comparison-litmus}.
We observe some speedup with the present model over the implementation of the
model of~\cite{mms12}.

\begin{table}[!h]
\begin{center}
\begin{tabular}{c|c|c|c}
tool & model & \# of tests & time in $s$ \\\hline
\prog{CBMC} & \cite{mms12} & $4450$ & $1944$ \\
\prog{CBMC} & present one & $4450$ & $1041$
\end{tabular}
\end{center}
\caption{Comparison of verification tools on litmus tests\label{fig:verif-comparison-litmus}}
\end{table}

We also compared the same tools, but on more fully-fledged examples, described
in detail in~\cite{akn13,akt13}: \prog{PgSQL} is an excerpt of the PostgreSQL database server
software
(\cf\url{http://archives.postgresql.org/pgsql-hackers/2011-08/msg00330.php});
\prog{RCU} is the Read-Copy-Update mechanism of the Linux
kernel~\cite{rcu:lwn}, and \prog{Apache} is a queue mechanism extracted from
the Apache HTTP server software. In each of the examples we added correctness
properties, described in~\cite{akt13}, as assertions to the original source
code.  
We observed that the verification times of these particular examples are not
affected by the choice of either of the two axiomatic models, as shown in
\mytab\ref{fig:verif-comparison-fledged}.

\begin{table}[!h]
\begin{center}
\begin{tabular}{c|c|c|c|c}
tool & model & \prog{PgSQL} & \prog{RCU} & \prog{Apache} \\\hline
\prog{CBMC} & \cite{mms12} & $1.6$ & $0.5$ & $2.0$ \\
\prog{CBMC} & present one & $1.6$ & $0.5$ & $2.0$ 
\end{tabular}
\end{center}
\caption{Comparison of verification tools on full-fledged examples\label{fig:verif-comparison-fledged}}
\end{table}

\section{A pragmatic perspective on our models}\label{sec:search}
\fixmeskip{jade@Michael: could you put in this section a stripped-down version
of pgsql, just like RCU?}
To conclude our paper, we put our modelling framework into perspective, in the
light of actual software. Quite pragmatically, we wonder
\DELETED{if}\NEW{whether} is it worth going through the effort of defining, on
the one hand, then studying or implementing, on the other hand, complex models
such as the Power and ARM models that we present in \mysec\ref{sec:power}. Are
there fragments of these models that are simpler to understand, and embrace
pretty much all the \DELETED{idioms}\NEW{patterns} that are used in actual
software? 

For example, there is a folklore notion that \textsf{iriw}
(\cf\myfig\ref{fig:iriw}) is very rarely used in practice.
If that is the case, do we need models that \DELETED{handle}\NEW{can explain} \textsf{iriw}? 

Conversely, one flaw of the model of~\cite{ams11} (and also of~\cite{bps12}) is
that it forbids the \DELETED{idiom}\NEW{pattern} \textsf{r+lwsync+sync} (\cf~\myfig\ref{fig:r}),
against the architect's intent~\cite{ssa11}. While designing the model that we
present in the current paper, we found that accounting for this
\DELETED{idiom}\NEW{pattern} increased the complexity of the model. If this
\DELETED{idiom}\NEW{pattern} is never used in practice, it might not be worth
inventing a model that accounts for it, if it makes the model much more
complex.
 
Thus we ask the following questions: what are the
\DELETED{idioms}\NEW{patterns} used in modern software? What are their
frequencies? 

Additionally, we would like to understand \DELETED{if}\NEW{whether} there are programming
\DELETED{idioms}\NEW{patterns} used in current software that are not accounted
for by our model. Are there programming \DELETED{idioms}\NEW{patterns} that are
not represented by one of the axioms of our model, \ie
\DELETED{\textsc{uniproc}}\NEW{\textsc{sc per location}}, \textsc{no thin air},
\textsc{\DELETED{causality}\NEW{observation}} or \textsc{propagation}, as given
in~\myfig\ref{fig:model}? 

Conversely, can we understand all the \DELETED{idioms}\NEW{patterns} used in
current software through the prism of, for example, our
\textsc{\DELETED{causality}\NEW{observation}} axiom, or is there an actual need
for the \textsc{propagation} axiom too?
Finally, we would like to understand to what extent do hardware anomalies, such
as the load-load hazard behaviour that we observed on ARM chips
(\cf\mysec\ref{sec:testing}) impair the behaviour of actual software?

To answer these questions, we resorted to the largest code base available to
us: an entire Linux distribution. 

\paragraph{What we analysed:} we picked the current stable release of the
Debian Linux distribution (version 7.1,
\url{http://www.debian.org/releases/stable/}), which contains more than $17\,000$
software packages (including the Linux kernel itself, server software such as
Apache or PostgreSQL, but also user-level software, such as Gimp or Vim).

David A.~Wheeler's SLOCCount tool (\url{http://www.dwheeler.com/sloccount/})
reports more than $400$ million lines of source code in this distribution. C
and C++ are still the predominant languages: we found more than $200$ million
lines of C and more than $129$ million lines of C++.

To search for \DELETED{idioms}\NEW{patterns}, we first gathered the packages
which possibly make use of concurrency.  That is, we selected the packages that
make use of either POSIX threads or Linux kernel threads anywhere in their C
code. This gave us $1590$ source packages to analyse; this represents $9.3\%$
of the full set of source packages.

The C language~\cite{c11} does not have an explicit notion of shared memory. Therefore,
to estimate the number of shared memory interactions, we looked for variables
with static storage duration (in the C11 standard sense~\cite[\S 6.2.4]{c11})
that were not marked thread local. We found a total of $2\,733\,750$ such
variables. In addition to these, our analysis needs to consider local
variables shared through global pointers and objects allocated on the heap to
obtain an overapproximation of the set of objects (in the C11 standard
sense~\cite[\S 3.15]{c11}) that may be shared between threads.
 
\paragraph{A word on C++:} the C++ memory model has recently received
considerable academic attention (see \eg~\cite{bos11,smo12,bdg13}).  Yet to
date even a plain text search in all source files for uses of the corresponding
\texttt{stdatomic.h} and \texttt{atomic} header files only reveals occurrences
in the source code of compilers, but not in any of the other source packages.

Thus practical assessment of our subset of the C++ memory model is necessarily
left for future work. At the same time, this result reinforces our impression
that we need to study hardware models to inform current concurrent programming. 

\subsection{Static \NEW{pattern} search} 

To look for \DELETED{idioms}\NEW{patterns} in Debian 7.1, we implemented a
static analysis in a new tool called \prog{mole}. This means that we are
looking for an overapproximation of the \DELETED{idioms}\NEW{patterns} used in
the program.  Building on the tool chain described in~\cite{akn13}, we use the
front-end \prog{goto-cc} \DELETED{to compile C programs to an intermediate
representation: \emph{goto-programs}. We then analyse these goto-programs
using}\NEW{and}
a variant of the \prog{goto-instrument} tool of~\cite{akn13}, with the new
option \prog{-{}-static-cycles}.  We distribute our tool \prog{mole}, along
with a documentation, at \url{http://diy.inria.fr/mole}. 

\subsubsection{\NEW{Preamble on the goto-* tools}}

\NEW{\prog{goto-cc} and \prog{goto-instrument} are part of the tool chain of
  CBMC~\cite{ckl04}, which is widely recognised for its
  maturity.\footnote{\url{http://www.research.ibm.com/haifa/conferences/hvc2011/award.shtml}}
  \prog{goto-cc} may act as compiler substitute as it accepts the same set of
  command line options as several C compilers, such as GCC.
  Instead of executables, however, \prog{goto-cc} compiles C programs to an
  intermediate representation shared by the tool chain around CBMC:
  \emph{goto-programs}.
  These goto-programs can be transformed and inspected
  using \prog{goto-instrument}. For instance, \prog{goto-instrument} can be
  applied to
  insert assertions of generic invariants such as valid pointer dereferencing
  or data race checks, or dump goto-programs as C code.
  Consequently we implemented the search described below in
\prog{goto-instrument}, adding the new option \prog{-{}-static-cycles}.}

\subsubsection{Cycles \label{sec:crit}} 

Note that an \DELETED{idiom}\NEW{pattern} like all the ones that we have presented in this paper
corresponds to a cycle of the relations of our model.  This is simply because
our model is defined in terms of irreflexivity and acyclicity checks. Thus
looking for \DELETED{idioms}\NEW{patterns} corresponds here to looking for cycles of relations.

\paragraph{Critical cycles}
Previous works~\cite{ss88,am11,bmm11,bdm13} show that a certain kind of cycles,
which we call \emph{critical cycles} (following~\cite{ss88}), characterises
many weak behaviours. Intuitively, a critical cycle violates SC in a minimal
way. 

We recall here the definition of a critical cycle (see~\cite{ss88} for more
details). Two events $x$ and $y$ are \emph{competing}, written $(x,y) \in \cmp$, if
they are from distinct processors, to the same location, and at least one of
them is a write (\eg in \textsf{iriw}, the write $a$ to $x$ on \myth{0} and the
read $b$ from $x$ on \myth{2}). A cycle $\sigma \subseteq \transc{({\cmp} \cup
{\po})}$ is critical when it satisfies the following two properties:
\begin{itemize}
\item \textsf{(i)} per thread, there are at most two memory accesses involved
in the cycle on this thread and these accesses have distinct locations, and
\item \textsf{(ii)} for a given memory location $\lo$, there are at most three
accesses relative to $\lo$, and these accesses are from distinct threads
($(w,w') \in \cmp$, $(w,r) \in \cmp$, $(r,w) \in \cmp$ or $\{(r,w),(w,r')\}
\subseteq \cmp$).  
\end{itemize}

All the executions that we give in \mysec\ref{sec:model} show critical cycles,
except for the \DELETED{\textsc{uniproc}}\NEW{\textsc{sc per location}} ones
(\cf\myfig\ref{fig:co}). Indeed a critical cycle has to involve more than one
memory location by definition.

\paragraph{Static critical cycles}
More \DELETED{constructively}\NEW{precisely}, our tool \prog{mole} looks for
cycles which:
\begin{itemize}
\item alternate program order $\po$ and competing accesses $\cmp$, 
\item traverse a thread only once (\cf~\textsf{(i)} above), and
\item involve at most three accesses per memory location (\cf~\textsf{(ii)} above).
\end{itemize}

Observe that the definition above \DELETED{does not always raise the
well-known}\NEW{is not limited to the well-known} 
\DELETED{idioms}\NEW{patterns} that we presented in \mysec\ref{sec:model}. Consider the two executions
\NEW{in \myfig\ref{fig:s}, both of which match the definition of a critical cycle given above.}
\begin{figure}[!h]
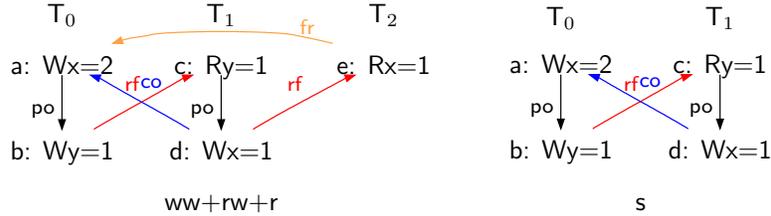

\begin{center}
\input{img/ww+rw+r\bw.pstex_t}\quad
\input{img/s\bw.pstex_t}
\end{center}
\vspace*{-3mm}
\caption{The \DELETED{idiom}\NEW{pattern} \textsf{s} (on the right), and an extended version of it (on the left) \label{fig:s}}
\end{figure}

On the left-hand side, the thread \myth{1} writes $1$ to location $x$ (\cf
event $d$); the thread \myth{2} reads this value from $x$ (\ie $(d,e) \in
\rf$), before the thread \myth{0} writes $2$ to $x$ (\ie $(e,a) \in \fr$). By
definition of $\fr$, this means that the write $d$ of value $1$ into $x$ by
\myth{1} is $\co$-before the write $a$ of value $2$ into $x$ by \myth{0}. This
is reflected by the execution on the right-hand side, where we simply omitted
the reading thread \myth{2}.

Thus to obtain our well-known \DELETED{idioms}\NEW{patterns}, such as the ones in \mysec\ref{sec:model}
and the \textsf{s} \DELETED{idiom}\NEW{pattern} on the right-hand side of \myfig\ref{fig:s}, we
implement the following reduction rules, which we apply to our cycles: 
\begin{itemize}
\item $\co;\co = \co$, which means that we only take the extremities of a chain
of coherence relations;
\item $\rf;\fr = \co$, which means that we omit the intermediate reading thread
in a sequence of read-from and from-read, just like in the \textsf{s} case
above;
\item $\fr;\co = \fr$ which means that we omit the intermediate writing thread
in a sequence of from-read and coherence.
\end{itemize}

We call the resulting cycles \emph{static critical cycles}.
 
Thus \prog{mole} looks for all the static critical cycles that it can find in
the goto-program given as argument. In addition, it also looks for
\DELETED{\textsc{uniproc}}\NEW{\textsc{sc per location}} cycles, \ie
\textsf{coWW, coRW1, coRW2, coWR} and \textsf{coRR} as shown in
\myfig\ref{fig:co}.

In the remainder of this section, we simply write \emph{cycles} for the cycles
gathered by \prog{mole}, \ie static critical cycles and
\DELETED{\textsc{uniproc}}\NEW{\textsc{sc per location}} cycles.

\subsubsection{Static search} 

Looking for \DELETED{idioms}\NEW{patterns} poses several challenges, which are also pervasive
in static data race analysis (\cf~\cite{kys07}):
\begin{itemize}
\item identify program fragments that may be run concurrently, in distinct threads; 
\item identify objects that are shared between these threads.
\end{itemize}

Finding shared objects may be further complicated by the presence of inline
assembly. We find inline assembly in $803$ of the packages to be analysed. At
present, \prog{mole} only interprets a subset of inline assembly deemed
relevant for concurrency, such as memory barriers, but ignores all other inline
assembly.

We can now explain how our \DELETED{idiom}\NEW{pattern} search works. Note that our approach does not
require analysis of whole, linked, programs -- which is essential to achieve
scalability to a code base this large. Our analysis proceeds as follows:
\begin{enumerate}
    \renewcommand{\labelenumi}{\arabic{enumi}.}
    \renewcommand{\theenumi}{\arabic{enumi}.}
\item identify candidate functions that could act as \emph{entry points} for a
  thread being spawned (an entry point is a function such that its first
  instruction will be scheduled for execution when creating a thread);
\item group these candidate entry points, as detailed below, according to shared
  objects accessed -- where we consider an object as \emph{shared} when it
  either has static storage duration (and is not marked thread local), or is referenced
  by a pointer that is shared;
\item assuming concurrent execution of the threads in each such group,
enumerate \DELETED{idioms}\NEW{patterns} using the implementation
from~\cite{akn13} with a flow-insensitive points-to analysis and, in order to
include \DELETED{\textsc{uniproc}}\NEW{\textsc{sc per location}} cycles, weaker
restrictions than when exclusively looking for critical cycles;
\item classify the candidates following their \DELETED{idioms}\NEW{patterns} (\ie using the litmus
naming scheme that we outlined in \myfig\ref{fig:gloss-litmus}) and the axioms
of the model. The categorisation according to axioms proceeds by testing the
sequence of relations occurring in a cycle against the axioms of
\myfig\ref{fig:model}; we detail this step below.
\end{enumerate}

Note that our analysis does not take into account program logic, \eg locks,
that may forbid the execution of a given cycle.  \NEW{If no execution of the
(concurrent) program includes a certain cycle, we call it a \emph{spurious}
cycle, and refer to others as \emph{genuine} cycles.  Note that this definition
is independent of fences or dependencies that may render a cycle forbidden for
a given (weak) memory model.  Our notion of genuine simply accounts for the
feasibility of a concurrent execution.} This overapproximation means that any
numbers of cycles given in this section cannot be taken as a quantitative
analysis of cycles that would be actually executed.

\NEW{With this approach, instead, we focus on not missing cycles rather than
avoiding the detection of spurious cycles. In this sense, the results are best
compared to compiler warnings.  Performing actual proofs of cycles being either
spurious or genuine is an undecidable problem. In principle we could thus only
do so in a best-effort manner, akin to all software verification tools aiming
at precise results.  In actual practice, however, the concurrent reachability
problem to be solved for each such cycle will be a formidable challenge for
current software verification tools, including several additional technical
difficulties (such as devising complex data structures as input values) as we
are looking at real-world software rather than stylised benchmarks.  With more
efficient tools such as the one of~\cite{akt13} we hope to improve on this
situation in future work, since with the tool of~\cite{akt13} we managed to
verify selected real-world concurrent systems code for the first time.}

We now explain these steps in further detail, and use the Linux
Read-Copy-Update (RCU) code~\cite{rcu:lwn} as an example.
\myfig\ref{fig:rcu-example-code} shows a code snippet, which was part of the
benchmarks that we used in~\cite{akt13}, employing RCU. The original code
contains several macros, which were expanded using the C preprocessor.

\begin{figure*}[!thp]
\begin{verbatim}
01 struct foo *gbl_foo;
02 
03 struct foo foo1, foo2;
04 
05 spinlock_t foo_mutex = (spinlock_t) { { .rlock = { .raw_lock = { 0 }, } } };
06 
06 void* foo_update_a(void* new_a)
07 {
08   struct foo *new_fp;
09   struct foo *old_fp;
10 
11   foo2.a=100;
12   new_fp = &foo2;
13   spin_lock(&foo_mutex);
14   old_fp = gbl_foo;
15   *new_fp = *old_fp;
16   new_fp->a = *(int*)new_a;
16 
17   ({ __asm__ __volatile__ ("lwsync" " " : : :"memory");
18      ((gbl_foo)) = (typeof(*(new_fp)) *)((new_fp)); });
19 
20   spin_unlock(&foo_mutex);
21   synchronize_rcu();
22   return 0;
23 }
24 
25 void* foo_get_a(void* ret)
26 {
26   int retval;
27   rcu_read_lock();
28   retval = ({ typeof(*(gbl_foo)) *_________p1 = 
29               (typeof(*(gbl_foo))* )(*(volatile typeof((gbl_foo)) *)&((gbl_foo)));
30               do { } while (0); ; do { } while(0);
31               ((typeof(*(gbl_foo)) *)(_________p1)); })->a;
32   rcu_read_unlock();
33   *(int*)ret=retval;
34   return 0;
35 }
36 
36 int main()
37 {
38   foo1.a=1;
39   gbl_foo=&foo1;
40   gbl_foo->a=1;
41 
42   int new_val=2;
43   pthread_create(0, 0, foo_update_a, &new_val);
44   static int a_value=1;
45   pthread_create(0, 0, foo_get_a, &a_value);
46 
46   assert(a_value==1 || a_value==2);
47 }
\end{verbatim}
\vspace*{-4mm}
\caption{Code example from RCU\label{fig:rcu-example-code}}
\end{figure*}

\paragraph{Finding entry points} To obtain an overapproximate set of \DELETED{idioms}\NEW{patterns} even
for, \eg library code, which does not have a defined entry point and thus may be
used in a concurrent context even when the code parts under scrutiny do contain
thread spawn instructions, we consider thread entry points as follows:
\begin{itemize}
\item explicit thread entries via POSIX or kernel thread create functions;
\item \DELETED{if we cannot find any of these,} any set of functions $f_1, \ldots, f_n$,
provided that $f_i$ is not (transitively) called from another function $f_j$ in
this set \NEW{and $f_i$ has external linkage (\cf \cite[\S 5.1.1.1]{c11})}.;
\item for mutually recursive functions an arbitrary function from this set of
  recursive functions.
\end{itemize}
For any function identified as entry point we create $3$ threads, thereby
accounting for multiple concurrent access to shared objects only used by a
single function, but also for cases where one of the called functions is
running in an additional thread.

For RCU, we see several functions (or function calls) in
\myfig\ref{fig:rcu-example-code}:
\texttt{main},
\texttt{foo\_get\_a},
\texttt{foo\_update\_a},
\texttt{spin\_lock},
\texttt{spin\_unlock},
\texttt{synchronize\_rcu},
\texttt{rcu\_read\_lock} and
\texttt{rcu\_read\_unlock}.
If we discard \texttt{main}, we no longer have a defined entry point nor POSIX
thread creation through \texttt{pthread\_create}. In this case, our algorithm
would consider \texttt{foo\_get\_a} and \texttt{foo\_update\_a}\ as the only
potential thread entry points, because all other functions are called from one
of these two, and there is no recursion.

\paragraph{Finding threads' groups} Then we form groups of threads using the
identified thread entry points. We group the functions $f_i$ and $f_j$ if and
only if the set of objects read or
written by $f_i$ or any of the functions (transitively) called by $f_i$ has a
non-empty intersection with the set for $f_j$. Note the transitivity in
this requirement: for functions $f_i$, $f_j$, $f_k$ with $f_i$ and $f_j$
sharing one object, and $f_j$ and $f_k$ sharing another object, all three
functions end up in one group. In general, however, we may obtain several
groups of such threads, which are then analysed individually.

\NEW{When determining shared objects, as noted above, pointer dereferencing has
to be taken into account. This requires the use of points-to analyses, for
which we showed that theoretically they can be sound~\cite{akl11}, even under
weak memory models. In practice, however, pointer arithmetic,
field-sensitivity, and interprocedural operation require a
performance-precision trade-off. In our experiments we use a flow-insensitive,
field-insensitive and interprocedural analysis. We acknowledge that we may thus
still be missing certain cycles due to the incurred incompleteness of the
points-to analysis.}

For RCU, \texttt{main}, \texttt{foo\_get\_a} and \texttt{foo\_update\_a} form a
group, because they jointly access the pointer \texttt{gbl\_foo} as well as the
global objects \texttt{foo1} and \texttt{foo2} through this pointer. Furthermore
\texttt{main} and \texttt{foo\_update\_a} share the local \texttt{new\_val}, and
\texttt{main} and \texttt{foo\_get\_a} share \texttt{a\_value}\NEW{, both of
which are communicated via a pointer}\DELETED{ (which has
static storage duration, but is communicated via a pointer)}.

\paragraph{Finding \DELETED{idioms}\NEW{patterns}} With the thread groups established, we enumerate
\DELETED{idioms}\NEW{patterns} as in~\cite{akn13}. We briefly recall this step here for completeness: 
\begin{itemize}
\item we first construct one control-flow graph (CFG) per thread;
\item then we add communication edges between shared memory accesses to the
same object, if at least one of these objects is a write (this is the $\cmp$
relation in the definition of critical cycles given at the beginning of this
section);
\item we enumerate all cycles amongst the CFGs and communication edges using
Tarjan's 1973 algorithm~\cite{DBLP:journals/siamcomp/Tarjan73}, resulting in a
set that also contains all critical cycles (but possibly more);
\item as final step we filter the set of cycles for those that satisfy the
conditions of static critical cycles or
\DELETED{\textsc{uniproc}}\NEW{\textsc{sc per location}} as described above.
\end{itemize}


Let us explain how we \NEW{may} find the \textsf{mp}
\DELETED{idiom}\NEW{pattern} (\cf~\myfig\ref{fig:mp} in \mysec\ref{sec:model})
in RCU.  The writing thread \myth{0} is given by the function
\texttt{foo\_update\_a}, the reading thread \myth{1} by the function
\texttt{foo\_get\_a}. Now for the code of the writing thread \myth{0}: in
\texttt{foo\_update\_a}, we write \texttt{foo2} at line~$11$, then we have an
$\lwsync$ at line~$17$, and a write to \texttt{gbl\_foo} at
line~$18$.\label{rcu:mp} 

For the code of the reading thread \myth{1}: the function \texttt{foo\_get\_a}
first copies the value of \texttt{gbl\_foo} at line~$29$ into the local
variable \texttt{\_\_\_\_\_\_\_\_\_p1}. Now, note that the write to
\texttt{gbl\_foo} at line~$18$ made \texttt{gbl\_foo} point to \texttt{foo2},
due to the assignment to \texttt{new\_fp} at line~$12$.

Thus dereferencing \texttt{\_\_\_\_\_\_\_\_\_p1} at line~$31$ causes a read
of the object \texttt{foo2} at that line. Observe that the dereferencing
introduces an address dependency between the read of \texttt{gbl\_foo} and the
read of \texttt{foo2} on \myth{1}.

\paragraph{Categorisation of cycles} As we said above, for each cycle that we
find, we apply a categorisation according to the axioms of our model (\cf
\myfig\ref{fig:model}). For the purpose of this categorisation we instantiate
our model for SC (\cf \myfig\ref{fig:sc+tso}). We use the sequence of relations
in a given cycle: for example for \textsf{mp}, this sequence is
$\lwsync;\rfe;\dep;\efr$.  We first test if the cycle is a
\DELETED{\textsc{uniproc}}\NEW{\textsc{sc per location}} cycle: we check if all
the relations in our input sequence are either $\poloc$ or $\com$. If, as for
\textsf{mp}, this is not the case, we proceed with the test for \textsc{no thin
air}. Here we check if all the relations in our sequence match $\hb$, \ie $\po
\cup \fences \cup \rfe$. As \textsf{mp} includes an $\efr$, the cycle cannot be
categorised as \textsc{no thin air}, and we proceed to
\textsc{\DELETED{causality}\NEW{observation}}.  Starting from $\efr$ we find
$\lwsync \in \prop$ (as $\prop = \po \cup \fences \cup \rf \cup \fr$ on SC),
and $\rfe;\dep \in \rstar{\hb}$. Thus we categorise the cycle as as a
\DELETED{causality}\NEW{observation} cycle. In the general case, we check for
\textsc{propagation} last.
 


\paragraph{Litmus tests:} we exercised \prog{mole} on the set of litmus tests that
we used for exercising \prog{CBMC} (\cf~\mysec\ref{sec:verif}). For each test
we find its general \DELETED{idiom}\NEW{pattern} (using the naming scheme that we presented
in~\mysec\ref{sec:model}): for example for \textsf{mp} we find the cycle
$\po;\rfe;\po;\efr$. Note that our search looks for memory barriers but does
not try to look for dependencies; this means that for the variant
\textsf{mp+lwfence+addr} of the \textsf{mp} \DELETED{idiom}\NEW{pattern}, we find the cycle
$\lwfence;\rfe;\po;\efr$, where the barrier appears but not the dependency.

\paragraph{Examples that we had studied manually before \!\!\!\!}
(see~\cite{akn13,akt13}) include Apache, PostgreSQL and RCU, as mentioned
in~\mysec\ref{sec:verif}. We analysed these examples with \prog{mole} to
confirm the \DELETED{idioms}\NEW{patterns} that we had found before.\footnote{In the following we
mention \DELETED{idioms}\NEW{patterns} that we have not displayed in the paper, and which do not follow
the convention outlined in \myfig\ref{fig:gloss-litmus}: \textsf{z6.[0-5]},
\textsf{3.2w}, or \textsf{irrwiw}. For the sake of brevity, we do not show them
in the paper, and refer the reader to the companion webpage:
\url{http://diy.inria.fr/doc/gen.html\#naming}.}

In Apache we find $5$ \DELETED{idioms}\NEW{patterns} distributed over $75$ cycles: $4 \times
\textsf{mp}$ (\cf\myfig\ref{fig:mp}); $1 \times \textsf{s}$
(\cf\myfig\ref{fig:s}); $28\times \textsf{coRW2}$, $25\times \textsf{coWR}$,
and $17\times \textsf{coRW1}$ (\cf\myfig\ref{fig:co}).

In PostgreSQL, we find $22$ different \DELETED{idioms}\NEW{patterns} distributed over $463$ cycles. We
give the details in \mytab\ref{fig:pgsql-cycles}.
\begin{table}[!h]
\begin{center}
\begin{tabular}{c|c}
\DELETED{idiom}\NEW{pattern} & \# cycles \\\hline
$\textsf{r}$ (\cf\myfig\ref{fig:r}) & $93$\\
$\textsf{w+rr+2w}$ & $68$ \\
$\textsf{w+rr+wr+ww}$ & $62$ \\
$\textsf{z6.4}$ & $54$ \\
$\textsf{sb}$ (\cf\myfig\ref{fig:sb}) & $37$ \\
$\textsf{2+2w}$ (\cf\myfig\ref{fig:2+2w}) & $25$ \\
$\textsf{w+rwc}$ (\cf\myfig\ref{fig:w+rwc}) & $23$ \\
$\textsf{mp}$ (\cf\myfig\ref{fig:mp}) & $16$ \\
$\textsf{w+rw}$ & $14$ \\
$\textsf{s}$ (\cf\myfig\ref{fig:s}) & $14$ \\
$\textsf{z6.5}$  & $6$ \\
$\textsf{w+rw+wr}$  & $6$ \\
$\textsf{w+rw+2w}$  & $4$ \\
$\textsf{z6.0}$  & $2$ \\
$\textsf{wrc}$ (\cf\myfig\ref{fig:wrc})  & $2$ \\
$\textsf{lb}$ (\cf\myfig\ref{fig:lb})  & $2$ \\
$\textsf{irrwiw}$  & $2$ \\
$\textsf{coWR}$ (\cf\myfig\ref{fig:co}) & $19$ \\
$\textsf{coWW}$ (\cf\myfig\ref{fig:co})  & $6$ \\
$\textsf{coRW1}$ (\cf\myfig\ref{fig:co})  & $4$ \\
$\textsf{coRW2}$ (\cf\myfig\ref{fig:co})  & $4$ \\
\end{tabular}
\end{center}
\caption{\DELETED{Idioms}\NEW{Patterns} in PostgreSQL \label{fig:pgsql-cycles}}
\end{table}


In RCU we find $9$ \DELETED{idioms}\NEW{patterns} in $23$ critical cycles, as
well as one \DELETED{\textsc{uniproc}}\NEW{\textsc{sc per location}} cycle.  We
give the details in \mytab\ref{fig:rcu-cycles}. For each
\DELETED{idiom}\NEW{pattern} we give \DELETED{an}\NEW{one} example cycle: we
refer to the excerpt in \myfig\ref{fig:rcu-example-code} to give the memory
locations and line numbers\footnote{Lines $1$ and $3$ result from
initialisation of objects with static storage duration, as prescribed by the
C11 standard~\cite[\S 6.7.9]{c11}.} it involves.
\NEW{Note that we list an additional example of \textsf{mp} in the table,
different from the one explained above.}
\begin{table}[!h]
\begin{center}
\begin{tabular}{c|c|c|c}
\DELETED{idiom}\NEW{pattern} & \# cycles & memory locations & line numbers \\\hline
\textsf{2+2w} (\cf\myfig\ref{fig:2+2w}) & $6$ &
  \texttt{foo2}, \texttt{gbl\_foo} & $1$, $3$, $16$\\
\textsf{3.2w} & $4$ & 
  \texttt{foo1}, \texttt{foo2}, \texttt{gbl\_foo} & $3$, $16$, $39$, $40$\\
\textsf{w+rr+ww+ww} & $3$ & 
  \texttt{foo1}, \texttt{foo2}, \texttt{gbl\_foo} & $3$, $14$, $15$, $16$, $39$\\
\textsf{z6.5} & $2$ &
  \texttt{foo1}, \texttt{foo2}, \texttt{gbl\_foo} & $3$, $11$, $14$, $39$, $40$\\
\textsf{r} (\cf\myfig\ref{fig:r}) & $2$ &
  \texttt{foo1}, \texttt{foo2} & $3$, $11$, $15$\\
\textsf{mp} (\cf\myfig\ref{fig:mp}) & $2$ &
  \texttt{foo1}, \texttt{gbl\_foo} & $14$, $15$, $38$, $39$\\
\textsf{w+rr+ww+wr} & $2$ & 
  \texttt{foo1}, \texttt{foo2}, \texttt{gbl\_foo} & $3$, $11$, $14$, $29$, $31$, $39$\\
\textsf{z6.3} & $1$ &
  \texttt{foo1}, \texttt{foo2}, \texttt{gbl\_foo} & $3$, $16$, $29$, $31$\\
\textsf{w+rr+2w} & $1$ &
  \texttt{foo1}, \texttt{gbl\_foo} & $29$, $31$, $38$, $39$\\
\textsf{coWW} (\cf\myfig\ref{fig:co}) & $1$ & \texttt{foo2} & $15$ $16$
\end{tabular}
\end{center}
\caption{\DELETED{Idioms}\NEW{Patterns} in RCU\label{fig:rcu-cycles}}
\end{table}

\subsection{Results for Debian 7.1}

We report on the results of running \prog{mole} on
\DELETED{$37\,724$}\NEW{$137\,163$} object files generated while compiling
source packages using \prog{goto-cc}, in \DELETED{$431$}\NEW{$1\,251$} source
packages of the Debian Linux distribution, release 7.1. \DELETED{We intend to
explore the entirety of the distribution; at the time of submission, we have
explored $27\%$ of it.}

We provide all the files compiled with \prog{goto-cc} at
\url{http://diy.inria.fr/mole}, and present our experimental data (\ie the
\DELETED{idioms}\NEW{patterns} per packages) at \url{http://diy.inria.fr/mole}\NEW{.}\DELETED{, and we
will update the web pages as new results become available.} \fixmeskip{jade: url
more detailed x2}

Our experiment runs on a system equipped with $8$ cores and $64$ GB of main
memory. In our setup, we set the time and memory bounds for each object file
subject to static cycle search to $15$ minutes and $16$ GB of RAM. \DELETED{To date, we
spent $53.7$ CPU hours in successful cycle search, but an additional
$6\,866$}\NEW{We spent more than $199$ CPU days in cycle search, yet $19\,930$}
runs did not finish within the above time and memory limits. More than $50\%$ of time are
spent within cycle enumeration for a given graph of CFGs and communication
edges, whereas only $12\%$ are spent in the points-to analysis. The remaining
time is consumed in generating the graph. \NEW{The resulting $79$ GB of raw data
were further processed to determine the results presented below.}

\subsubsection{General results}
We give here some general overview of our experiments.  We detected a total of
\DELETED{$15\,683\,669$}\NEW{$86\,206\,201$} critical cycles, plus
\DELETED{$858\,788$}\NEW{$11\,295\,809$}
\DELETED{\textsc{uniproc}}\NEW{\textsc{sc per location}} cycles.  Amongst
these, we find \DELETED{$548$}\NEW{$551$} different
\DELETED{idioms}\NEW{patterns}.  \myfig\ref{fig:top-ten-shapes} gives the
thirty most frequent \DELETED{idioms}\NEW{patterns}. 

\begin{figure}[!h]
  \centering
\scalebox{.7}{  \includegraphics{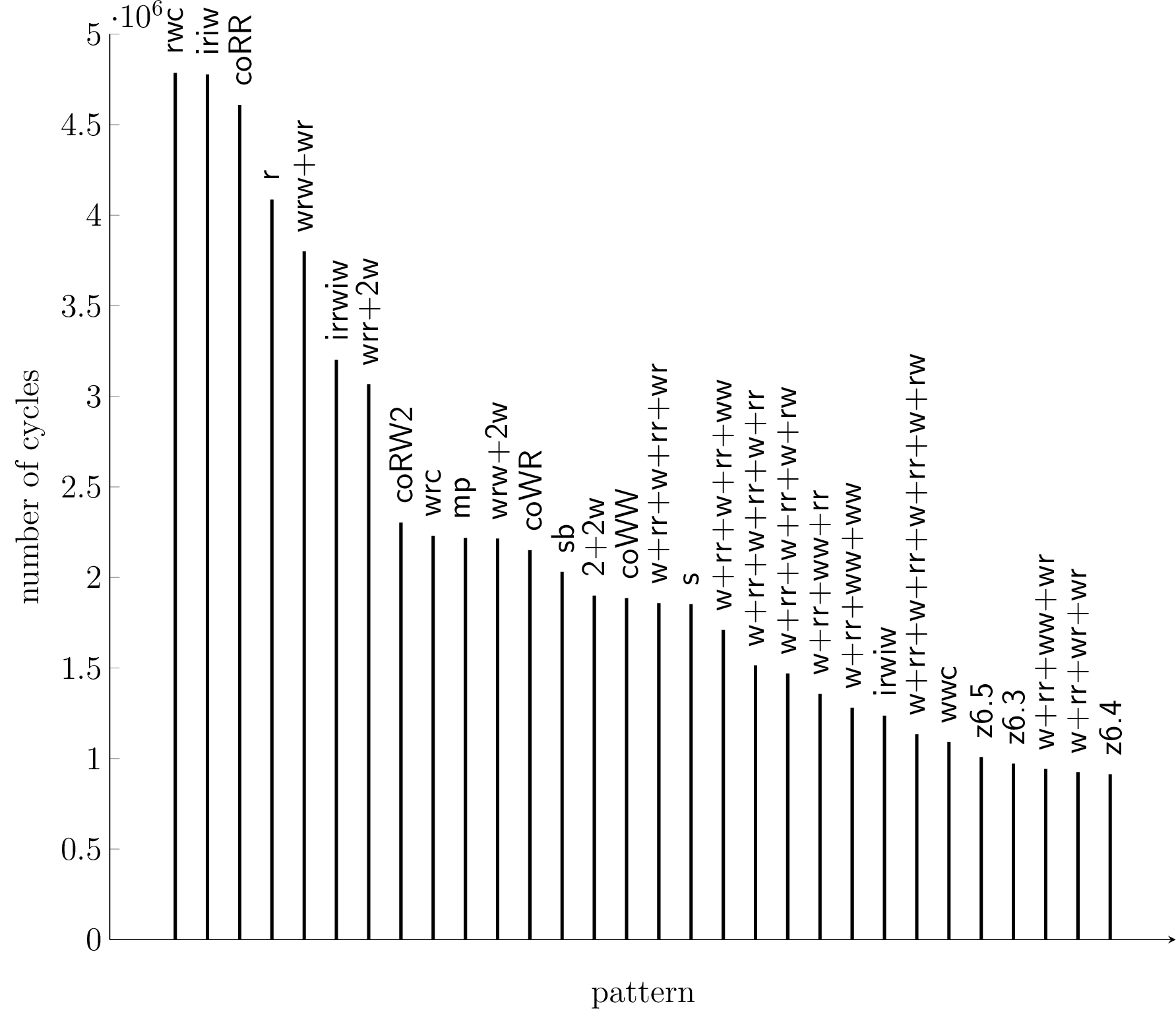}}
  \caption{Thirty most frequent \DELETED{idioms}\NEW{patterns} \UPDATENOTE{figure updated}}
  \label{fig:top-ten-shapes}
\end{figure}

\DELETED{The source package with most cycles is \prog{criticalmass} (a 3D video
game, url{http://packages.debian.org/wheezy/criticalmass},
$1\,337\,299$ cycles) with the most frequently occurring shapes
\textsf{ww+ww+ww+ww+ww} ($406\,082$), \textsf{ww+ww+ww+ww+ww+ww} ($361\,031$)
and \textsf{4.2w} ($289\,013$). The source package with the widest variety of
cycles is \prog{binutils-msp430} (tools to manipulate object files for the
MSP430 architecture, url{http://packages.debian.org/wheezy/binutils-msp430},
$421$ cycles); its most frequent shapes are \textsf{z6.5} ($90\,031$),
\textsf{wrr+2w} ($80\,702$) and \textsf{r} ($77\,227$).}

\NEW{The source package with most cycles is \prog{mlterm} (a multilingual
terminal emulator, \url{http://packages.debian.org/wheezy/mlterm},
$4\,261\,646$ cycles) with the most frequently occurring patterns
\textsf{iriw} ($296\,219$), \textsf{w+rr+w+rr+w+rw} ($279\,528$) and
\textsf{w+rr+w+rw+w+rw} ($218\,061$) The source package with the widest
variety of cycles is \prog{ayttm} (an instant messaging client,
\url{http://packages.debian.org/wheezy/ayttm}, $238$ different patterns); its
most frequent patterns are \textsf{z6.4} ($162\,469$), \textsf{z6.5} ($146\,005$)
and \textsf{r} ($90\,613$).}

We now give an account of what kind of \DELETED{idioms}\NEW{patterns} occur for a given functionality.
By functionality we mean what the package is meant for, \eg web servers
(\prog{httpd}), mail clients (\prog{mail}), video games (\prog{games}) or
system libraries (\prog{libs}). For each functionality, \mytab\ref{fig:func}
gives the number of packages (\eg $11$ for \prog{httpd}), the three most
frequent \DELETED{idioms}\NEW{patterns} within that functionality with their number of occurrences (\eg
$30\,506$ for \textsf{wrr+2w} in \prog{httpd}) and typical packages with the number
of cycles contained in that package (\eg \DELETED{$161\,687$}\NEW{$70\,283$} for \prog{apache2}).

\begin{table}[!h]
\begin{center}
\begin{tabular}{c|p{.5\linewidth}|p{.28\linewidth}}
function & \DELETED{idioms}\NEW{patterns} & typical packages \\\hline
\prog{httpd} ($11$) & \textsf{wrr+2w} ($30\,506$), \textsf{mp} ($27\,618$), \textsf{rwc} ($13\,324$) & \prog{libapache2-mod-perl2} ($120\,869$), \prog{apache2} ($70\,283$), \prog{webfs} ($27\,260$) \\
\prog{mail} ($24$) & \textsf{w+rr+w+rw+ww} ($75\,768$), \textsf{w+rr+w+rr+ww} ($50\,842$), \textsf{w+rr+w+rr+w+rw} ($45\,496$) & \prog{opendkim} ($702\,534$), \prog{citadel} ($337\,492$), \prog{alpine} ($105\,524$) \\
\prog{games} ($57$) & \textsf{2+2w} ($198\,734$), \textsf{r} ($138\,961$), \textsf{w+rr+w+rr+wr} ($134\,066$) & \prog{spring} ($1\,298\,838$), \prog{gcompris} ($559\,905$), \prog{liquidwar} ($257\,093$) \\
\prog{libs} ($266$) & \textsf{iriw} ($468\,053$), \textsf{wrr+2w} ($387\,521$), \textsf{irrwiw} ($375\,836$) & \prog{ecore} ($1\,774\,858$), \prog{libselinux} ($469\,645$), \prog{psqlodbc} ($433\,282$) \\
\end{tabular}
\end{center}
\caption{\DELETED{Idioms}\NEW{Patterns} per functionality\label{fig:func} \UPDATENOTE{table updated}} 
\end{table}

\subsubsection{Summary per axiom}

\mytab\ref{fig:ax} gives a summary of the \DELETED{idioms}\NEW{patterns} we found, organised per axioms of
our model (\cf~\mysec\ref{sec:model}).
\NEW{We chose a classification with respect to SC, \ie we fixed $\prop$ to be
defined as shown in \myfig\ref{fig:sc+tso}.}
For each axiom we also give some
typical examples of packages that feature \DELETED{idioms}\NEW{patterns} relative to this axiom.
\begin{table}[!h]
\begin{center}
\begin{tabular}{c|c|p{.65\linewidth}}
axiom & \# \DELETED{idioms}\NEW{patterns} & typical packages \\\hline
\DELETED{\textsc{uniproc}}\NEW{\textsc{sc per location}} & $11\,295\,809$ & \prog{vips} ($412\,558$), \prog{gauche} ($391\,180$), \prog{python2.7} ($276\,991$) \\
\textsc{no thin air} & $445\,723$ & \prog{vim} ($36\,461$), \prog{python2.6} ($25\,583$), \prog{python2.7} ($16\,213$) \\
\textsc{\DELETED{causality}\NEW{observation}} & $5\,786\,239$ & \prog{mlterm} ($285\,408$), \prog{python2.6} ($183\,761$), \prog{vim} ($159\,319$) \\
\textsc{propagation} & $79\,974\,239$ & \prog{isc-dhcp} ($891\,673$), \prog{cdo} ($889\,532$), \prog{vim} ($878\,289$) \\
\end{tabular}
\end{center}
\caption{\DELETED{Idioms}\NEW{Patterns} per axiom \label{fig:ax} \UPDATENOTE{table updated}}
\end{table}

We now give a summary of the \DELETED{idioms}\NEW{patterns} we found, organised
by axioms.  Several distinct \DELETED{idioms}\NEW{patterns} can correspond to
the same axiom, \eg \textsf{mp}, \textsf{wrc} and \textsf{isa2} all correspond
to the \textsc{\DELETED{causality}\NEW{observation}} axiom
(\cf~\mysec\ref{sec:model}). For the sake of brevity,
we do not list all the \DELETED{$548$}\NEW{$551$}
\DELETED{idioms}\NEW{patterns}. \myfig\ref{fig:pies} gives one pie chart of
\DELETED{idioms}\NEW{patterns} per axiom.  \begin{figure}[!h]
\begin{center}\let\tst\relax \begin{tabular}{c c} \scalebox{.7}{
\tst{\includegraphics{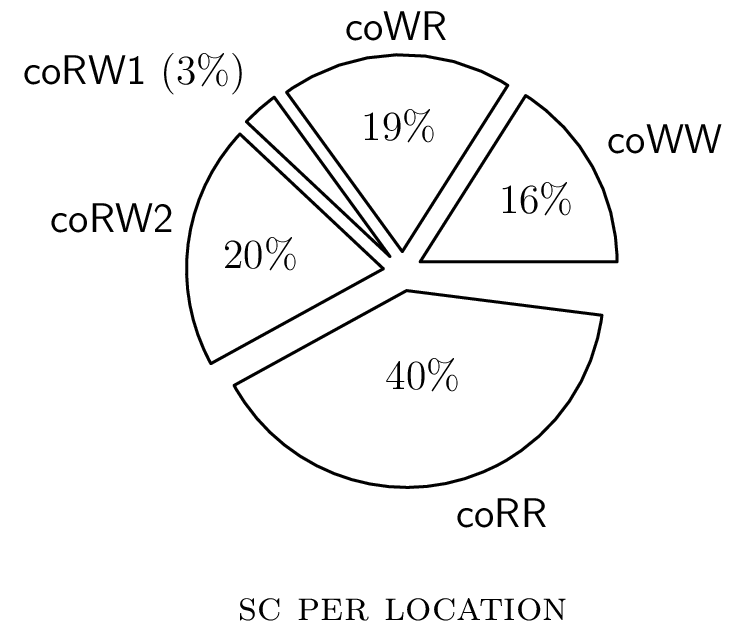}}} & \scalebox{.7}{
\tst{\includegraphics{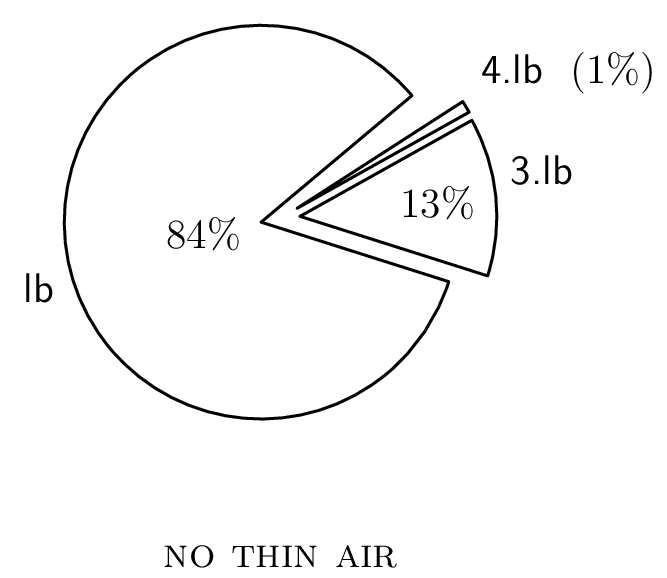}}} \\ \scalebox{.7}{
\tst{\includegraphics{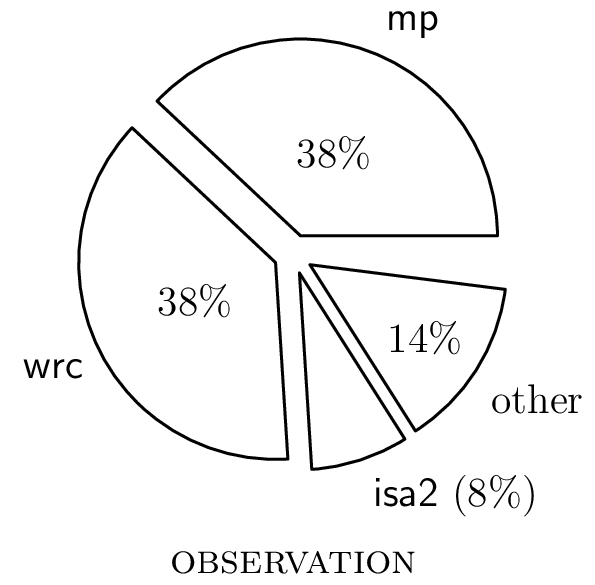}}} & \scalebox{.7}{
\tst{\includegraphics{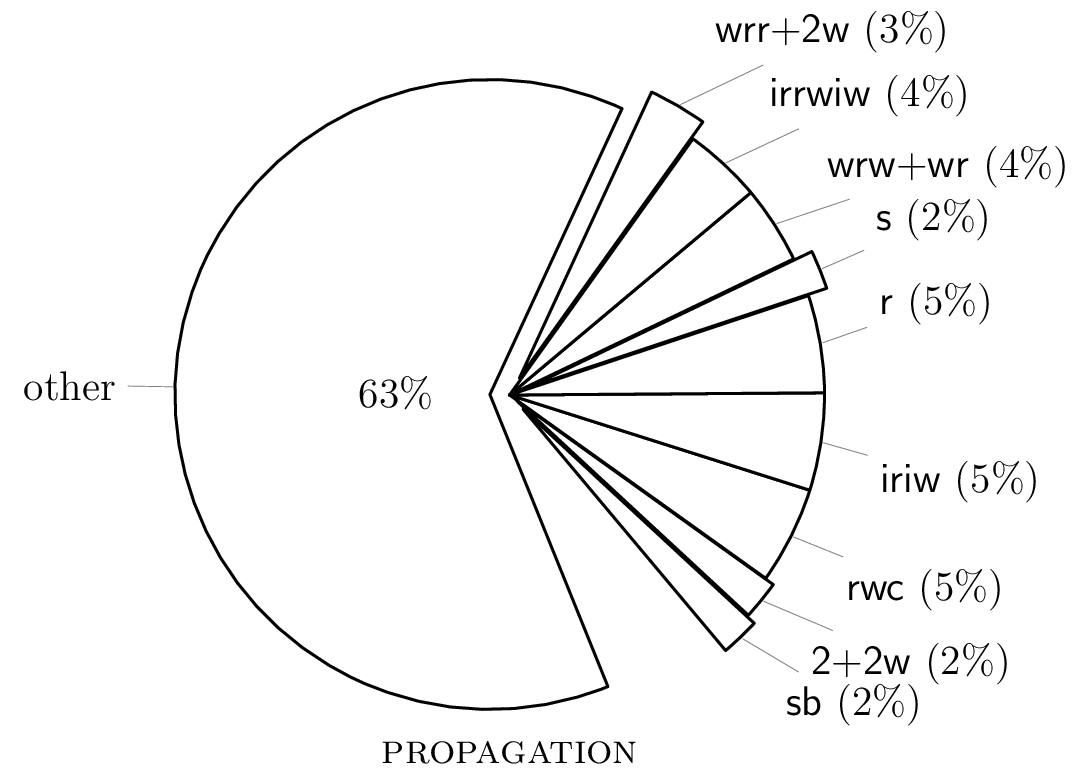}}} \end{tabular} \end{center}
\caption{Proportions of \DELETED{idioms}\NEW{patterns} per axiom
\label{fig:pies} \UPDATENOTE{figure updated}} \end{figure}

\paragraph{Observations} We did find \DELETED{$13\,600$}\NEW{$4\,775\,091$}
occurrences of \textsf{iriw} (\cf\myfig\ref{fig:iriw}), which represents
\DELETED{$0.082\%$}\NEW{$4.897\%$} of all the \DELETED{idioms}\NEW{static
cycles} that we have found. \DELETED{This probably confirms the folklore about
\prog{iriw} being very rarely used.}\NEW{As such it is the second most frequent
pattern detected, which appears to invalidate the folklore claim that
\textsf{iriw} is rarely used. It should be noted, however, that these static
cycles need not correspond
to genuine ones that are actually observed in executions, as discussed further
below.}

We found \DELETED{$1089236$}\NEW{$4\,083\,639$} occurrences of \textsf{r}
(\cf\myfig\ref{fig:r}), which represents \DELETED{$6.584\%$}\NEW{$4.188\%$} of
the \DELETED{idioms}\NEW{static cycles} that we have found. Observe that \textsf{r} appears in PostgreSQL
(\cf\mytab\ref{fig:pgsql-cycles}) and RCU (\cf\mytab\ref{fig:rcu-cycles}), as
well as in the thirty most frequent patterns (\cf\myfig\ref{fig:top-ten-shapes}).
This seems to suggest that a good model needs to handle this \DELETED{idiom}\NEW{pattern} properly.

We also found \DELETED{$924$}\NEW{$4\,606\,915$} occurrences of \textsf{coRR},
which corresponds to the acknowledged ARM bug that we presented
in~\mysec\ref{sec:testing}. This represents \DELETED{$0.005\%$}\NEW{$4.725\%$}
of all the \DELETED{idioms}\NEW{static cycles} that we have found.
Additionally, we found \DELETED{$20151$}\NEW{$2\,300\,724$} occurrences of
\textsf{coRW2}, which corresponds to the \DELETED{striking}
violation\DELETED{s} of \DELETED{\textsc{uniproc}}\NEW{\textsc{sc per
location}}\NEW{,} \DELETED{that we} observed on ARM machines
\DELETED{(\cf}\NEW{that we show in }\myfig\ref{fig:arm-corw-bugs}\DELETED{)}.
This represents \DELETED{$0.121\%$}\NEW{$2.360\%$} of all the
\DELETED{idioms}\NEW{static cycles} that we have found. These two percentages
perhaps \DELETED{temperate}\NEW{nuance} the severity of the ARM anomalies
that we expose in \mysec\ref{sec:testing}. 

We believe that our experiments with \prog{mole} provide results that could be
used by programmers or static analysis tools to identify where weak memory may
come into play and ensure that it does not introduce unexpected behaviours.
Moreover, we think that the data that \prog{mole} gathers can be useful to 
both hardware designers and software programmers.

\begin{new}While we do provide quantitative data, we would like to stress that, at
present, we have little information on how many of the detected cycles are
actually genuine.  Many of the considered cycles may be spurious, either
because of additional synchronisation mechanisms such as locks, or simply
because the considered program fragments are not executed concurrently in any
concrete execution.  Thus, as said above, at present the results are best
understood as warnings similar to those emitted by compilers.
In future work we will both work towards the detection of
spurious cycles, but also aim at studying particular software design patterns
that may give rise to the most frequently observed patterns of our models.

We nevertheless performed manual spot tests of arbitrarily selected
cycles in the packages \prog{4store}, \prog{acct} and \prog{acedb}.
For instance, the \textsc{sc per location} patterns in the package
\prog{acct} appear genuine,
because the involved
functions could well be called concurrently as they belong to a memory
allocation library.
An analysis looking at the entire application at once would be required to
determine whether this is the case. For other cases, however, it may not at all
be possible to rule out such concurrent operations: libraries, for instance,
may be used by arbitrary code. In those cases only locks (or other equivalent
mutual exclusion primitives) would guarantee data-race free (and thus
weak-memory insensitive) operation. The same rationale applies for other
examples that we looked at: while our static analysis considers this case in
order to achieve the required safe overapproximation, the code involved in some
of the \textsf{iriw} cycles in the package \prog{acedb} or \prog{4store} is not
obviously executed concurrently at present. Consequently these examples of
\textsf{iriw} might be spurious, but we note that no locks or fences are in
place to guarantee this.

For the RCU and PostgreSQL examples presented in this paper we use harnesses
that perform concurrent execution. For RCU this mimics the intended usage
scenario of RCU (concurrent readers and writers),
and in the case of PostgreSQL this was modelled after a regression
test\footnote{\NEW{See the attachment at}
\url{http://www.postgresql.org/message-id/24241.1312739269@sss.pgh.pa.us}}
built by PostgreSQL's developers.
Consequently we are able to tell apart genuine and spurious cycles in those
cases.

For PostgreSQL, the \textsc{sc per location} patterns (\textsf{coWW},
\textsf{coWR}, \textsf{coRW1}, \textsf{coRW2}, listed in
\mytab\ref{fig:pgsql-cycles}) and the critical cycles described in detail
in~\cite{akn13} are genuine: one instance of \textsf{lb} (amongst the $2$
listed in \mytab\ref{fig:pgsql-cycles}) and one instance of \textsf{mp}
(amongst the $16$ listed). 

For RCU, the \textsf{coWW} cycle liste in \mytab\ref{fig:rcu-cycles} and the
instance of \textsf{mp} described above (see top of page \pageref{rcu:mp}) are
genuine -- but note that the latter contains synchronisation using $\lwsync$,
which means that the cycle is forbidden on Power.  

All other cycles are spurious, because the cycle enumeration based on Tarjan's
1973 algorithm does not take into account
the ordering of events implied by spawning threads. For example, we report
instances of \textsf{mp} in RCU over lines $38$ and $39$ in function
\texttt{main} as first thread, and lines $14$ and $15$ in function
\texttt{foo\_update\_a} as second, and seemingly concurrent, thread. As that
second thread, however, is only spawned after execution of lines $38$ and $39$,
no such concurrency is possible.  \end{new}

\section{Conclusion}\label{sec:ccl}

To close this paper, we recapitulate the criteria that we listed in the
introduction, and explain how we address each of them.

\paragraph{\textsf{Stylistic proximity of models}: } the framework that we presented
embraces a wide variety of hardware models, including SC, x86-TSO, Power and
ARM. We also explained how to instantiate our framework to produce a
significant fragment of the C++ memory model, and we leave the definition of
the complete model (in particular including consume atomics) for future work.

\paragraph{\textsf{Concision \!\!\!}} is demonstrated
in~\myfig\ref{fig:herd-ppc-model}, which contains the \emph{unabridged}
specification of our Power model. 

\paragraph{\textsf{Efficient simulation and verification}: } the performance of our
tools, our new simulator \prog{herd}
(\cf~\myfig\ref{fig:simulation-tools-comparison}), and the bounded
model-checker \prog{CBMC} adapted to our new Power model
(\cf~\myfig\ref{fig:verif-comparison-litmus}),
confirm (following~\cite{mms12,akt13}) that the modelling style is crucial.

\paragraph{\textsf{Soundness \wrt hardware}: } to the best of our knowledge, our Power
and ARM models are to this date not invalidated by hardware, except for the
$\narmreluctant$ surprising ARM behaviours detailed in \mysec\ref{sec:testing}.
Moreover, we keep on running experiments regularly, which we record at
\url{http://diy.inria.fr/cats}.  

\paragraph{\textsf{Adaptability of the model \!\!\!}} was demonstrated by the ease
with which we were able to modify our model to reflect the subtle
\textsf{mp+dmb+fri-rfi-ctrlisb} behaviour (\cf~\mysec\ref{sec:testing}). 

\paragraph{\textsf{Architectural intent}: } to the best of our knowledge, our Power
model does not contradict the architectural intent, in that we build on the
model of~\cite{ssa11}, which should reflect said intent, and that we have
regular contacts with hardware designers.

For ARM, we model the \textsf{mp+dmb+fri-rfi-ctrlisb} behaviour which is
claimed to be intended by ARM designers. 

\paragraph{\textsf{Account for what programmers do}: } with our new tool \prog{mole}, we
explored the C code base of the Debian Linux distribution version 7.1
(about $200$ millions lines of code) to collect statistics of concurrency
\DELETED{idioms}\NEW{patterns} occurring in real world code.  

Just like our experiments on hardware, we keep on running our experiments on
Debian regularly; we record them at \url{http://diy.inria.fr/mole}.  \fixmeskip{jade: url more detailed} 

\medskip
\centerline{***}

As future work, on the modelling side, we will integrate the semantics of the \textsf{lwarx} and
\textsf{stwcx} Power instructions (and their ARM equivalents \textsf{ldrex} and
\textsf{strex}), which are used to implement locking or compare-and-swap
primitives. We expect their integration to be relatively straight forward: the
model of~\cite{smo12} uses the concept of a write reaching coherence point to
describe them, a notion that we have in our model as well.

On the rationalist side, it remains to be seen if our model is well-suited for
proofs of programs: we regard our experiments \wrt verification of programs as
preliminary.

 \begin{acks}
We thank Nikos Gorogiannis for suggesting that the extension of the input files
for \prog{herd} should be \prog{.cat}. We thank Carsten Fuhs (even more so
since we forgot to thank him in~\cite{akt13}) and Matthew Hague for their
patient and careful comments on a draft. \NEW{We thank Mark Batty, Viktor
Vafeiadis and Tyler Sorensen for comments on a draft. We thank our reviewers
for their careful reading, comments and suggestions.} We thank Arthur Guillon
for his help with the simulator of~\cite{bps12}.  We thank Susmit Sarkar, Peter
Sewell and Derek Williams for discussions on the Power model(s). Finally, this
paper would not have been the same without the last year of discussions on
related topics with Richard Bornat, Alexey Gotsman, Peter O'Hearn and Matthew
Parkinson.  \end{acks}


\bibliographystyle{ACM-Reference-Format-Journals}
\bibliography{aes,jade}

\end{document}